\documentclass[rmp,aps,epsfig,amssymb,twocolumn]{revtex4}

\usepackage{amsmath}
\usepackage{amstext}
\usepackage{latexsym}
\usepackage[dvips]{graphicx}
\usepackage{psfrag}

\def\be{\begin{eqnarray}}
\def\bea{\begin{eqnarray}}
\def\bma{\begin{mathletters}}
\def\ee{\end{eqnarray}}
\def\eea{\end{eqnarray}}
\def\ema{\end{mathletters}}

\newcommand{\nc}{\newcommand}
\nc{\rnc}{\renewcommand} \nc{\ket}[1]{\left | \, #1 \right
\rangle} \nc{\bra}[1]{\left \langle #1 \, \right |}
\nc{\proj}[1]{\ket{#1}\bra{#1}} \rnc{\vec}{\mathbf}
\nc{\braket}[2]{\langle\, #1\,|\,#2\,\rangle}
\nc{\half}{\frac{1}{2}}
\nc{\vfigure}[3]{
\begin{figure}[th]
\centerline{\psfig{file=#1.eps,width=#2}}
\vspace*{8pt}
\caption{#3}
\end{figure}}
\nc{\vpstexfigure}[3]{
\vfigure{#1}{#2}{#3}}
\nc{\prj}{\mathcal{P}} \nc{\hilb}{\mathcal{H}}
\nc{\pth}{\mathcal{C}} \nc{\inprod}[2]{\braket{#1}{#2}}
\nc{\upket}{\ket{\uparrow}} \nc{\downket}{\ket{\downarrow}}
\nc{\upbra}{\bra{\uparrow}} \nc{\downbra}{\bra{\downarrow}}

\def\CC{{\rm\kern.24em \vrule width.04em height1.46ex depth-.07ex
\kern-.30em C}}
\def\P{{\rm I\kern-.25em P}}
\def\N{{\rm I\kern-.25em N}}
\def\RR{{\rm
         \vrule width.04em height1.58ex depth-.0ex
         \kern-.04em R}}
\def\id{{\rm 1\kern-.26em l}}
\def\ZZ{{\sf Z\kern-.44em Z}}
\def\e{{\rm e}}

\def\eps{\varepsilon}
\def\trace{{\rm tr}\;}

\def\d{{\rm d}}
\def\up{\uparrow}
\def\down{\downarrow}

\newcommand{\Matrix}[2]{\left( \begin{array}{#1} #2 \end{array}
  \right)}
\newcommand{\diag}{{\rm diag}\;}

\newenvironment{eqblock}[2]{\beq\label{#2}\begin{array}{#1}}{\end{array}
                                \eeq}
\newenvironment{neqblock}[1]{\[\begin{array}{#1}}{\end{array}\]}

\newcommand{\beqb}{\begin{eqblock}}
\newcommand{\eeqb}{\end{eqblock}} 
\newcommand{\nbeqb}{\begin{neqblock}}
\newcommand{\neeqb}{\end{neqblock}} 
\newcommand{\bigfrac}[2]{\mbox {${\displaystyle \frac{ #1 }{ #2 }}$}}
\newcommand{\beq}{\begin{equation}}
\newcommand{\beqa}{\begin{eqnarray}}
\newcommand{\eeq}{\end{equation}}
\newcommand{\eeqa}{\end{eqnarray}}
\newcommand{\nbeqa}{\begin{eqnarray*}}
\newcommand{\neeqa}{\end{eqnarray*}}
\newcommand{\Expect}[3]{\bra{#1} #2 \ket{#3}}
\newcommand{\expect}[1]{\left\langle #1 \right\rangle}

\begin{document}
\title{Entanglement in Many-Body Systems}
\author{Luigi Amico}
\affiliation{MATIS-CNR-INFM $\&$ Dipartimento di Metodologie Fisiche e
    		Chimiche (DMFCI),\\ viale A. Doria 6, 95125 Catania, Italy}
\author{Rosario Fazio}
\affiliation{International School for Advanced Studies (SISSA)
        	via  Beirut 2-4,  I-34014 Trieste, Italy\\
		and\\
		NEST-CNR-INFM $\&$ Scuola Normale Superiore, I-56126 Pisa, Italy}
\author{Andreas Osterloh}
\affiliation{Institut f\"ur Theoretische Physik, Leibniz Universit\"at Hannover, 30167
		Hannover, Germany}
\author{Vlatko Vedral}
\affiliation{The School of Physics and Astronomy, University of Leeds, Leeds 
		LS29JT, United Kingdom \\
and Center for Quantum Technologies, National University of Singapore, 3 Science Drive 2, Singapore 117543}


\begin{abstract}
The recent interest in aspects common to quantum information and condensed 
matter has prompted a flory of activity at the border of these disciplines 
that were far distant untill few years ago. Numerous interesting 
questions have been addressed so far. Here we review an important part of this 
field, the properties of the entanglement in many-body systems. We discuss the 
zero and finite temperature properties of entanglement in interacting spin, 
fermion and boson model systems. Both bipartite and multipartite entanglement 
will be considered.  In equilibrium we show how entanglement is tightly connected 
to the characteristics of the phase diagram. The behavior of entanglement can 
be related, via certain witnesses, to thermodynamic quantities thus offering 
interesting possibilities for an experimental test. Out of equilibrium we discuss 
how to generate and manipulate entangled states by means of many-body Hamiltonians. 

\end{abstract}


\maketitle

\tableofcontents


\section{Introduction}
\label{intro}
Entanglement expresses the ``spooky'' non-locality inherent to quantum mechanics~\cite{Bell87}.
Because of that, it gave rise to severe skepticisms since the early days of 
quantum mechanics.
It was only after the seminal contribution of John Bell that the fundamental 
questions related to the existence of entangled states could 
be tested experimentally. In fact, under fairly general assumptions, Bell derived a set 
of inequalities for correlated measurements of two physical observables that any
local theory should obey.  The overwhelming majority of experiments done so far 
are in favor of quantum mechanics thus demonstrating that quantum entanglement 
is physical reality~\cite{Peres93}.~\footnote{There are states that do not 
violate Bell inequalities and nevertheless are entangled~\cite{Methot06}.}

Entanglement has gained renewed interest with the development of quantum 
information science~\cite{nielsen}. In its framework, quantum entanglement 
is viewed at as a precious resource in quantum information processing. 
It is e.g. believed to be the main ingredient of the quantum speed-up in 
quantum computation and communication. Moreover several quantum protocols, as 
teleportation~\cite{Bennett93} just to mention an important example, can be 
realized exclusively with the help of entangled states.

The role of entanglement as a resource in quantum information has stimulated intensive 
research trying to unveil both its qualitative and quantitative 
aspects~\cite{Bruss01,bengtsson,Th-Eisert,Wootters01,Plenio98,Plenio07,Vedral02,horodecki07}. 
To this end, necessary criteria for any entanglement measure to be fulfilled 
have been elaborated and lead to the notion of an entanglement 
monotone~\cite{MONOTONES} allowing to attach a precise number to the entanglement 
encoded in a given state. There is a substantial bulk of work for bipartite systems, 
in particular for the case of qubits. Many criteria have been proposed to 
distinguish separable from entangled pure states, as for example 
the Schmidt rank and the von Neumann entropy. The success in the bipartite case 
for qubits asked for extensions to the multipartite case, but 
the situation proved to be far more complicated: different classes 
of entanglement occur, which are inequivalent not only under deterministic local 
operations and classical communication, but even under their 
stochastic analogue~\cite{SLOCC}. 

In the last few years it has become evident that quantum information may lead 
to further insight into other areas of physics as statistical mechanics and quantum 
field theory~\cite{preskill00}. 
The attention of the quantum information community to systems intensively 
studied in condensed matter has stimulated an exciting cross-fertilization 
between the two areas. 
Methods developed in quantum information have proved to be extremely 
useful in the analysis of the state of many-body systems. 
At $T=0$ many-body systems are most often described by complex ground state wave function 
which contain all the correlations that  give rise to the various phases 
of matter (superconductivity, ferromagnetism, quantum hall systems, $\ldots$). 
Traditionally many-body systems have been studied by looking for example at their response 
to external perturbations, various order parameters and excitation spectrum. 
The study of the ground state of many-body systems with methods developed in 
quantum information may unveil new properties.  
At the same time experience built up over the years in condensed matter is 
helping in finding new protocols for quantum computation and communication: 
A quantum computer is  a many-body system where, differently from 
``traditional ones'', the Hamiltonian can be controlled and manipulated.

The amount of work at the interface between statistical mechanics and quantum 
information has grown tremendously during the last few years,
shining light on many different aspects of both subjects.
In particular, there has been an extensive analysis of entanglement in 
quantum critical models~\cite{OstNat,Osborne02,GVidal03}. 
Tools from quantum information theory also provided important support for
numerical methods, as the density matrix renormalization group or the 
design of new efficient simulation strategies for many-body systems 
(see for example~\onlinecite{dmrg1,dmrg2,dmrg3}). 
Spin networks have been proposed as quantum channels~\cite{bose03} by 
exploiting the collective dynamics of their low lying excitations for 
transporting quantum information. 

Despite being at its infancy, this new area of research has grown so fast 
that a description of the whole field is beyond the scope of a single review.
Many interesting facets of this branch of research will therefore remain 
untouched here. In this review we will only discuss the properties of entanglement
in many-body systems. The models which will be considered include interacting 
spin networks, itinerant fermions, harmonic and bosonic systems. 
All of them are of paramount interest in condensed matter physics. 

This review is organized as follows. In the next Section we give a brief 
overview on the concepts and measures of entanglement,
with particular attention to those measures that we will use later on. 
In Section~\ref{mod} we then proceed with a brief introduction to 
several models of interacting many-body systems which will be 
subject of the review. 
We will discuss various aspects of quantum correlations 
starting from the pairwise entanglement, Section~\ref{pairgr}, we then 
proceed with the properties of block entropy, Section~\ref{block-entropy} , 
and localizable entanglement, Section~\ref{local}. 
In these three Sections it is especially relevant the connection between 
entanglement and quantum phase transitions.
The effect of a finite temperature is considered in Section~\ref{th}. The 
characterization of entanglement in many-body systems requires also the 
understanding of multipartite entanglement. This topic  will be 
reviewed in Section~\ref{multip}. From the point of view of quantum information 
processing, dynamical properties of entanglement are important as well. They 
will be addressed in Section~\ref{dyn}.
The conclusions, the outlook and a very short panorama of what is left out 
from this review are presented in the concluding Section.

\section{Measures of entanglement}
\label{meas}

The problem of measuring entanglement is a vast and lively field of 
research in its own. Numerous different methods have been proposed for its quantification.
In this Section we do not attempt to give an exhaustive
review of the field.
Rather we want to introduce those measures that are largely being used to 
quantify entanglement in many-body systems. Comprehensive overviews of 
entanglement measures can be found in~\cite{bengtsson,Bruss01,Th-Eisert,
horodecki07,Plenio98,Plenio07,Vedral02,Wootters01}
In this context, we also outline a method of detecting 
entanglement, based on entanglement witnesses.

\subsection{Bipartite entanglement in pure states}
\label{measure-pairwise-pure}

Bipartite entanglement of pure states is conceptually well understood, 
although quantifying it for local dimensions higher than two still bears 
theoretical challenges~\cite{ORDERING,horodecki07}. 
A pure bipartite state is not entangled if and only 
if it can be written as a tensor product of pure states of the parts.
Moreover for every pure bipartite 
state $\ket{\psi_{AB}}$ (with the two parts, $A$ and $B$),
two orthonormal bases $\{\ket{\psi_{A,i}}\}$ and $\{\ket{\phi_{B,j}}\}$ exist
such that $\ket{\psi_{AB}}$ can be written as
\beq
\ket{\psi_{AB}}=\sum_{i}\alpha_{i}\ket{\psi_{A,i}}\ket{\phi_{B,i}}
\label{Schmidt}
\eeq
where $\alpha_i$ are positive coefficients.
This decomposition is called the Schmidt decomposition
and the particular basis coincide with the eigenbasis of the corresponding
reduced density operators
$
\rho_{B/A} = \trace_{A/B}(\ket{\psi_{AB}})=\sum_{i} \alpha_{i}^2\ket{\psi_{B/A,i}}
\bra{\psi_{B/A,i}}  
$
The density operators $\rho_A$ and $\rho_B$ have common 
spectrum, in particular they are equally mixed. 
Since only product states lead to pure reduced density matrices,
a measure for their mixedness points a way towards quantifying entanglement in this 
case. Given the state $\ket{\psi_{AB}}$, we can thus take its Schmidt 
decomposition, Eq.(\ref{Schmidt}), and use a suitable function of the 
$\alpha_i$ to quantify the entanglement. 

An entanglement measure $E$ is fixed uniquely 
after imposing the following conditions:
1)  $E$ is invariant under local unitary operations ($\Rightarrow$ 
    $E$ is indeed a function of the $\alpha_i$'s only);
2)  $E$ is continuous (in a certain sense also in the asymptotic 
    limit of infinite copies of the state; see e.g. Ref.~\cite{Plenio07});
3)  $E$ is additive, when several copies of the system are present:
    $E(\ket{\psi_{AB}}\otimes \ket{\phi_{AB}}) = 
    E(\ket{\psi_{AB}})+E(\ket{\phi_{AB}})$.
The unique measure of entanglement satisfying 
all the above conditions is the von Neumann entropy of 
the reduced density matrices
\begin{equation}
S(\rho_A)= S(\rho_B)= -\sum_i \alpha_i^2 \log(\alpha_i^2) \; ,
\end{equation}
this is just the Shannon entropy of the moduli squared of the
Schmidt coefficients. In other words: under the above regularity conditions,
the answer on the question of how entangled a bipartite pure state is, 
is given by the von Neumann entropy of (either of)
the reduced density matrices. 
The amount of entanglement is generally difficult to define once we are 
away from bipartite states, but in several cases we can still gain some 
insight into many-party entanglement if one considers different 
bipartitions of a multipartite system. 
 
It is worth to notice that 
a variety of purity measures are admissible when 
the third condition on additivity is omitted. In principle, there are 
infinitely many measures for the mixedness of a density matrix; 
two of them will typically lead to a different ordering when the Hilbert 
space of the parts has a dimension larger than two. \\
In contrast, if we trace out one of two qubits in a pure state, 
the corresponding reduced density matrix $\rho_A$ contains only a single 
independent and unitarily invariant parameter: its eigenvalue $\leq 1/2$.
This implies that 
each monotonic function $[0,1/2]\mapsto [0,1]$ of this eigenvalue
can be used as an entanglement measure. 
Though, also here an infinity of different mixedness measures
exists, all lead to the same ordering of states with respect to their 
entanglement, and in this sense all are equivalent.
A relevant example is the (one-) tangle~\cite{Coffman00}
\begin{equation}
\tau_1[\rho_A] = 4 {\rm det} \rho_A\; .
\label{onetangle}
\end{equation}
By expressing $\rho_A$ in terms of spin expectation values, it follows that 
$
\tau_1[\rho_A]={1}-
4 (\expect{S^x}^2+\expect{S^y}^2+\expect{S^z}^2) 
$
where $\expect{S^{\alpha}} = tr_A(\rho_A S^{\alpha})$ and $S^{\alpha}=\frac{1}{2} \sigma^\alpha$, 
$\sigma^\alpha$ $\{\alpha=x,y,z\}$ being the Pauli matrices,
For a pure state $\ket{\psi_{AB}}$ of two qubits the relation 
$
\tau_1\equiv
|\bra{\psi^*}\sigma^y_{A} \otimes \sigma^y_{B}\ket{\psi}|^2=:C[\ket{\psi_{AB}}]^2=:\tau_2
$
applies, where $C$ is the
concurrence~\cite{Hill97,Wootters98} for pure states of two qubits, 
a measure of pairwise entanglement (see next Section),
and $*$ indicates the complex conjugation in the eigenbasis of $\sigma^z$.
The von Neumann entropy can be expressed as a function of the (one-) tangle
$\displaystyle{
S[\rho_A]=h\left(\frac{1}{2}\left(1+\sqrt{1-\tau_1[\rho_A]}\right) \right)}
$
where
$h(x)=: -x \log_2 x - (1-x) \log_2 (1-x)$ is the binary entropy.


\subsection{Pairwise qubit entanglement in mixed states}
\label{measure-pairwise-mixed}

Subsystems of a many-body (pure) state will generally be in a mixed state.
In this case different way of quantifying entanglement can be introduced.
Three important representatives are the entanglement cost $E_C$,
the distillable entanglement $E_D$ 
(both defined in Ref.~\cite{Bennett96}) and the 
entanglement of formation $E_F$~\cite{BennettDiVincenzo96}.
Whereas $E_D$ and $E_C$ are asymptotic limits of multi-copy extraction
probabilities of Bell states and creation from such states, 
the entanglement of formation is the amount of pure state entanglement 
needed to create a single copy of the mixed state. 
Although recent progress have been achieved~\cite{AdditiveEoF}, the full additivity of the $E_F$ for bipartite systems has not been     
               established yet (see e.g.~\cite{VidalECost02}).

The conceptual difficulty behind the calculation of $E_F$ lies
in the infinite number of possible decompositions of a density matrix.
Therefore, even knowing how to quantify bipartite 
entanglement in pure states, we cannot simply apply this knowledge 
to mixed states in terms of an average over the mixtures of 
pure state entanglement. 
The problem is that two decompositions of the same density matrix 
usually lead to a different average entanglement. 
Which one do we choose? 
It turns out that we must take the minimum over all possible decompositions, 
simply because if there is a decomposition where the average is zero, 
then this state can be created locally without need of any entangled pure
state, and therefore $E_F=0$. 
The same conclusion can be drawn from the requirement that entanglement 
must not increase on average by means of local operations including 
classical communication (LOCC). 

The entanglement of formation of a state ${\rho}$ is 
therefore defined as 
\begin{eqnarray}
E_{F}(\rho):= \min \sum_j p_j S(\rho_{A,j}) \; , \label{ef}
\end{eqnarray}
where the minimum is taken over all realizations of the state 
$\rho_{AB} = \sum_j p_j |{\psi_j}\rangle\langle{\psi_j}|$, and $S(\rho_{A,j})$ 
is the von Neumann entropy of the reduced density matrix 
$\rho_{A,j}:=\trace_B \ket{\psi_j}\bra{\psi_j}$.
Eq.(\ref{ef}) is the so called {\em convex roof} (also the expression
{\em convex hull} is found in the literature) of the entanglement of 
formation for pure states, and a decomposition leading to this
convex roof value is called an {\em optimal decomposition}.

For systems of two qubits, an analytic expression for $E_F$ does exist
and it is given by
\begin{equation}
\label{EoF}
E_F (\rho) = -\sum_{\sigma = \pm}\frac{\sqrt{1+\sigma 
C^2(\rho)}}{2}\ln\frac{\sqrt{1+\sigma C^2(\rho)}}{2}
\end{equation}
where $C(\rho)$ is the so called concurrence~\cite{Wootters98,Wootters01},
the convex roof of the pure state concurrence, 
which has been defined in the previous section. 
Its convex roof extension is encoded in the positive Hermitean matrix 
$
 R\equiv\sqrt{\rho}\tilde{\rho}\sqrt{\rho} =\sqrt{\rho}(\sigma^y\otimes\sigma^y)
\rho^*(\sigma^y\otimes\sigma^y)\sqrt{\rho}\; ,
$
with eigenvalues $\lambda_1^2\geq \dots \geq \lambda_4^2$ 
in the following way
\begin{equation}
  \label{eq:concurrence}
  C = \mbox{max}\{\lambda_1-\lambda_2-\lambda_3-\lambda_4,0\}\; .
\end{equation}
As the entanglement of formation is a monotonous function of the concurrence, 
also $C$ itself or its square $\tau_2$ - called 
also the 2-tangle - can be used as entanglement measures. 
This is possible due to a curious peculiarity of two-qubit systems: namely 
that a continuous variety of optimal decompositions exist~\cite{Wootters98}.
The concurrence $C$ and the tangle $\tau_1$ both range 
from $0$ (no entanglement) to $1$.

By virtue of (\ref{eq:concurrence}), the concurrence in a spin-1/2 chain can be computed
in terms of up to two-point spin correlation functions.  
As an example we consider a case where the model has a 
parity symmetry, it is translational invariant and the Hamiltonian is real; 
the concurrence in this case reads
\begin{equation}
\label{C-of-corrs}
C_{ij}=2\max\left\{0,C_{ij}^I,C_{ij}^{II}\right\}\;.
\end{equation}
where
$
C^I_{ij}=|g_{ij}^{xx}+g_{ij}^{yy}|-\sqrt{\left(1/4+g^{zz}_{ij}\right)^2-M_z^2}
$
and 
$
C^{II}_{ij}=|g_{ij}^{xx}-g_{ij}^{yy}|+g_{ij}^{zz}-1/4
$, 
with $g_{ij}^{\alpha\alpha}=\langle S^\alpha_i S^\alpha_{j} \rangle$
and $M_z=\langle S^z \rangle$.
A state with dominant fidelity of parallel and anti-parallel Bell states is 
characterized by dominant $C^I$ and $C^{II}$, respectively. 
This was shown in~\cite{Fubini06}, 
where the concurrence was expressed in terms of the {\em fully entangled 
fraction} as defined in~\cite{BennettDiVincenzo96}.
A systematic  analysis of the relation 
between the concurrence (together with the 3-tangle, see section 
\ref{measure-multipartite}) and the 
correlation functions has been presented in~\cite{glaser}.

The importance of the tangle and the concurrence is due to the {\em monogamy} 
inequality derived in~\cite{Coffman00} for three qubits. 
This inequality has been 
proved to hold also for n-qubits system~\cite{Osborne06}. In the case of 
many-qubits (the tangle may depend on the site $i$ that is considered)
it reads 
\begin{equation}
\label{ckw}
        \sum_{j\neq i} C^2_{ij} \leq  \tau_{1,i} \; .
\end{equation}
The so called {\em residual tangle} $\tau_{1,i}-\sum_{j\neq i} C^2_{ij}$, is 
a measure for multipartite entanglement
not stored in pairs of qubits only.
We finally mention that the antilinear form of the pure state
concurrence was the key
for the first explicit construction of a convex roof, and hence its 
extension to mixed states~\cite{Hill97,Wootters98,Uhlmann00}. 

Another measure of entanglement we mention is the relative entropy 
of entanglement~\cite{RelativeEntropy}. It can be applied to any number of 
qubits in principle (or any dimension of the local Hilbert space).
It is formally defined as
$
E({\sigma}):= \min_{\rho \in {\cal D}}\,\,\, S(\sigma || \rho)
$,
where 
$S(\sigma || \rho) = \trace \sigma \left[\ln \sigma - \ln \rho \right]$
is the quantum relative entropy. This relative entropy of entanglement
quantifies the entanglement in $\sigma$ by its distance from the set ${\cal D}$
of separable states (since ${\cal D}$ is compact, 
the minimum is assumed always). 
The main difficulty in computing this measure is to find the 
disentangled state closest to $\rho$. This is in general an open 
problem, even for two qubits. In the presence of certain symmetries 
- which is the case for e.g. eigenstates of certain models - an analytical 
access is possible. In these cases, the relative entropy of entanglement  
becomes a very useful tool. The relative entropy reduces to the 
entanglement entropy in the case of pure bi-partite states; this also 
means that its convex roof extension coincides with the 
entanglement of formation, and is readily deduced from the 
concurrence~\cite{Wootters98}. 

We close this summary on the pairwise entanglement by commenting on the 
notion on the quantum mutual information. Groisman {\em et al} quantified 
the work necessary to erase the total correlations existing in a bipartite 
system~\cite{Groisman05}. The entanglement can be erased by a 
suitable random ensemble of unitary transformations acting on one of the 
parts, but a certain amount of classical correlation among the two 
partners may survive. The work necessary to erase all correlations 
is given by the quantum mutual information
\begin{equation}
{\cal I}_{AB}=S(\rho_A)+S(\rho_B)-S(\rho_{AB})
\label{mutual}
\end{equation}

\subsection{Localizable entanglement}
\label{locsection}

A different approach to entanglement in many-body systems arises from the 
quest to swap or transmute different types of multipartite entanglement
into pairwise entanglement between two parties by means of 
generalized measures on the rest of the system. In a system 
of interacting spins on a lattice one could then try to maximize the 
entanglement between two spins (at positions $i$ and $j$) by performing 
measurements on all the others. The system is then partitioned in three 
regions: the sites $i$, $j$  and the rest of the lattice.
This concentrated pairwise entanglement can then
be used  e.g. for quantum information processing.
A standard example is that the three qubit Greenberger-Horne-Zeilinger 
(GHZ) state 
$(1/\sqrt{2})(\ket{000}+\ket{111})$ 
after a projective measure in $x$-direction on one of the sites 
is transformed into a Bell state.

The concept of localizable entanglement has been introduced 
in~\cite{verstraete04,popp05}. 
It is defined as the maximal amount of entanglement that can be 
localized, on average, by doing local measurements in the rest of the 
system. In the case of $N$ parties, the possible outcomes of the measurements 
on the remaining $N-2$ particles are pure states $|\psi_s \rangle$ 
with corresponding probabilities $p_s$. The localizable entanglement 
$E_{loc}$ on the sites $i$ and $j$ is defined as the maximum of 
the average entanglement over all possible outcome states $\ket{\psi_s}_{ij}$
\begin{equation}
E_{loc}(i,j) = \mbox{sup}_{\cal{E}} \sum_s p_s E(\ket{\psi_s}_{ij})
\end{equation}
where ${\cal E}$ is the set of all possible outcomes 
$(p_s, |\psi_s \rangle)$ of the measurements, and $E$ represents 
the chosen measure of entanglement of a pure state of two qubits 
(e.g. the concurrence). 
Although very difficult to compute, lower and upper bounds have 
been found which allow to deduce a number of consequences for this quantity.  

An upper bound to the localizable entanglement is given by the 
entanglement of assistance~\cite{laustsen03} obtained from 
localizable entanglement when also global and joint measurements 
were allowed on the $N-2$ spins .
A lower bound of the localizable entanglement comes from the 
following theorem~\cite{verstraete04}:
{\em Given a (pure or mixed) state of N qubits with reduced 
correlations $Q_{ij}^{\alpha,\beta} = 
\langle S_{i}^{\alpha}S_{j}^{\beta}\rangle - 
\langle S_{i}^{\alpha}\rangle \langle S_{j}^{\beta}\rangle$ 
between the  spins $i$ and $j$ and 
directions $\alpha$ and $\beta$ then there always exist directions in 
which one can measure the other spins such that this correlation do not 
decrease, on average.}
It then follows that a lower bound to localizable entanglement is fixed 
by the maximal correlation function between the two parties 
(one of the various spin-spin correlation functions 
$Q_{ij}^{\alpha,\beta}$)\footnote{It has been argued recently~\cite{Gour05,Gour06} 
that in order to extend the entanglement of assistance and 
the localizable entanglement to being an entanglement monotone~\cite{MONOTONES} one should
admit also local operations including classical communication on the 
extracted two spins, this was named entanglement of collaboration.}.

\subsection{Entanglement witnesses} 

It is important to realize that not just the quantification of 
many-party entanglement is a difficult task; it is an open 
problem to tell in general, whether a state of $n$ parties is separable or not. 
It is therefore of great value to have a tool that is able to merely
certify if a certain state is entangled. An entanglement witness $W$ 
is a Hermitean operator which is able to detect entanglement in a state. 
The basic idea is
that the expectation value of the witness $W$ for the state $\rho$
under consideration exceeds certain bounds only when $\rho$ is entangled. 
An expectation value of $W$ within this bound however
does not guarantee that the state is separable. Nonetheless, this is a very 
appealing method also from an experimental point of view, since it is 
sometimes possible to relate the presence of the entanglement to the 
measurement of few observables.

Simple geometric ideas help to explain the  witness 
operator $W$ at work. Let $\mathcal{T}$ be the set of all density matrices 
and let $\mathcal{E}$ and $\mathcal{D}$ be the subsets of entangled and 
separable states, respectively. 
The convexity of $\mathcal{D}$ is a key property for witnessing 
entanglement
The entanglement witness is then an operator 
defining a hyper-plane which separates a given entangled state from the set of 
separable states. The main scope of this geometric approach is then 
to optimize the witness operator~\cite{LewensteinOptWit} or to replace 
the hyper-plane by a curved manifold, tangent to the set of separable 
states~\cite{witguehne} (for other geometric aspects of entanglement 
see~\cite{Klyachko,bengtsson,Leinaas06}).
We have the freedom to choose $W$ such that 
$
\trace(\rho_D W)\leq 0
$ 
for all disentangled states $\rho_D\in{\cal D}$.
Then,  
$
\trace(\rho W)>0
$ 
implies that $\rho$ is entangled.
A caveat is that the concept of a witness is not invariant 
under local unitary operations (see e.g.~\onlinecite{FoolingWits}).

Entanglement witnesses are a special case of a more general concept, 
namely that of positive maps. These are injective superoperators
on the subset of positive operators.
When we now think of superoperators that act non-trivially only on part
of the system (on operators that act non trivially only on a sub-Hilbert space),
then we may ask the question whether a positive map on the subspace is 
also positive when acting on the whole space. Maps that remain positive 
also on the extended space are called {\em completely positive maps}.
The Hermitean time evolution of a density matrix is an example for a 
completely positive map. Positive but {\em not} completely positive 
maps are important for entanglement theory. There is a remarkable isomorphism 
between positive maps and Hermitean operators~\cite{Jamiolkowski}. 
This can be used to prove a key theorem~\cite{Horodecki96}:
{\em A state $\rho_{AB}$ is entangled if and only if
a positive map $\Lambda$ exists (not completely positive) such that }
$
(\id_A \otimes \Lambda_B) \rho_{AB} < 0
$.
For a two dimensional local Hilbert space the situation simplifies
considerably in that any positive map $P$ can be written as
$
P=CP_1 + CP_2 T_B \; ,
$
where $CP_1$ and $CP_2$ are completely positive maps and $T_B$ is a
transposition operation on subspace $B$. This decomposition tells that for a 
system of two qubits the lack of complete positivity in a positive
map is due to a partial transposition. This partial transposition 
clearly leads to a positive operator if the state is a tensor 
product of the parts. In fact, also the opposite is true: 
a state of two qubits $\rho_{AB}$ is separable if and only if 
$\rho_{AB}^{T_B} \geq 0$ that is, its partial transposition is positive. 
This is very simple to test and it is known as the Peres-Horodecki 
criterion~\cite{Peres96,Horodecki96}. The properties of entangled 
states under partial transposition lead to a measure of entanglement
known as the {\em negativity}. The negativity $N_{AB}$  
of a bipartite state is defined as the absolute value of the sum of 
the negative eigenvalues of $\rho_{AB}^{T_A}$. The 
{\em logarithmic negativity} is then defined as
\begin{equation}
     E_N  = \log_22 (2N_{AB}  + 1).
\end{equation}
For bipartite states of two qubits, $\rho_{AB}^{T_A}$ has at most one 
negative eigenvalue~\cite{Sanpera98}. For general multipartite and 
higher local dimension this is only a sufficient condition for the 
presence of entanglement. There exist entangled states with a
positive partial transpose known as bound entangled 
states~\cite{acinbe,Horodeckibe}.

\subsection{Multipartite entanglement measures}
\label{measure-multipartite}

Both the classification of entanglement and its quantification
are at a preliminary stage even for distinguishable particles 
(see however~\onlinecite{Duer00,Miyake02,VerstraeteDMV02,
Briand03,BriandLT03,OS04,OS05,Luque05,mandilara} and references therein).
We restrict ourselves to those approaches which have been applied 
so far for the study of condensed matter systems discussed in this review. 
It has already been mentioned that several quantities  are useful as indicators for multipartite entanglement
when the whole system is in a pure state.
The entropy of entanglement is an example for such a quantity and several  
works use multipartite measures constructed from and related to
it (see e.g.~\onlinecite{Coffman00,Wallach,Viola03,Scott04,oliveira,Love06}).
These measures are of 'collective' nature - 
in contrast to 'selective' measures -
in the sense that they give 
indication on a global correlation without discerning among the different 
entanglement classes encoded in the state of the system.

The geometric measure of entanglement
quantifies the entanglement of a 
pure state through the minimal distance of the state from the set of 
pure product states~\cite{RelativeEntropy,wei03}
\begin{equation}
   E_{g}(\Psi) = - \log_2 \max_{\Phi}\mid \langle \Psi | \Phi \rangle \mid ^2
\label{weimeas}
\end{equation}
where the maximum is on all product states $\Phi$. 
As discussed in detail in~\cite{wei03}, the previous definition
is an entanglement monotone if the convex-roof extension to mixed states is
taken. It is zero for separable states 
and rises up to unity for e.g. the maximally entangled n-particle GHZ states. 
The difficult task in its evaluation is the maximization 
over all possible separable states and of course the convex roof extension to mixed states. 
Despite these complications, a clever use of the symmetries of the problem 
renders this task accessible by substantially reducing the 
number of parameters (see Section~\ref{multip}).

Another example for the collective measures of multipartite
entanglement as mentioned in the beginning of this section
are the measures introduced by Meyer and Wallach~\cite{Wallach} 
and by Barnum {\em et al}~\cite{Viola03,Barnum04}. 
In the case of qubit system the $Q$-measure of
Meyer and Wallach is the average purity (which is the average one-tangle 
in~\cite{Coffman00}) of the state~\cite{Wallach,Brennen03,Barnum04}
\begin{equation}
  E_{gl} = 2 - \frac{2}{N} \sum_{j=1}^N \mbox{Tr} \rho_j^2 \;\; .
\label{GE}
\end{equation}

The notion of {\em generalized entanglement} introduced 
in~\cite{Viola03,Barnum04} relaxes the typically chosen 
partition into local subsystems in real space. 
The generalized entanglement measure used by
Barnum {\em et al.} is the purity relative to a distinguished 
Lie algebra ${\cal A}$ of observables. 
For the state $| \psi \rangle$ it is defined as 
\begin{equation}
  P_{\cal A} = \mbox{Tr} \left\{ \left[ {\cal P}_{{\cal A}} | 
\psi \rangle \langle \psi | \right]^2 \right\} 
\end{equation}
where ${\cal P}_{\cal A}$ is the projection map 
$\rho \rightarrow {\cal P}_{\cal A}(\rho)$. 
If the set of observables is defined by the operator basis 
$\left\{ A_1, A_2,\dots, A_L\right\}$ then
$  P_{\cal A} = \sum_{i=1}^L \langle A_{i}\rangle ^2
$
from which the reduction to Eq.(\ref{GE}) in the case of all local observables 
is evident.
This conceptually corresponds to a redefinition of {\em locality} as induced by
the distinguished observable set, beyond the archetype of partition in the real space.
It defines an observer dependent concept of entanglement
adapted to e.g. experimentally accessible or physically relevant observables. 
In this case, the generalized entanglement coincides with the global 
entanglement of Meyer and Wallach.  

Another approach pursued is the generalization of the concurrence.
For the quantification of pairwise entanglement in
higher dimensional local Hilbert spaces, the concept of
concurrence vectors has been formulated~\cite{Audenaart01,Badziag02}
besides the I-concurrence~\cite{Rungta}. 
A concurrence vector was also proposed for multipartite systems of 
qubits~\cite{Akhtarshenas05}. 
It consists in applying the pure state concurrence formula to a mixed
two-site reduced density matrix. It coincides with the true concurrence
if and only if the eigenbasis of the density matrices provide 
 optimal decompositions. 
 
The {\em n-tangle} is a straightforward extension of the concurrence 
to multipartite states
as the overlap of the state with its time-reversed~\cite{Wong00}.
It vanishes identically for an odd number of qubits, but 
an entanglement  monotone is obtained for an even number of qubits.
It detects products of even-site entangled states in addition to certain 
genuine multipartite entangled states: it detects the multipartite 
GHZ or cat state, but not for example the four qubit cluster state. 

Three classes of states inequivalent under SLOCC (Stochastic LOCC) exist for four qubits~\cite{OS04,OS05}. 
Representatives are the GHZ state, the 
celebrated cluster state and a third state, which is also measured by
the 4-qubit Hyperdeterminant. Class selective measures are constructed from two basic 
elements, namely the operator $\sigma_y$ employed for the concurrence, 
and the operator 
$\sigma_\mu\cdot \sigma^\mu:=
\id\cdot\id - \sigma^x\otimes \sigma^x - \sigma^z\cdot \sigma^z$
where the $\bullet$ is a tensor product indicating that
the two operators are acting on different copies of the same qubit.
Both are invariant under $sl(2,\CC)$ operations on the qubit.
The 3-tangle is then expressed as
$
\tau_3[\psi]=\bra{\psi*}\cdot\bra{\psi*}
\sigma_\mu\cdot \sigma_\nu\cdot \sigma_\lambda\otimes
\sigma^\mu\otimes \sigma^\nu\otimes \sigma^\lambda \ket{\psi}\cdot\ket{\psi}
$
The multilinearity, however, makes it problematic to employ
the procedure of convex roof construction presented 
in~\cite{Wootters98,Uhlmann00}
for general mixtures. 

Finally we mention the approach pursued in~\cite{guehne05} (see also~\onlinecite{Sharma06})
where different bounds on the average energy of a given system were obtained 
for different types of n-particle quantum correlated states.
A violation of these bounds then implies the presence of 
multipartite entanglement in the system.
The starting point of G\"uhne {\em et al.} is the notion of 
{\em n-separability} and {\em k-producibility} which admit to discriminate 
particular types of n-particle correlations present in the system. 
A pure state $\mid \psi \rangle$ of a quantum systems of N parties is 
said to be n-separable if it is possible to find a partition of 
the system for which $\mid \psi \rangle = |\phi_1 \rangle |\phi_2 
\rangle \cdots |\phi_n \rangle$.
A pure state $\mid \psi \rangle$ can be 
produced by k-party entanglement (i.e. it is k-producible) if we can write
$\mid \psi \rangle = |\phi_1 \rangle |\phi_2 \rangle \cdots |\phi_m \rangle$
where the $|\phi_i \rangle$ are states of maximally k parties; 
by definition $m \ge N/k$. 
It implies that it is sufficient to generate specific 
k-party entanglement to construct the desired state. 
Both these indicators for multipartite entanglement are collective,
since they are based on the factorizability of a given many particle
state into smaller parts. k-separability and -producibility both
do not distinguish between different k-particle entanglement classes
(as e.g. the k-particle W-states and different k-particle 
graph states~\cite{hein04}, like the GHZ state).

\subsection{Indistinguishable particles}
\label{indistparts}

For indistinguishable 
particles the wave function is \mbox{(anti-)~symmetrized} and therefore the 
definition of entangled states as given in the previous Section does not 
apply. In particular, it does not make sense to consider each individual 
particle as parts of the partition of the system.
Having agreed upon a definition of entanglement,
concepts as entanglement cost or distillation remain perfectly valid.
Following~\cite{Ghirardi02,Ghirardi03} one can address the 
problem of defining entanglement in an ensemble of indistinguishable 
particles by seeing if one can attribute to each of the subsystems
a complete set of measurable properties, e.g. momenta for free pointless particles.
Quantum states satisfying the above requirement represent the separable states 
for indistinguishable particles.

There is another crucial difference 
between the entanglement of (indistinguishable) spin-1/2 particles and that of 
qubits. Let us therefore consider two fermions on two sites. Whereas the Hilbert 
space ${\cal H}_s$ of a two-site spin lattice has dimension
${\rm dim}\, {\cal H}_s= 4$, the Hilbert space ${\cal H}_f$ for two 
fermions on the same lattice has dimension ${\rm dim}\, {\cal H}_f=6$. 
This is due to the possibility that both fermions, with opposite spins, 
can be located at the  same lattice site.
When choosing the following numbering of the states
$\ket{1}=f^\dagger_1\ket{0}=:c^\dagger_{L,\up}\ket{0}$, 
$\ket{2}=f^\dagger_2\ket{0}=:c^\dagger_{L,\down}\ket{0}$, 
$\ket{3}=f^\dagger_3\ket{0}=:c^\dagger_{R,\up}\ket{0}$,
$\ket{4}=f^\dagger_4\ket{0}=:c^\dagger_{R,\down}\ket{0}$
and the definition 
$\ket{i,j} = f^\dagger_i f^\dagger_j\ket{0}$, there are Bell states 
analogous to those occurring for distinguishable
particles $(\ket{1,3}\pm\ket{2,4})/\sqrt{2}$ and 
$(\ket{1,4}\pm\ket{2,3})/\sqrt{2}$. There are however new 
entangled states, as $(\ket{1,2}\pm\ket{3,4})/\sqrt{2}$,
where both fermions take the same position.
The local Hilbert space is made of four states 
labelled by the occupation number and the spin, if singly occupied. 
The {\em site-entanglement} of indistinguishable particles is then defined
as the entanglement of the corresponding Fock states. 
It can be measured e.g. by the local von Neumann entropy. 
This quantity is the analogue to the one-tangle for 
qubits, but the local Hilbert space dimension is $4$ due to the possibility of
having empty and doubly occupied sites. 
Also the quantum mutual information~\cite{Groisman05},
see Eq.(\ref{mutual}), can be defined in this way,
quantifying the total amount (classical and quantum) of
correlations stored in a given state of a second quantized system. 


Although from a mathematical point of view the entanglement 
of indistinguishable particles can be quantified, 
the major part of the literature on second quantized systems 
that we discuss in this review considers the 
site-entanglement described above or the entanglement of degrees 
of freedom, singled out from a suitable set 
of local quantum numbers (e.g. the spin of the particle at site $i$).
In both cases, entanglement measures for distinguishable particles (see 
Sections~\ref{sec-noninteracting} and \ref{hubbardsent}) can be used. 
With this respect, this Section has a different scope 
than the others on the quantification of entanglement; although most of the 
discussion which follows will not be used later on, we believe that it 
will be of interest for further studies of entanglement in 
itinerant many-body systems.

\subsubsection{Two Fermion entanglement}
\label{ferment}

Due to the antisymmetry under particle exchange, there is no 
Schmidt decomposition for Fermions. Nevertheless,
a Fermionic analogue to the Schmidt rank, which classifies entanglement
in bipartite systems of distinguishable particles does exist: the
so called {\em Slater rank}.
A generic state of two-electrons on two lattice sites can be written as
$
\ket{\omega}:=\sum_{i,j=1}^4\omega_{i,j}\ket{i,j}
$
where $\omega$ is a $4\times 4$ matrix which can be assumed antisymmetric
and normalized as $\trace \omega^\dagger\omega=\frac{1}{2}$.
Since here the local entities whose entanglement
shall be studied, are the particles, unitary transformations act
on the $4$-dimensional single particle Hilbert space. 
Due to the indistinguishability of the particles, the transformation must 
be the same for each of the particles.
Given a unitary transformation $U\in {\rm SU}(4)$ such that
$f^{'}_j:=U_{jk}f^{}_k$, the transformed state is given by
$\ket{\omega'}$ where $\omega':=U\omega U^T$.
The above unitary transformation preserves the antisymmetry of
$\omega$ and can transform every pure state of two 
spin-1/2 particles on two sites into a state corresponding to 
the normal form of $\omega$.
In fact, every two-particle state within a $D$-dimensional single
particle Hilbert space can be transformed into the normal form
$\omega_s = {\rm diag}\{Z_1,\dots,Z_r,Z_0\}$
where $Z_j = i z_j \sigma_y $ and 
$(Z_0)_{ij} = 0$ for  $i,j\in\{1,\dots,D-2r\}$. In the previous expression
$r$ is then called the {\em Slater rank} of the pure 
Fermion state~\cite{Schliemann01,SchLoss01,Eckert02}.
A pure Fermion state is entangled 
if and only if its Slater rank is larger than $1$.
It is important to notice that the above concept of entanglement only
depends on the dimension of the Hilbert space accessible to each of the
particles (this includes indistinguishable particles
on a single $D$-level system).

For electrons on an $L$-site lattice the ``local'' Hilbert space dimension
is $2L$, and the question, whether a pure state living in a 
$2L$-dimensional single particle
Hilbert space has full Slater rank, can be answered
by considering the Pfaffian of $\omega$~\cite{Caianello52,Muir60}
\beq\label{Pfaffian}
\sum_{\pi\in {\cal S}_{2L}^<} {\rm sign}(\pi)\prod_{j=1}^L \omega_{\pi(2j-1),\pi(2j)}
\eeq
which is non-zero only if $\omega$ has full Slater rank $L$.
In the above definition ${\cal S}_{2L}^<$ denotes those elements $\pi$ of 
the symmetric group ${\cal S}_{2L}$ with 
ordered pairs, i.e. $\pi(2m-1)<\pi(2m)$ for all $m\leq L$
and $\pi(2k-1)<\pi(2m-1)$ for $k<m$.
Notice that relaxing the restriction to ${\cal S}_{2L}^<$
just leads to a combinatorial factor of $2^L L!$ by virtue of the
antisymmetry of $\omega$ and hence can we write
\beq\label{eps-pfaff}
{\rm pf}[\omega]=\frac{1}{2^L L!} \sum_{j_1,\dots,j_{2L}=1}^{2L}
      \eps^{j_1,\dots,j_{2L}}\omega_{j_1,j_2}\dots\omega_{j_{2L-1},j_{2L}}
\eeq
where $\eps^{j_1,\dots,j_{2L}}$ is the fully antisymmetric tensor
with $\eps^{1,2,\dots,2L}=1$.
There is a simple relation between the Pfaffian and
the determinant of an antisymmetric even-dimensional matrix:
${\rm pf}[\omega]^2=\det[\omega]$.

For the simplest case of two spin-1/2 Fermions on two lattice sites
the Pfaffian reads 
${\rm pf}[\omega]=\omega_{1,2}\omega_{3,4}-\omega_{1,3}\omega_{2,4}+
\omega_{1,4}\omega_{2,3}$.
Normalized in order to range in the interval $[0,1]$ this has been
called the Fermionic concurrence 
${\cal C}[\ket{\omega}]$~\cite{Schliemann01,SchLoss01,Eckert02} 
\beq\label{ferm-Conc}
{\cal C}[\ket{\omega}]=|\braket{\tilde{\omega}}{\omega}|=8|{\rm pf}[\omega]|
\eeq
where
$
\tilde{\omega}:=\frac{1}{2}\eps^{ijkl}\omega_{k,l}^*
$
has been termed the {\em dual} to $\omega$. Then,
$\ket{\tilde{\omega}}=:{\cal D}\ket{\omega}$
is the analog to
the conjugated state in~\cite{Hill97,Wootters98,Uhlmann00}
leading to the concurrence for qubits.
It is important to notice that the Pfaffian in Eq.(\ref{Pfaffian})
is invariant under the complexification of ${\rm su}(2L)$,
since it is the expectation value of an antilinear operator,
namely the conjugation ${\cal D}$ for the state $\ket{\omega}$.
Since this invariant is a bilinear expression in the state coefficients,
its convex roof is readily obtained~\cite{Uhlmann00}
by means of the positive eigenvalues
$\lambda_i^2$ of the $6\times 6$ matrix
$
R=\sqrt{\rho}{\cal D}\rho{\cal D}\sqrt{\rho}
$.
The conjugation $\cal D$, expressed in the basis 
\mbox{$\{\ket{1,2},\ket{1,3},\ket{1,4},\ket{2,3},\ket{2,4},\ket{3,4}\}$}
takes the form ${\cal D}_0\;{\cal C}$,
where ${\cal C}$ is the complex conjugation and the only non-zero elements 
of ${\cal D}_0$ are $({\cal D}_0)_{16}=1$, $({\cal D}_0)_{25}=-1$, $({\cal D}_0)_{34}=1$,
$({\cal D}_0)_{43}=1$, $({\cal D}_0)_{52}=-1$, and $({\cal D}_0)_{61}=1$.
Notice that the center part of this matrix  is precisely
$\sigma_y\otimes\sigma_y$ and indeed corresponds to
the Hilbert space of two qubits. The remaining part of the Hilbert space
gives rise to an entanglement of different values for the 
occupation number. 
This type of entanglement has been referred to as the 
{\em fluffy bunny}~\cite{WisemanFluffyBunny,CiracFluffyBunny} in the 
literature. 

For a single particle Hilbert space with dimension larger than $4$
one encounters similar complications as for two distinguishable
particles on a bipartite lattice and local Hilbert space dimension
larger than $2$, i.e. for two {\em qudits}.
This is because different classes of entanglement occur, which
are characterized by different Slater rank as opposed to 
their classification by different Schmidt rank for distinguishable particles.
The Slater rank can be obtained by looking at Pfaffian minors~\cite{Muir60}:
if the Slater rank is $r$, all Pfaffian minors of dimension larger than $2r$
are identically zero.

\subsubsection{Multipartite Entanglement for Fermions}

For indistinguishable particles the only classification available
up to now is to check whether or not a pure state has Slater rank 
one. Eckert {\em et al.} formulated two recursive 
lemmata~\cite{Eckert02} that can be summarized as follows: 
let an $N$-electron state be contracted
with $N-2$ arbitrary single electron states encoded in the vectors $\vec{a}^j$
as $\vec{a}^j_kf^\dagger_k\ket{0}$ ($j=1,\dots,N-2$ and sum convention)
to a two-electron state. Then the Pfaffian of the two-electron state
is zero if and only if the original state (and hence all intermediate
states in a successive contraction) has Slater rank one.
This means that all 4-dimensional Pfaffian minors of $\omega$ are zero.

Instead of the Pfaffian of $\omega$, also the single-particle reduced 
density matrix can be considered, and its von Neumann entropy as a 
measure for the quantum entanglement has been analyzed 
in~\cite{Lizeng01,Paskauskas01}. 
It is important to remind that for distinguishable particles the local reduced 
density matrix has rank one if and only if the original state were a product.
This is no longer true for indistinguishable particles.
For an $N$-particle pure state with Slater rank one
the rank of the single-particle reduced density matrix
coincides with the number of particles, $N$.
A subtlety is that a measure of entanglement is 
obtained after subtraction of the constant value of the von 
Neumann entropy of a disentangled state. This must be taken into
account also for the extension of the measure to mixed states.

\subsubsection{``Entanglement of particles''}
\label{measures-entanglement-particles}

Entanglement in the presence of super selection rules (SSR) induced by 
particle conservation has been discussed in Refs.~\cite{Bartlett03,Wiseman03,
shuchprl,shuch}. The main difference in the concept of 
{\em entanglement of particles}~\cite{Wiseman03} from the entanglement
of indistinguishable particles as described in the preceding section
(but also to that obtained from the reduced density matrix of e.g. spin
degrees of freedom of indistinguishable particles)
consists in the projection of the Hilbert space onto a subspace
of fixed particle numbers for either part of a bipartition of the system.
The bipartition is typically chosen to be space-like, as motivated from
experimentalists or detectors sitting at distinct positions.
E.g. two experimentalists, in order to detect the
entanglement between two indistinguishable particles, must have
one particle each in their laboratory. 

This difference induced by particle number superselection 
is very subtle and shows up if multiple occupancies occur 
at single sites for Fermions with some inner degrees of freedom,
as the spin. Their contribution is finite 
for finite discrete lattices and will generally scale to zero 
in the thermodynamic limit with vanishing lattice spacing. 
Therefore both concepts of spin 
entanglement of two distant particles coincide in this limit.
Significant differences are to be expected only for finite non-dilute
systems.
It must be noted that the same restrictions imposed by SSR which 
change considerably the
concept of entanglement quantitatively and qualitatively,
on the other hand enable otherwise impossible protocols of quantum 
information processing~\cite{shuchprl,shuch} which are based on variances 
about the observable fixed by superselection.

Wiseman and Vaccaro project an $N$-particle state $\ket{\psi_N}$ 
onto all possible subspaces, where the two parties have a well defined 
number $(n_A,n_B=N-n_A)$ of particles in their laboratory~\cite{Wiseman03}. 
Let $\ket{\psi[n_A]}$ be the respective projection, and let $p_{n_A}$ 
be the weight 
$\braket{\psi[n_A]}{\psi[n_A]}/\braket{\psi_N}{\psi_N}$ of this projection.
Then the entanglement of particles $E_p$ is defined as
\beq
E_p[\ket{\psi_n}] = \sum_n p_n E_M[\psi[n_A]]
\eeq
where $E_M$ is some measure of entanglement for distinguishable particles.
Although this certainly represents a definition of entanglement
appealing for experimental issues, it is sensitive only to situations,
where e.g. the two initially indistinguishable particles
eventually are separated and can be examined one-by-one by Alice and Bob.
Consequently, ``local operations'' have been defined in~\cite{Wiseman03}
as those performed by Alice and Bob in their laboratory after having
measured the number of particles\footnote{
As a potential difference between the entanglement of photons as
opposed to that of massive bosonic particles, 
it has been claimed that certain superselection rules 
may hold for massive particles only.
One such claim is that we would in practice not be able to build
coherent superpositions of states containing a different number of massive 
particles (for a recent discussion see~\onlinecite{bartlett07}). 
This superselection rule would, for instance, prohibit creating a 
superposition of a Hydrogen atom and a Hydrogen molecule. 
However, the origin and validity of any superselection rule 
remains a very much debated subject. The arguments pro superselection rules 
usually involve some symmetry considerations, or some decoherence mechanism. 
On the other hand, it turns out that if we allow most general operations in 
quantum mechanics, we no longer encounter any superselection restrictions. 
Recent work~\onlinecite{terra,dowling06b} 
shows that it should be possible to coherently 
superpose massive particles and to
observe a violation of certain Bell inequalities~\onlinecite{terra}
also for this case.}. 

Verstraete and Cirac pointed out that the presence of SSR gives rise to a 
new resource which has to be quantified. They have proposed to replace 
the quantity $E_p$ with the {\em SSR-entanglement of formation}. This is 
defined as 
$$
E_f^{(SSR)} [\ket{\psi_N}]  = \min_{p_n,\psi_n} \sum_n p_n E_M[\psi_n]
$$ 
where the minimization is performed over all those decomposition of 
the density matrix where the $|\psi \rangle_n$ are eigenstates of the total number of 
particles~\cite{shuchprl,shuch}.

\subsubsection{Entanglement for Bosons}

The quantification and classification of boson entanglement is very close
in spirit to that of Fermions as described in Section \ref{ferment}.
In the bosonic case the matrix $\omega$ introduced in the previous section is 
symmetric under permutations of the particle numbers. Consequently, for any 
two-particle state of indistinguishable bosons, $\omega$ can be diagonalized 
by means of unitary transformations of the single particle basis. This leads 
to the Schmidt decomposition for bosons~\cite{Eckert02}.
An curious feature distinguishing this case from the entanglement measures 
of distinguishable particles is that the Schmidt decomposition is not 
unique. In fact, any two equal Schmidt coefficients 
admit for a unitary transformation of the two corresponding basis states, such that 
the superposition of the two doubly occupied states can be written as a 
symmetrized state of two orthogonal states~\cite{Lizeng01,Ghirardi05}.
This is the reason why it is not directly the Schmidt rank, but rather the 
reduced Schmidt rank - obtained after having removed all double
degeneracies of the Schmidt decomposition - that 
determines whether or not a state is entangled. 
This non-uniqueness of the Schmidt rank is also responsible for
the ambiguity of the von Neumann entropy or other purity measures
of the single particle reduced density matrix as an entanglement
measure for Bosons~\cite{Ghirardi05}.

With $z_i$ being the Schmidt 
coefficients with degeneracy $g_i$, the reduced Schmidt rank is at most 
$\frac{g_i}{2}+2\left\{\frac{g_i}{2}\right\}$, where $\{.\}$ denotes the 
non-integer part. As a consequence, a Schmidt rank larger than two implies 
the presence of entanglement. Schmidt rank $2$ with degenerate Schmidt 
coefficients can be written as a symmetrized product of orthogonal
states and consequently is disentangled~\cite{Ghirardi05}.
This features is also present in the $N$-boson case, where in presence of 
up to $N$-fold degenerate Schmidt coefficients the corresponding state can 
be rewritten as a symmetrization of a product. 

For bipartite systems $\omega$ has full Schmidt rank if $\det \omega\neq 0$. 
A Schmidt rank $1$ can be verified by the same contraction technique 
described for the Fermion case in the previous section, where the Pfaffian 
must be replaced by the determinant. This applies to both the bipartite and the
multipartite case~\cite{Eckert02}.

\subsection{Entanglement in harmonic systems}
\label{entharmon}

In this section we concentrate on the entanglement between distinct 
modes of harmonic oscillators (see~\cite{CV-Review,illuminatirev} for 
recent reviews on the subject). The entanglement  in this case is termed 
as {\em continuous variable entanglement} in the literature (to be distinguished from the 
entanglement of indistinguishable bosonic particles; 
see Section~\ref{indistparts}). 

Dealing with higher dimensional space of the local degrees of freedom 
generally involves complications which are not tamable within
the current knowledge about entanglement. The Peres-Horodecki criterion, 
just to mention an important example, is not sufficient 
already for two three-level systems, $3\times 3$. 
The situation simplifies if only so called {\em Gaussian states} of the 
harmonic oscillator modes are considered. 
This restriction makes the infinite dimensional case 
even conceptually simpler than the finite dimensional counterparts.
In order to explain what Gaussian 
states are, we introduce the Wigner 
distribution function $W(p,q)$~\cite{Wigner32}. For a single degree 
of freedom it is defined from the density operator $\rho$ as
\beq\label{Wigner}
W(r,p):=\frac{1}{\pi\hbar}\int_{-\infty}^\infty \d r'\; \bra{r+r'}\rho\ket{r-r'}
\e^{\frac{2i}{\hbar}pr'} \; ,
\eeq
where $r$ and $p$ are conjugate position and momentum variables of the
degree of freedom. 
The connection between bosonic operators $\hat{a}$, $\hat{a}^\dagger$
and phase space operators $\hat{r}$, $\hat{p}$ is
$\hat{a}=(\hat{r}+i\hat{p})/\sqrt{2}$, 
$\hat{a}^\dagger=(\hat{a})^\dagger=(\hat{r}-i\hat{p})/\sqrt{2}$.
More degrees of freedom are taken into account in a 
straight forward manner. A (mixed) state $\rho$ is then called Gaussian 
when its Wigner distribution function is Gaussian. 
Examples for such states are coherent pure states $\ket{\alpha}$,   
$\hat{a}\ket{\alpha}=\alpha \ket{\alpha}$ with $\alpha\in\CC$,
and arbitrary mixtures of coherent states
$\rho=\int \d^2\alpha\; P(\alpha) \ket{\alpha}\bra{\alpha}$,
determined by the so called {\em $P$-distribution} $P(z)$.
Such states are termed classical if the Wigner function and the
$P$-distribution are non-negative~(see \cite{Simon00}).
 
The key quality of Gaussian states is that they are completely
classified by second moments, which are encoded in the symmetric so called 
{\em (co-)variance matrix} with the uncertainties of 
the phase space coordinates as entries.
For two bosonic modes the phase space is four-dimensional
and the covariance matrix $V$ is defined as
\beq
V_{\alpha\beta}:=\expect{\{\Delta\hat{\xi}_\alpha,\Delta\hat{\xi}_\beta\}}=
\int\d^4 \xi\; \Delta\xi_\alpha\Delta\xi_\beta W(\{\xi_\gamma \})\ ,
\label{covariantm}
\eeq
where the curly brackets on the left hand side indicate the anti-commutator.
The components of $\hat{\xi}_\alpha$, $\alpha=1,\dots,4$ are 
$(\hat{r}_1, \hat{p}_1, \hat{r}_2, \hat{p}_2)$ and  
$\Delta \hat{\xi}_\alpha:=\hat{\xi}_\alpha-\langle \hat{\xi}_\alpha \rangle$;
the averages $\langle . \rangle$ is taken with respect to the 
given two-mode density matrix $\rho$, or equivalently, 
using the Wigner distribution of $\rho$.
Then, the canonical commutation relations assume the compact 
form
$[\hat{\xi}_\alpha,\hat{\xi}_\beta]=i\hbar\Omega_{\alpha,\beta}$
with $\Omega=i\sigma_y \otimes \id$.
When expressed in terms of $V$, the Heisenberg uncertainty relation
can be invoked in invariant form with respect to canonical transformations
as $\det V\geq \frac{\hbar^2}{4}$ (see e.g. \onlinecite{Simon94}).
The set of the real linear canonical 
transformation generates the symplectic group $Sp(2n,\RR)$ that plays 
an important role in the theory.
Being a symplectic matrix, $V$ can be brought in its diagonal form 
$V_n$ by means of symplectic transformations.
The elements on the diagonal are then called the {\em symplectic eigenvalues}
of $V$.
An analysis of $V_n\Omega$ has unveiled 
an even more powerful invariant form of the Heisenberg uncertainty
principle, 
$
V+\frac{i\hbar}{2}\Omega\geq 0
$,
where the positive semi-definiteness means that
all symplectic eigenvalues are non-negative.
The uncertainty relation can hence be cast directly in terms of the 
symplectic eigenvalues of the covariance matrix $V$, 
which are the absolute values of the eigenvalues of $-i\Omega V$. 

Some of the aspects of the harmonic systems can be disclosed by 
recognizing that the Gaussian structure of the bosonic states can 
be thought as a certain limit of the algebraic structure of the qubits
in the sense that  ${\rm Sp}(2,\RR)\simeq SL(2,\CC)$. 
The latter is the invariance group relevant for qubit entanglement 
classification and quantification~\cite{Duer00,VerstraeteDM03,OS04}.

We now introduce the notion of bipartite entanglement for Gaussian states.
In complete analogy to the finite-dimensional case, a state 
is termed separable if it is a mixture of product states.
In particular, all classical states, i.e.
\beq\label{separable}
\rho=\int\d^2 z_1\, \d^2 z_2\; P(z_1,z_2) 
                   \ket{z_1}\bra{z_1}\otimes \ket{z_2}\bra{z_2}
\eeq
with positive $P(z_1,z_2)$ are separable.

It was Simon~\cite{Simon00} that first proved the Peres-Horodecki 
Positive Partial Transpose criterion being necessary and sufficient
for entanglement of two harmonic oscillator modes, again in complete 
analogy to a system of two qubits.
The effect of the transposition of the density matrix is a
sign change in the momentum variables of the Wigner function (\ref{Wigner}).
Consequently, a partial transposition induces a sign change of 
those momenta in the phase space vector, where the transposition 
should act on.
For an entangled state, the partial transposition $\tilde{V}$ of
its covariance matrix $V$ might then have symplectic eigenvalues smaller
than $\hbar/2$.
This can be detected by the logarithmic negativity as defined
from the symplectic (doubly degenerate) eigenvalues 
$\{\tilde{c}_i\, ;\, i=1\dots n\}$ of $\tilde{V}/\hbar$~\cite{Vidal02}
\beq\label{logneg}
E_{N}(V)=-\sum_{i=1}^n \log_2 (2\tilde{c}_i)\quad .
\eeq 
These important results paved the way towards a systematic analysis 
of multipartite systems of distinguishable bosonic modes.


\section{Model systems}
\label{mod}
This section is devoted to the basic properties 
of the model systems that will be analyzed in the rest of the review 
(in several cases we concentrate on one-dimensional systems).

\subsection{Spin models}
Interacting spin models~\cite{spin-textbook,Schollwock} provide a paradigm 
to describe a wide range of many-body systems. 
They account for the effective interactions in a variety of very 
different physical contexts ranging from high energy to nuclear 
physics~\cite{Polyakov77,Belitsky}. In 
condensed matter beside describing the properties of magnetic compounds 
(see~\onlinecite{Matsumoto} for a recent survey), they capture several aspects 
of high-temperature superconductors, quantum Hall systems, heavy fermions,
just to mention few important examples.
Hamiltonians for interacting  spins   can be realized artificially in
Josephson junctions arrays~\cite{FAZIO-JJA} or with neutral atoms 
loaded in  optical lattices~\cite{Jane,duan03,Cirac-spin}. 
Interacting spins are paradigm systems for 
quantum information processing~\cite{nielsen}.

\subsubsection{Spin-$1/2$ models with short range interactions}
\label{models-spin-short}

A model Hamiltonian for a set of localized spins interacting 
with nearest neighbor exchange coupling on a $d$-dimensional lattice is    
\begin{eqnarray}
  {\cal {H}}(\gamma,\Delta,h_z/J) {=} & \frac{J}{2}& \sum_{\langle i,j \rangle} 
                              \left[(1+\gamma) S_i^x S_{j}^x +
                              (1-\gamma) S_i^y S_{j}^y \right] 
                              \nonumber \\
                            + & J &\Delta \sum_{\langle i,j \rangle} 
                            S_i^z S_{j}^z - h_z \sum_{i} S_i^z \; .
\label{general-spin}
\end{eqnarray}
In the previous expression $i,j$ are lattice points, 
$\langle \cdot \rangle$ constraints the sum over nearest neighbors 
and $S_i^\alpha$ ($\alpha=x,y,z$) are spin-$1/2$ operators. 
A positive (negative) exchange coupling $J$ favors antiferromagnetic (ferromagnetic) 
ordering in the $xy-$plane. The parameters $\gamma$ and $\Delta$ account for the anisotropy in 
the exchange coupling in $z$ direction, and 
$h_z$ is the  transverse magnetic field. There are only 
very few exact results concerning ${\cal {H}}(\gamma,\Delta,h_z/J)$  
in dimension $d>1$. 
The ground state of Eq.(\ref{general-spin}) is in general entangled. It exists however,
for any value of the coupling constants $\gamma$ and $\Delta$, $J>0$ a point in $d=1,2$ (for 
bipartite lattices) where the ground state is factorized~\cite{Kurmann82,Firenze05}. 
It occurs at the so called factorizing field $h_{\rm f}$ given by 
$
                     h_{\rm f} = \frac{z}{2}J \sqrt{(1+\Delta)^2 - (\gamma/2)^2}
$ 
where $z$ is the coordination number. 

In $d=1$ the model is exactly solvable in several important cases. 
In the next two paragraphs we discuss  the anisotropic quantum $XY$ model 
( $\Delta =0$ and $  0\leq\gamma\leq 1$)  and the $XXZ$ model ($\gamma =0$).
Also the $XYZ$-model in zero field, $\gamma \ne 0, \Delta \ne 0$ can be solved 
exactly but it will not be discussed here 
(see~\onlinecite{Takahashibook99} for a review).

\paragraph{$\Delta=0$: Quantum $XY$-model}
\label{models-qIsing} 

The  quantum Ising 
model corresponds to  $\gamma =1$ while the (isotropic) $XX$-model is obtained for 
$\gamma = 0$. In the  isotropic case the model possesses an additional symmetry 
resulting in the conservation of the magnetization along the $z$-axis. 
For any value of the anisotropy the model can be solved exactly~\cite{LIEB,PFEUTY,McCOY}.   
By first applying  the Jordan-Wigner transformation 
$
     c_k = e^{i \pi \sum_{j=1}^{k-1} \sigma_j^+ \sigma_j^-} \, \sigma_k^-
$
(with $\sigma^{\pm}=(1/2) (\sigma^x\pm i \sigma^y)$) the $XY$ model can be 
transformed onto a free fermion Hamiltonian
\begin{equation}
     H = \sum_{i,j} \left[ c_i^\dagger A_{i,j} c_j +
     \frac{1}{2} ( c_i^\dagger B_{i,j} c_j^\dagger + \mathrm{h.c.}) \right] +
     \frac{1}{2} \sum_i A_{i,i} \;\; .
\label{eq:quadratic}
\end{equation}
In the previous equations $c_i, c_i^\dagger$ are the annihilation and creation operators
for the spinless Jordan-Wigner fermions.
The two matrices {\bf A}, {\bf B} are defined as
$     
A_{j,k}  =  - J \big( \delta_{k,j+1} + \delta_{j,k+1} \big) - h_z \delta_{j,k}
$
and
$
     B_{j,k}   =  - \gamma J \big( \delta_{k,j+1} - \delta_{j,k+1} \big)
$.
For the case of periodic boundary conditions on the spins, an  
extra boundary term appears in the fermionic Hamiltonian which depends on the 
parity of the total number of fermions $N_F$. 
Notice that although $N_F$ does not commute with 
the Hamiltonian the parity of $N_F$ is conserved. 
A generic quadratic form, like Eq.~(\ref{eq:quadratic}), can be diagonalized in terms 
of the normal-mode spinless Fermi operators by first going to the Fourier space and 
then performing a Bogoliubov transformation.

The properties of the Hamiltonian are governed by 
the dimensionless coupling constant $\lambda=J/2h$.
In the interval  $0<\gamma\le 1$ the system undergoes a second order quantum phase
transition at the critical value $\lambda_c=1$. The order parameter is
the magnetization in $x$-direction, $\langle S^x\rangle $, which is different 
from zero for $\lambda >1$. In the ordered phase the ground state has a two-fold 
degereracy reflecting a global phase flip symmetry of the system. 
The magnetization along the $z$-direction,
$\langle S^z\rangle $, is different from zero for any value of $\lambda$, 
but presents a singular behavior of its first derivative at the transition. 
In the whole interval $0<\gamma\le 1$ the transition belongs to the Ising universality class. 
For $\gamma=0$ the quantum phase transition is of the 
Berezinskii-Kosterlitz-Thouless type. 

As it was discussed in Section~\ref{measure-pairwise-pure} and \ref{measure-pairwise-mixed}
one- and two site- entanglement
measures can be related to various equal-time spin correlation functions
 (in some important cases also the block entropy can be reduced to 
the evaluation of two-point correlators) 
$
       M^{\alpha}_l(t) =\langle \psi | S_l^\alpha(t) |\psi  \rangle
$
and
$
g^{\alpha\beta}_{lm}(t) =
\langle \psi | S_l^\alpha(t) S_m^\beta(t)| \psi  \rangle
$.
These correlators have been calculated for this class of models in the case of thermal
equilibrium~\cite{LIEB,PFEUTY,McCOY}. These can be recast in the form of
Pfaffians that for stationary states reduce to 
Toeplitz determinants (i.e. determinants, whose entries depend only on 
the difference of their row and column number). 
it can be demonstrated that the equal time correlation 
functions can be expressed as a {\it sum} of Pfaffians~\cite{Amico-Osterloh}.

\paragraph{$\gamma=0$: $XXZ$ model}
\label{XXZ}
The two isotropic points $\Delta=1$ and $\Delta=-1$ describe the antiferromagnetic 
and ferromagnetic chains respectively. In one dimension the $XXZ$ Heisenberg model 
can be solved exactly by the Bethe Ansatz technique~\cite{Bethe31} 
(see e.g.~\onlinecite{Takahashibook99}) and the  correlation 
functions can be expressed in terms of certain determinants 
(see~\onlinecite{Korepin-book} for a review). 
Correlation functions, especially for intermediate distances, are in general 
difficult to evaluate, although important steps in this direction have been 
made~\cite{Mallet,Korepin-inverse}.

The zero temperature phase diagram of the XXZ model in zero magnetic field 
shows a gapless phase in the interval $-1\le \Delta < 1$. Outside this interval
the excitations are gapped. The two phases are separated by a 
Berezinskii-Kosterlitz-Thouless  phase transition at $\Delta=1$ while at $\Delta=-1$ 
the transition is of the first order. 
In the presence of the  external magnetic field a finite  energy gap appears 
in the spectrum. The universality class of the transition is not 
affected, as a result of the conservation of the total spin-$z$ 
component~\cite{Takahashibook99}.  

When one moves away from one dimension, exact results are rare. Nevertheless is it now 
established that the ground state of a two-dimensional antiferromagnet 
possesses N\'eel long range order~\cite{Dagotto,Manousakis}.

\subsubsection{Spin-$1/2$ models with infinite range interaction}

In this case each spin interacts with all the other spins in the system with the 
same coupling strength
$
  {\cal {H}} {=}-{{J}\over{2}}   \sum_{ij} \left[ 
                         S_i^x S_{j}^x +
                        \gamma S_i^y S_{j}^y \right]- 
                        \sum_{i} \vec{h}_i \cdot \vec{S}_i~.
$
For site-{\em independent} magnetic field $h_i^\alpha=h_\alpha \, \forall i$, 
this model was originally proposed by  
Lipkin, Meshkov and  Glick (LMG)~\cite{LIPKIN1,LIPKIN2,LIPKIN3} to describe a 
collective motion in nuclei.
In this case the dynamics of the system can be described in terms of a collective 
spin $S_\alpha=\sum_jS^\alpha_j$. The previous Hamiltonian reduces to  
\begin{equation}
{\cal {H}}=- {{J}\over{2}} [(S^x)^2+\gamma (S^y)^2]- \vec{h} \cdot \vec{S}\;\; .
\label{LMG}
\end{equation}
Since the Hamiltonian commutes with the Casimir operator ${\bf S}^2$ 
the eigenstates can be labeled by the representation $S$ of the collective 
spin algebra, at most linear in the number $N$ of spins; 
this reduces (from $2^N$ to $N/2$) the complexity of the problem. A further 
simplification is achieved at the  supersymmetric point
corresponding  to  $J^2\gamma=4 h_z$, where  the Hamiltonian can be 
factorized in two terms linear in the collective spin~\cite{Unanyan-SUSY}; 
then the  ground state can be obtained explicitly. 
For a ferromagnetic coupling ($J>0$) and $h_x=h_y=0$ the system undergoes a second 
order  quantum phase transition at $\lambda_c=1$, characterized by mean 
field critical indices~\cite{PFEUTY-LONG}. The average
magnetization (for any $\gamma$) $m_z=\langle S_z\rangle/N$ saturates 
for $\lambda \le \lambda_c$ while it is suppressed for $\lambda >\lambda_c$.
For $h_y=0$, $h_z < 1$ and $\gamma=0$ the model exhibits a first order transition at 
$h_x=0$~\cite{arias} while for an antiferromagnetic coupling and $h_y=0$
a first order phase transition at $h_z=0$ occurs, where the magnetization 
saturates  abruptly at the same value $m^z=1/2$ for  any $\gamma$'s. 

The model Hamiltonian introduced at the beginning of this section embraces an important class 
of interacting fermion systems as well. By interpreting the non homogeneous magnetic 
field as  a set of energy levels $(h_z)_i\equiv -\epsilon_i$, for $h_x=h_y=0$ and 
$\gamma =1$, it expresses 
the BCS model. This can be realized by noticing that the operators 
$
S^-_j := c_{j\uparrow} c_{j\downarrow}, \;\; S^+_j:=(S^-_j)^\dagger \;\;
S^z_j := (c^\dagger_{j\uparrow} c_{j\uparrow}+c^\dagger_{j\downarrow} c_{j\downarrow}-1)/2$ 
span the $su(2)$ algebra in the 
representation $1/2$.
In the fermion language the Hamiltonian reads
$
          {\cal {H}}_{BCS} = \sum_{j,\sigma=\{\uparrow,\downarrow\}}\epsilon_j 
                             c^{\dagger}_{j\sigma}c_{j\sigma}
                           - \frac{J}{2} \sum_{ij} c^{\dagger}_{j\uparrow}
                               c^{\dagger}_{j\downarrow}
                               c_{i\downarrow}c_{i\uparrow} \; .
$

Both the LMG and the BCS type  models can be solved exactly by Bethe 
Ansatz~\cite{Richardson,Richardson-Sherman}  (see 
also~\onlinecite{Sierra} for a review) as they are quasi classical descendants 
of the six vertex model~\cite{AMIC0-OFF,DILORENZO,ORTIZ}.

\subsubsection{Frustrated spin-$1/2$ models}

Frustration arises in systems where certain local constraints prevent the 
system from reaching a local energy minimum. The constraints can be of geometric 
nature (for example the topology of the underlying lattice) or of dynamical 
nature (two terms in the Hamiltonian tending to favor incompatible 
configurations).  A classical example of the first type is that of 
an antiferromagnet in a triangular lattice with Ising interaction. 
At a quantum mechanical level this 
phenomenon can result in the appearance of ground state degeneracies.
The equilibrium and dynamical properties of frustrated systems have been 
extensively studied in the literature~\cite{diep} both in classical and 
quantum systems. 

A prototype of frustrated models in one dimension is the antiferromagnetic 
Heisenberg model with nearest and  next-nearest neighbor interactions.
This class of models were discussed originally to study the spin-Peierls transition
\cite{Schollwock}. The Hamiltonian reads 
\begin{equation}
H_{\alpha}=J \sum_{i=1}^N (\vec{S}_i\cdot \vec{S}_{i+1}+
        \alpha \vec{S}_i\cdot \vec{S}_{i+2}) 
\label{frustrated-Heisenberg}
\end{equation}
Analytical calculations~\cite{Haldane82}, corroborated by numerical 
result~\cite{Okamoto} indicate that at $\alpha\approx 1/4$ there is a quantum phase 
transition to a dimerized 2-fold degenerate ground state, where 
singlets are arranged on doubled lattice constant distances. Such a phase 
is characterized by a finite gap in the low lying excitation spectrum.  

The Majumdar-Ghosh model~\cite{Majumdar69a,Majumdar69b,Majumdar70} is 
obtained from  Eq.(\ref{frustrated-Heisenberg}) for $\alpha=1/2$. The exact ground 
state can be solved by means of matrix product states (see next Section) and it 
is shown to be disordered. It is a doubly degenerate valence bond state made of 
nearest neighbor spin singlets. Although all two-point correlation function vanish, 
a finite four-spin correlation function does reflect an ordered dimerization.

\subsubsection{Spin-1 models}
\label{models-spin1}

Spin-$1$ systems where originally considered to study the quantum dynamics of 
magnetic solitons in antiferromagnets  with single ion anisotropy~\cite{Mikeska}.
In one dimension, half-integer and integer spin chains have very different 
properties~\cite{HALDANE-CONJ1,HALDANE-CONJ2}. Long range order which is established
in the ground state of systems with half-integer spin~\cite{LIEB}, 
may be  washed out for integer spins. In this latter case, the system has a gap in the 
excitation spectrum. A paradigm model of interacting 
spin-$1$ systems is
\begin{equation}
H=\sum_{i=0}^{N}\vec{S}_i \cdot \vec{S}_{i+1}+\beta (\vec{S}_i \cdot \vec{S}_{i+1})^2
\label{haldane-phase}
\end{equation} 
The resulting  gapped phase arises because of the presence of
zero as an eigenvalue of $S^z_i$; the corresponding 
eigenstates represent a spin excitation that can move freely in the chain, 
ultimately disordering the ground state of the system~\cite{Mikeska,Gomez-Santos}.
The so called string order parameter was proposed to 
capture the resulting 'floating' N\'eel order, made of alternating spins 
$|\uparrow\rangle$,  $|\downarrow \rangle$ with strings of $|0\rangle$'s
in between~\cite{denijs} 
\begin{equation}
                 \displaystyle{O_{string}^\alpha=
                  \lim_{R\rightarrow \infty}
                   \langle S^\alpha_i (\prod_{k=i+1}^{i+R-1}
                  e^{i\pi  S^\alpha_k})S^\alpha_{i+R}\rangle } \;\; .
\label{string}
\end{equation}
The ground state of physical systems described by Hamiltonians of the form of 
Eq.(\ref{haldane-phase}) has been studied in great details~\cite{Schollwock}. 
Various phase transitions have been found between antiferromagnetic phases, 
Haldane phases, and a phase characterized by a large density of vanishing 
weights ($S^z_i=0$) along the chain. 

\begin{figure}
\centering
\includegraphics[width=8cm]{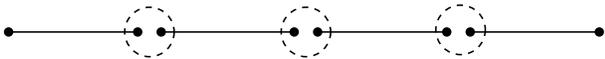}
\caption{A cartoon of the nearest neighbour valence bond state, exact ground 
state of the spin-$1$ model in 
Eq.(\ref{haldane-phase}) for $\beta=1/3$ (AKLT-model). The ground state 
is constructed regarding every $S=1$ in the lattice 
sites  as made of a pair of $S=1/2$, and projecting out the singlet state. 
The singlets are then formed taking pairs of $S=1/2$ in 
nearest neighbor sites. } 
\label{rvb}
\end{figure}

\paragraph{The Affleck-Kennedy-Lieb-Tasaki (AKLT) Model.}
Some features of the phenomenology 
leading to the  destruction of the antiferromagnetic  order  can be put on a 
firm ground for  $\beta=1/3$ (AKLT model), where the ground state of the Hamiltonian in 
Eq.(\ref{haldane-phase}) is known exactly~\cite{AFFLECK}. In this case it was 
proved that the ground state is constituted  by a sea of nearest neighbour  
valence bond states, separated from the first excitation by 
a finite gap with exponentially decaying correlation functions. 
Such a state is sketched in Fig.\ref{rvb}. 
In fact it  is a Matrix Product State (MPS), i.e. it belongs to 
the class of states which can be expressed in the form 
\begin{equation}
  \mid \psi_{MPS}\rangle = \sum_{s_1, \ldots, s_N}^{D} \mbox{Tr} A^{s_1}_1\ldots A^{s_N}_N
                        \mid s_1, \ldots, s_N\rangle  
\label{mps}
\end{equation}
where the matrices $(A^{s_i}_k)_{lm}$ parametrize the state; 
$|s_i>$ denotes a local basis of the $D$-dimensional Hilbert space; the trace contracts the indices $l,m$ labelling bond states of the auxiliary system 
(namely the spin $1/2$ for the AKLT model). 
The dimensions of $A$ depends on the 
particular state considered, if the state is only slightly entangled then the dimension
of $A$ is bounded by some $D_{MPS}$. MPS, first discussed 
in~\cite{fannes},  appear naturally in the Density Matrix Renormalization Group (DMRG)~\cite{ostlund}.
In one-dimensional non-critical systems they describe faithfully the ground state. 
Infact, as shown by Vidal, 
matrix product states constitute an efficient representation of slightly 
entangled states~\cite{dmrg1}. 

\subsection{Strongly correlated fermionic models}
\label{models-second}
The prototype model of interacting fermions on a lattice is the Hubbard 
model~\cite{HUBBARD-BOOK}
\begin{equation}
                  {\cal H}=  -t \sum_{\langle ij \rangle} 
                     [c_{i,\sigma}^\dagger c_{j,\sigma}+h.c.] +
                    U \sum_i n_{i,\uparrow} n_{i,\downarrow} -\mu N
\label{Hubbardmod}
\end{equation}
where  $c_{i,\sigma}$, $c_{i,\sigma}^\dagger$ are fermionic operators: 
$\{c_{i,\sigma}, c_{j,\sigma'}^\dagger\}=\delta_{i,j}\delta_{\sigma \sigma'}$.  The 
coupling constant $U$ describes the on-site repulsion, $t$ is the hopping amplitude.

The Hubbard model possess an $u(1)\oplus su(2)$ symmetry expressing the 
charge conservation: $u(1)={\rm span}\{N=\sum_{j\sigma} n_{j\sigma}\}$ 
and the invariance under spin rotation: 
$su(2)={\rm span}
\{S^z=\sum_{j} (n_{j\uparrow}- n_{j\downarrow}), S^+=\sum_{j}c_{j,\uparrow}^\dagger 
c_{j,\downarrow},S^-=(S^+)^\dagger \}$. Such a symmetry allows one to employ the total 
charge and magnetization as good quantum numbers.
At half filling $n=N/L=1$ ($\mu=U/2$) the symmetry is enlarged to 
$so(4)=su(2)\oplus su(2)$ by the generator 
$\eta=\sum_{j}(-)^j c_{j,\uparrow}c_{j,\downarrow}$ together with its hermitean 
conjugate~\cite{so4}. It was demonstrated that 
$|\Psi\rangle =(\eta)^N|gs\rangle$ are eigenstates of the Hubbard model 
(in any dimension),  characterized by off-diagonal-long-range-order
 via the mechanism of the so called $\eta$-pairing~\cite{Yang89}.

In one dimension the Hubbard model undergoes a Mott transition at $U=0$ of 
the Berezinskii-Kosterlitz-Thouless type. By means of the 
Bethe Ansatz solution~\cite{Lieb-Wu} it can be demonstrated how the bare electrons 
decay in charge and spin excitations. The phenomenon of spin-charge separation 
occurs at low energies away from half filling. 
For repulsive interaction the half-filled band is gapped in the charge sector; 
while the spin excitations  remain gapless. The mirror-inverted situation occurs for 
attractive interaction where the gap is in the spin excitations instead
(see~\onlinecite{HUBBARD-BOOK} for a recent review). 
    
The Hubbard model in a magnetic field was proved to exhibit two quantum critical points 
at $h^{a\pm}_c=4(|U|\pm 1)$ and half filling
for $U<0$, while there is one at $h^r_c=4(\sqrt{t^2+U^2}-U)$ for $U>0$~\cite{YangK00}.

If a nearest neighbor Coulomb repulsion $V \sum_{\sigma,\sigma',j} 
n_{j\sigma}n_{j+1\sigma'}$ is taken into account in Eq.(\ref{Hubbardmod}),
a spin density wave and a charge density wave  phase appear.
A transition to a phase separation  of high density and
low density regions (see e.g. Ref.~\onlinecite{Clay99}) is also present. 

The bond charge extended Hubbard model, originally proposed
in the context of high $T_c$ superconductivity~\cite{HIRSCH}, include further  correlations in the 
hopping process already  involved in (\ref{Hubbardmod}).  The Hamiltonian 
reads
\begin{equation}
                  {\cal H}=U \sum_i^L n_{i,\uparrow} n_{i,\downarrow}
                  -t[1-x(n_{i,-\sigma}+n_{i+1,-\sigma})]c_{i,\sigma}^\dagger c_{i+1,\sigma}+h.c. 
\label{H-fermions}
\end{equation}
(for $x=0$ the Eq.(\ref{H-fermions}) coincides with the Hubbard model (\ref{Hubbardmod})). 
For $x\neq 0$ the hopping amplitudes are modulated by the  occupancy of the sites 
involved in the processes of tunneling. Because of the particle-hole symmetry, $x$ 
can be restricted in $[0,1]$ without loss of generality. 
For $x=1$  the correlated 
hopping term commutes with the interaction. In this case the exact ground state 
was shown to exhibit a 
variety of quantum phase transitions between insulators and superconducting 
regimes, controlled by the Coulomb repulsion parameter $U$. 
For $x=1$ the phase diagram is shown in Section \ref{pairgr}, Fig.\ref{Anfossi1-2}. 
At $U/t=4$ and $n=1$, a 
superconductor-insulator quantum phase transition occurs; for $-4\le U/t\le 4 $ 
the ground state is characterized by off-diagonal  
long-range order; the low lying excitations are gapless. For $U/t=-4$ a further 
quantum critical point projects the ground state into  the Hilbert subspace spanned by 
singly and doubly occupied states~\cite{Arrachea94,Schadschneider}.
For intermediate $x$ the model has not been solved exactly. 
Numerical calculations indicate a superconducting-insulator transition controlled 
by $U$ and parametrized by $x$. Specifically, for $0\le x\le 1/2$  
the phase is gapped at any non vanishing $U$; for   $1/2< x\le1$ the onset to a 
superconducting phase was evidenced at some finite $U$~\cite{AnfossiBoschi05}.
 
\subsection{Spin-boson models}

A prototype model in this class is that of  a quantum system coupled to a  bath 
of harmonic oscillators (see ~\onlinecite{WEISS} for a review of open quantum 
mechanical systems) known also as the Caldeira-Leggett model. 
In this case the quantum system is a two level system. 
This class of models was intensely 
investigated  to study the quantum-to-classical transition and the corresponding loss 
of quantum coherence~\cite{Zurek}. 

The spin-boson Hamiltonian has the form
\begin{equation}
H_{sb}=-\frac{\delta}{2} S^x+\sum_n \omega_n (a^\dagger_n a_n +\frac{1}{2}) +
\frac{1}{2} S^z\sum_n\lambda_n (a^\dagger_n+a_n)\; ,
\label{spin-boson}
\end{equation}
it can be demonstrated to be equivalent to the anisotropic Kondo model~\cite{Andersonkondo,Guinea}.
The coupling constants $\{\lambda_n\}$ fix the spectral density of the bath:
$
\displaystyle{J(\omega) =(\pi/2})\sum_n\lambda_n^2 \delta(\omega-\omega_n)/\omega_n
$.
At low energy the spectral function can be represented as a power-law: 
$J(\omega)\propto 2\alpha \omega^s \Lambda_0^{1-s}$ where $\alpha$ is 
the parameter controlling the spin-bosons interaction and
$\Lambda_0$ is a ultraviolet cutoff frequency. The power $s$ characterizes the  bath. 
For $s=1$ the bath is called {\em Ohmic}, in this case the model has a second 
order quantum phase transition at $\alpha=1$ from under-damped to over-damped 
oscillations (where the value of the spin is frozen). The value 
$\alpha=1/2$ identifies a cross over regime where the two-level system 
is driven from coherent to incoherent oscillations. If the bath is {\em super-Ohmic}  
($s>1$), the quantum critical point is washed out, while a  cross over occurs 
at $\alpha\sim \log (\Lambda_0/\delta)$.
For {\em sub-Ohmic} baths ($s<1$), several studies indicate the existence of 
a quantum critical point~\cite{Spohn}. The question however is not completely 
settled~\cite{Kehrein,Bulla,Stauber}.

An interesting case is also that of a spin interacting with a single bosonic mode,
$\lambda_n=\lambda \delta_{n,0}$.
\begin{equation}
H =-\frac{\delta}{2} S^x+\omega_0 (a^\dagger_0 a_0 +\frac{1}{2}) +
\frac{\lambda_0}{2} S^z (a^\dagger_0+a_0)
\label{spin-singleboson}
\end{equation}
Such model describes for example an atom interacting with a monochromatic 
electromagnetic field~\cite{COHEN} via a dipole force~\cite{JAYNES}.   
Recently, the dynamics corresponding to (\ref{spin-singleboson}) 
was intensely 
studied in relation to ion traps~\cite{ION-TRAPS} and quantum 
computation~\cite{RECENT-EXP}. The model defined in Eq.(\ref{spin-singleboson}) 
with $S=1/2$ (Jaynes-Cummings model) was generalized and solved exactly to 
consider generic spin~\cite{Tavis}  in order to discuss the
super-radiance phenomenon in cavity-QED.

\subsection{Harmonic lattices}
\label{harmodels}
The Hamiltonian for a lattice of coupled harmonic oscillators
(short: harmonic lattice) can be expressed in terms of 
the phase space vector $\vec{\xi}^T=(q_1,\dots,q_n;p_1,\dots,p_n)$ as
\beq
\label{HamEisert}
H=\vec{\xi}^T\Matrix{cc}{\frac{m}{2}\omega^2 \mathbb{U} &0\\0&\frac{1}{2m}\id_n}\vec{\xi}
\eeq
where $\mathbb{U}$ is the $n\times n$ interaction matrix for the coordinates. 
If the system is translational invariant the matrix $\mathbb{U}$ is a Toeplitz matrix with 
periodic boundaries, also called {\em circulant matrix}~\cite{HornJohnson}.
In the case of finite range interaction of the form
$
\sum_r\sum_{k=1}^n K_r(q_{k+r}-q_k)^2
$
and assuming periodic boundary conditions, its entries are
$\mathbb{U}_{j,j}=1+2\sum_r \alpha_r$ and $\mathbb{U}_{j,j+r}=-\alpha_r$
with $\alpha_r = 2K_r/m\omega^2$.
Since the Hamiltonian (\ref{HamEisert}) is quadratic in the canonical variables 
its  dynamical algebra  is $sp(2n,\RR)$. Then the diagonalization can be achieved by    
$RHR^{-1}$ where $R=\otimes_{\alpha=1}^n \exp{(i \theta_\alpha G_\alpha)}$ 
where $G_\alpha$ is the generic Hermitean element of $sp(2,\RR)$. 

As we discussed in Section~\ref{entharmon} the key quantity that characterizes 
the properties of harmonic systems is the covariance matrix defined in 
Eq.(\ref{covariantm}).
For the resulting decoupled harmonic oscillators 
it is ${\rm diag}\{(m\sqrt{\eta_1}\omega)^{-1},\dots,(m\sqrt{\eta_n}\omega)^{-1};
m\sqrt{\eta_1}\omega,\dots,m\sqrt{\eta_n}\omega)$, where
$\eta_j$ are the eigenvalues of $\mathbb{U}$.
Employing the virial theorem for harmonic oscillators, 
the covariance matrix for a thermal state with inverse temperature
$\beta=1/k_B T$ can be calculated as well
\beq\label{covarianceThS}
V=\Matrix{cc}{(m\omega \sqrt{\mathbb{U}})^{-1}\mathbb{N}_{\beta} &0\\
0&(m\omega \sqrt{\mathbb{U}})\mathbb{N}_{\beta}}\; .
\eeq
where $\mathbb{N}_{\beta} = \id_n+2/(\exp{(\beta \omega \sqrt{\mathbb{U}})}-\id_n)$
The range of the position or momentum correlation functions is related to the 
low lying spectrum of the Hamiltonian. For a gapped systems the correlations 
decay exponentially. The absence of a gap (some eigenvalues of $\mathbb{U}$ tend to zero 
for an infinite system) 
leads to critical behavior of the system and characteristic long ranged correlations.
A rigorous and detailed discussion of the relations 
between the gap in the energy spectrum and the properties of the correlations 
can be found in~\cite{cramer06b}.

\section{Pairwise entanglement}
\label{pairgr}
At $T=0$ many-body systems are most often described by complex ground state wave functions 
which contain all the correlations that  give rise to the various phases 
of matter (superconductivity, ferromagnetism, quantum hall systems, $\ldots$). 
Traditionally many-body systems have been studied by looking for example at their response 
to external perturbations, various order parameters and excitation spectrum. 
The study of the ground state of many-body systems with methods developed in 
quantum information may unveil new properties.  In this Section we classify 
the properties of the ground state of a many-body system according to its 
entanglement. We concentrate on spin systems. Spin variables constitute a good example 
of distinguishable objects, for which the problem of entanglement quantification is 
most developed. 
We will discuss various aspects starting from the pairwise entanglement, we then 
proceed with the properties of block entropy and localizable entanglement. 
Most of 
the calculations are for one-dimensional systems where exact results are available. 
Section~\ref{ddim} will overview the status in the d-dimensional case.  
Multipartite entanglement in the ground state will be discussed later in Section~\ref{multip}.

\subsection{Pairwise entanglement in spin chains}

\subsubsection{Concurrence and magnetic order}
\label{concu.spin}

The study of entanglement in interacting spin systems was initiated with the works on 
isotropic Heisenberg rings~\cite{Gunlycke01,Arnesen01,O'Connors01}. 
O'Connors and Wootters aimed at finding the maximum pairwise entanglement
that can be realized in a chain of $N$ qubits with periodic boundary conditions. 
Starting from the assumption that the state maximizing the nearest 
neighbor concurrence $C(1)$ were an eigenstate of the $z$ component 
of the total spin~\cite{Ishizaka00,Verstraete00,Munro01}  the problem 
was recast to an optimization  procedure similar in spirit to the coordinate 
Bethe ansatz~\cite{Bethe31}:
\footnote{Such a method relies in 
the existence of a 'non interacting picture' where the wave function of the 
system can be written as a finite sum of plane waves; the ansatz is successful 
for a very special form of the scattering among such non interacting pictures.}:
the search for the optimal state was restricted to those cases which  excluded 
the possibility to find two nearest neighbor up spins. 
For fixed number of spins $N$ and $p$ spins up, the state can 
be written as 
$
|\psi \rangle = \sum_{1 \le i_1<\cdots <i_p \le N}
b_{i_1\cdots i_p} |i_1 \cdots i_{p} \rangle
$
($b$ are the coefficients and $i_j$ are the positions of the 
up-spins)
therefore mapping the spin state onto a particle state such that the positions of the 
$p$ particles correspond to those of the up spins. The maximum concurrence within 
this class of states could be related to the ground state of this  
gas of free spinless particles with the result
\begin{equation}
C(1)=- \frac{1}{N} E_{gs}= -\frac{2 \sin{\frac{\pi p}{N-p}}}{\sin{\frac{\pi }{N-p}}}
\label{conc.optim}  
\end{equation} 
Eq.(\ref{conc.optim}) gives a lower bound for the maximal attainable concurrence.
The isotropic antiferromagnetic chain was considered as the physical system closest 
to a perfectly 
dimerized system (classically,  with alternating up and down spins).
It was noticed  however  that the concurrence of the ground state of the antiferromagnetic 
chain is actually smaller than the value of the ferromagnetic chain, 
indicating that the situation is more complicated~\cite{O'Connors01}. 
In order to clarify this point, a couple of simple examples are useful. 
For a system of $N=2$ spins the ground state is a singlet. However for general $N$ (we assume 
an even number of sites) the ground state is not made of nearest-neighbor 
singlets (Resonant Valence Bond (RVB) state). For example the $N=4$ the ground state is
$
        |gs\rangle = (1/\sqrt{6})[2|0100 \rangle+  2 |1000\rangle-|1001 \rangle  
                     -|0110 \rangle - |0011 \rangle - |1100 \rangle ]
$,
different from the product of two singlets. It can be seen that the effect of 
the last two components of the state  is to reduce the concurrence with
respect to its maximum attainable value. Given  the simple relation Eq.(\ref{conc.optim})
between the nearest neighbor concurrence and the ground state energy,
the deviation from the RVB state can be quantified by looking at the difference from the exact 
ground state energy corresponding to the maximum concurrence. This maximum value is 
reached within the set  of eigenstates with zero total magnetization (the 'balanced' states 
in~\onlinecite{O'Connors01}), 
indicating that the concurrence is maximized only on the restricted Hilbert space of  
z-rotationally invariant  states. 
Indications on how to optimize the concurrence were discussed in~\cite{Meyer04,Hiesmayr06}. 
The solution to the problem for $N\rightarrow \infty$ was recently given 
in~\cite{Poulsen06}.
It turns out that the  states with  nearest-neighbors aligned spins (not included 
in~\onlinecite{O'Connors01}), correspond to a 'density-density' interaction in the gas of 
the spinless particles considered above, that hence are important for 
the analysis. 
(in the analogy of the coordinate Bethe ansatz method, 
this provides the 'interacting picture').
Following the ideas of~\cite{Wolf03}, the problem to find the optimum concurrence was shown 
to be equivalent to that of finding the ground state energy of an  effective spin 
Hamiltonian, namely the $XXZ$ model in an external magnetic field. 
The optimal concurrence is found in the gapless regime of the spin model with a magnetization 
$M_z=1-2p/N$. It was further demonstrated that states considered in~\cite{O'Connors01} 
actually maximize the concurrence for $M_z > 1/3$ (for $0\le M_z\le 1/3$ the states 
contain nearest neighbor up-spins). 

The concurrence, beyond nearest-neighbors, in isotropic Heisenberg antiferromagnets 
in an external magnetic field was discussed in~\cite{Arnesen01,wang02,Fubini06}. The 
combined effect of the magnetic field and the 
anisotropy in Heisenberg magnets was studied  in~\cite{jin04} making use of the exact 
results existing for the one-dimensional $XXZ$ model. 
It turns out that the concurrence increases with
the anisotropy  $\Delta$~\cite{kartsev}. 
For strong magnetic fields the entanglement vanishes (the order is ferromagnetic); for large
values of the anisotropy $\Delta$ the state is a classical Neel state with Ising order.  
Except for these cases, quantum fluctuations in the ground state lead to entangled 
ground states.

As we discussed in Sec.~\ref{mod}, in low dimensional spin system there exists a 
particular choice of the coupling constants for which the ground state is 
factorized~\cite{Kurmann82}. This is a special point also from the perspective 
of investigating the entanglement in the ground state.
Several works were devoted to 
the characterization of the entanglement close to the {\em factorizing point}. 
It turns out that the point at which the state of the system becomes separable marks an exchange 
of parallel and anti-parallel sector in the ground state concurrence,
see  Eq.(\ref{C-of-corrs}). 
As this phenomenon involves a global (long-range) reorganization of the state of 
the system, the range of the concurrence diverges. 
(We notice that several definitions of characteristic lengths associated to entanglement decay exist).
The concurrence is often observed to vanish when the two sites are 
farther than $R$ sites apart: the distance $R$ is then taken as the range of the concurrence.
For the $XY$ model it was found that this range is
\begin{equation}
R\propto \left (\ln \frac{1-\gamma}{1+\gamma}\right )^{-1} \ln |\lambda^{-1}-
\lambda_f^{-1}|^{-1}\; .
\end{equation}     
The divergence of $R$ suggests, as a consequence of the monogamy of the 
entanglement~\cite{Coffman00,Osborne06}, that the role of pairwise
entanglement is 
enhanced while approaching the separable point~\cite{Firenze04,Roscilde.jltp,Firenze05}.
In fact, for the Ising model (i.e. $\gamma=1$), one finds that the ratio 
$\tau_2/\tau_1 \rightarrow 1$ when the 
magnetic field approaches the factorizing field $h_{\rm f}$~\cite{Amico06}. 
For $\gamma\neq 1$ and $h_{\rm f}<h_z < h_c$ it was found that $\tau_2/\tau_1$ 
monotonically increases for $h_z\rightarrow h_{\rm f}^+$ and that the 
value $(\tau_2/\tau_1)|_{h_{\rm f}^+}$ increases with $\gamma\to 1$.
The existence of the factorizing  has been also pointed out in other one 
dimensional systems both for short~\cite{Amico06,Firenze04,Roscilde.jltp} and long range 
interactions~\cite{DusVidPRL}. In all these cases the range of the two-site entanglement 
diverges. 
The range of the concurrence was also studied for the $XXZ$~\cite{jin04} where it was 
shown to vary as 
\begin{equation}
R=|{{2A(h_z)}\over{1-4 M_z^2}}|^\theta 
\end{equation}
The exponent $\theta=2$ for finite fields, while it is $\theta=1$ for $h=0$; the 
coefficient $A(h)$ is known exactly in the paramagnetic 
phase~\cite{Lukyanov97,Lukyanov99,Lukyanov03} 
(vanishing magnetization) and in the saturation limit~\cite{VaidyaPRL,VaidyaJMP} 
For generic  $h$ it was calculated numerically in ~\cite{Furusaki04}. 
For the isotropic Heisenberg antiferromagnet, $R=1$~\cite{GuLinLi03}.

In all the previous cases the increasing in the range of the pairwise entanglement 
means that all the pairs at distance smaller than $R$ share a finite amount of 
entanglement (as quantified by the concurrence). There are one-dimensional spin 
systems where the pairwise entanglement has qualitative different features as a 
function of the distance between the sites. An example is the
{\em long-distance entanglement} observed in~\cite{Venuti06b}.  
Given a measure of entanglement $E(\rho_{ij})$, Campos Venuti {\em et al} showed that 
it is possible that $E(\rho_{ij}) \ne 0$ when $|i-j| \to \infty$ in the ground state. 
Long-distance entanglement can be realized in various one-dimensional models as in the 
dimerized frustrated Heisenberg models or in the AKLT model.  For these two models the 
entanglement is highly non uniform and it is mainly concentrated in the end-to-end pair
of the chain. 

Spontaneous symmetry breaking can influence the entanglement in the 
ground state. To see this, it is convenient to introduce the 
``thermal ground state'' 
$
\rho_0=\frac{1}{2} \left (|gs^o\rangle\langle gs^o| +
|gs^e\rangle\langle gs^e|\right )
=\frac{1}{2} \left (|gs^-\rangle\langle gs^-| +
|gs^+\rangle\langle gs^+|\right )
$
which is the $T\to 0$ limit of the thermal state.
In the previous expression  $gs^+$ and $gs^-$ are the symmetry broken states which
give the correct order parameter of the model.
They are superpositions
of the degenerate parity eigenstates $gs^o$ and $gs^e$.
Being convex, the concurrence in 
$gs^{\pm}$ will be larger than in $gs^{o/e}$~\cite{oster06}. 
The opposite is true for the concave entropy of entanglement 
(see Ref.~\cite{Osborne02} for the single spin von Neumann entropy). 
The spontaneous parity symmetry 
breaking does not affect the concurrence in the ground state as long as it
coincides with $C^{I}$, Eq.(\ref{C-of-corrs}): that is, if
the spins are entangled in an antiferromagnetic way~\cite{Syl}.
For the quantum Ising model, the concurrence coincides with $C^{I}$ for
all values of the magnetic field, and therefore, the concurrence is unaffected
by the symmetry breaking, the hallmark of the present QPT.
For generic anisotropies $\gamma$ instead,
also the parallel entanglement $C^{II}$ is observed precisely for 
magnetic fields smaller than the factorizing field~\cite{OsterlohSPIE};
this interval excludes the critical point. 
This changes at $\gamma=0$, where the concurrence
indeed shows an infinite range. 
Below the critical field, the concurrence is enhanced by the parity 
symmetry breaking~\cite{oster06}


\subsubsection{Pairwise entanglement and quantum phase transitions}
\label{ground-critical}

A great number of papers have been devoted to the study  entanglement close to 
quantum phase transition (QPTs). QPT occur at zero temperature. They are induced 
by the change of an external parameter or coupling constant~\cite{Sachdev99}. 
Examples are transitions occurring in quantum Hall systems, localization,  the 
superconductor-insulator transition in  two-dimensional systems. Close to the 
quantum critical point the system is characterized by a diverging correlation 
lenght $\xi$ which is responsible for the singular behavior of different physical 
observables.  The behavior of correlation functions however is not necessarily related 
to the behavior of quantum correlations present in the system. This question seem 
particularly interesting as quantum phase transitions are associated with drastic 
modifications of the ground state. 

The critical properties in the entanglement we are 
going to summarize below allow for a screening of the qualitative change 
of the state of the system experiencing a quantum phase transition. In order to 
avoid possible confusion, it is worth to stress that the study of entanglement close 
to quantum critical points does not provide new understanding to the scaling theory of 
quantum phase transitions. Rather it may be useful in a deeper characterization of 
the ground state wave function of the many-body system undergoing a phase transition.
In this respect it is important to understand, for instance, how the entanglement depends 
on the order of the transition, or what is the role of the range of the interaction to 
establish the entanglement in the ground state. 
In this section we discuss exclusively the pairwise entanglement 
while in the next section we approach the same problem by looking at the block 
entropy~\footnote{QPTs were also studied  
by looking at quantum fidelity~\cite{cozzini1,cozzini2} or the effect of single 
bit operations~\cite{giampaolo1,giampaolo2}}.

\begin{figure}\centering
\includegraphics[width=8cm]{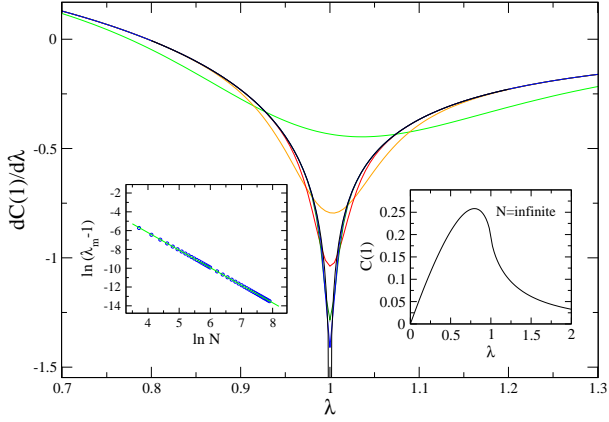}
\vspace*{0.3cm}
\caption{
The derivative of the nearest neighbor concurrence 
as a function of the reduced coupling strength. The  curves correspond to different 
lattice sizes. On increasing the system size, the minimum gets more pronounced
and the position of the minimum tends as  (see the left side inset) 
towards the critical point where for an infinite system a logarithmic divergence 
is present. The right hand side inset shows the behavior of the concurrence 
for the infinite system. The maximum is not related to the critical 
properties of the Ising model. [From \protect\cite{OstNat}]}
\label{a-derivative-gs}
\end{figure}

Pairwise entanglement close to  quantum phase transitions was originally analyzed 
in~\cite{Osborne02,OstNat} for the Ising model in one dimension. Below we 
summarize their results in this specific case. The concurrence tends to zero for  
$\lambda\gg 1$ and $\lambda \ll 1$, the ground state  of the system is fully polarized along 
the $x$-axes ($z$-axes). Moreover the concurrence is zero unless the two sites 
are at most next-nearest neighbors, we therefore discuss only the nearest neighbor 
concurrence $C(1)$ (see however Section~\ref{concu.spin} for cases where there is a 
longer-range pairwise entanglement). 
The concurrence itself is a smooth function of the coupling with a 
maximum close to the critical  point (see the right inset of Fig.\ref{a-derivative-gs}); it was argued that the maximum in the 
pairwise entanglement  does not  occur at the quantum critical point because 
of the monogamy property  
(it is the global 
entanglement that  should be maximal at the critical point).    
The critical properties of the ground state are captured  by the derivatives of
the concurrence as a function of $\lambda$. The results for systems of different 
size (including the  thermodynamic limit) are shown in Fig.\ref{a-derivative-gs}.
For the infinite chain $\partial_\lambda C(1) $ 
diverges on approaching the critical value as 
\begin{equation}
\partial_\lambda C(1) \sim \frac{8}{3 \pi^2} \ln|\lambda-\lambda_c| \;\; .
\label{nn-concurrence}
\end{equation}      				
For finite system the precursors of the critical behavior 
can be analyzed by means of finite size scaling. In the critical region the concurrence 
depends only on the combination $N^{1/\nu}(\lambda-\lambda_m)$ where  
$\nu$ is the critical exponent governing the divergence of the correlation lenght and 
$\lambda_m$ is the position of the minimum
(see the left inset of Fig.\ref{a-derivative-gs}). 
In the case of log divergence the scaling ansatz has to be adapted and takes the 
form
$
\partial_\lambda C(1)(N,\lambda) -
\partial_\lambda C(1)(N,\lambda_0) 
\sim
Q[N^{1/\nu}\delta_m\lambda] - Q[N^{1/\nu}\delta_m\lambda_0]
$
where $\lambda_0$ is some non critical value, $\delta_m (\lambda) = \lambda-\lambda_m$ 
and $Q(x)\sim Q(\infty) \ln x$ (for large x). 
Similar results have been obtained for the $XY$ universality class~\cite{OstNat}. 
Although the concurrence describes short-range properties, nevertheless scaling 
behavior typical of continuous phase transition emerges.

For this class of models the concurrence coincides with $C^{I}$ in Eq.(\ref{C-of-corrs}) 
indicating that the spins can only be entangled in a antiparallel way (this is a peculiar 
case of $\gamma=1$; for generic anisotropies also the parallel entanglement 
is observed). 
The analysis of the finite size scaling  in the, so called, period-$2$ and 
period-$3$ chains where the exchange coupling varies every second and third lattice 
sites respectively, leads to the same 
scaling laws in the concurrence~\cite{Zhang05}. 

\begin{figure}
\centering
\includegraphics[width=7cm]{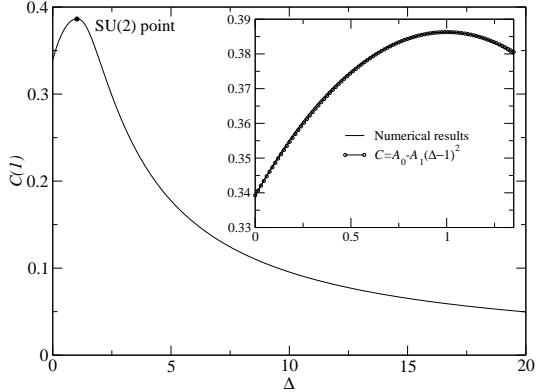}
\caption{Nearest neighbor concurrence of the $XXZ$ model. [From~\protect\cite{GuLinLi03}]}
\label{xxz-gs}
\end{figure}

The concurrence was found to be discontinuous at  the first order ferromagnetic 
transition $\Delta=-1$ in the $XXZ$ chain~\cite{GuLinLi03}(see \onlinecite{glaser} 
for explicit formulas relating the concurrence and correlators for the XXZ model in various regimes). This  result can be understood 
in terms of the sudden change of the wave function occurring because of 
the level crossing characterizing these type of quantum critical points.
The behavior of the two-site entanglement at the continuous quantum
critical point of the Kosterlitz-Thouless type $\Delta=1$ separating the $XY$ 
and the antiferromagnetic phases is more complex. In this case the nearest neighbor 
concurrence (that is the only non vanishing one) reaches a maximum as shown in 
Fig.\ref{xxz-gs}. Further understanding on such behavior can be achieved by analyzing  
the symmetries of the model. At the antiferromagnetic point the ground state 
is an $su(2)$ singlet where nearest neighbor spins tend to form 
singlets; away from $\Delta=1$, 
this behavior is 'deformed' and the system has the tendency to 
to reach a state  of the type $\otimes|\phi_q^j\rangle$ made of '$q$-deformed singlets' 
corresponding to the quantum algebra $su_q(2)$ with $2\Delta=q+q^{-1}$~\cite{Pasquier90}. 
This allows to rephrase the existence of the maximum in  the concurrence 
as the loss of entanglement associated to the q-deformed symmetry of the 
system away from $\Delta=1$ (note that $q$-singlets  are less entangled than 
the undeformed ones). This behavior can be traced back to the properties of 
the finite size spectrum~\cite{GuTianLin-spectra}.  
In fact, at $\Delta=1$ the concurrence can be related to the eigenenergies.
The maximum arises since both the transverse and longitudinal orders are power law 
decaying at this critical point, and therefore the excited states contribute to 
$C(1)$ maximally.   

Studies of finite size energy spectrum of  other models like the dimerized Heisenberg 
chain~\cite{Sun05} and  Majumdar-Ghosh model (Eq.(\ref{frustrated-Heisenberg}) 
with $\alpha=1/2$) show how level crossings in the energy spectrum affect 
the behavior of the bipartite entanglement occurring at the quantum phase 
transition~\cite{GuTianLin-spectra}. 

\paragraph{LMG model}
Because of the symmetry of the LMG models (see Eq.\ref{LMG}) any two 
spins are entangled in the same way. The concurrence $C$ is independent on the 
two site indices, it can be obtained exploiting the explicit expression 
of the eigenstates. Due to the monogamy of entanglement the result must be rescaled 
by the coordination number, $C_{R}=(N-1) C$, to have a finite value in the thermodynamic. 
For the ferromagnetic model~\cite{Vidal-Mosseri03}, 
it was proved that  close to the  continuous QPT, 
$\lambda=1$  characterizing the ferromagnetic LMG model, 
the derivative of the concurrence  diverges, but, differently from 
Ising case, with a power law. It is interesting that 
$C_{R} $ can be related to, the so called, spin squeezing parameter 
$\Sigma=2 \sqrt{\Delta S_{n_\perp}}$~\cite{Wang03}, measuring the spin 
fluctuations in a quantum correlated state (he subscript $n_\perp$ indicates
 a perpendicular axes to $\langle \vec{S}\rangle$). The relation reads
$\Sigma=\sqrt{1-C_R}$
\begin{figure}[ht]
\includegraphics[width=8.3cm]{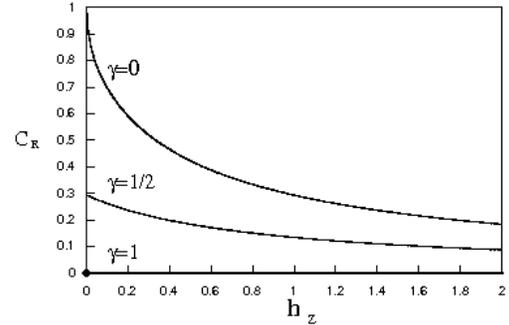}
\caption{The rescaled concurrence of the antiferromagnetic LMG model.
The first order transition occurs at $h=0$. [From \cite{Vidal-Mosseri03}]}
\label{Range-gs}
\end{figure}
According to~\cite{Lewenstein98} the two-spins reduced density operator can be decomposed 
in a separable part and a pure entangled state $\rho_e$ with a certain weight $\Lambda$. 
Such a decomposition leads to the relation $C(\rho) = (1-\Lambda) C(\rho_e) $.
Critical spin fluctuations are related to the concurrence of the pure state $C(\rho_e)$  
while the diverging correlation  lenght is related to the weight $\Lambda$~\cite{shimizu}.
The analysis of critical entanglement at the first order quantum critical point of the 
antiferromagnetic LMG model shows that~\cite{Vidal-Mosseri03} the discontinuity
is observed directly in the concurrence also for spin interacting with  a long-range,
see Fig.\ref{Range-gs}.

\paragraph{Pairwise entanglement in spin-boson models}

\begin{figure}
\centerline{
\includegraphics[clip=false,width=7cm]{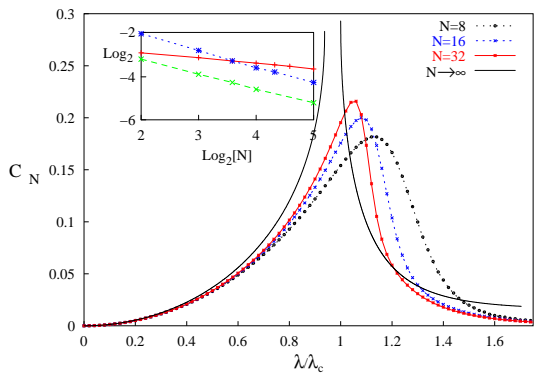}
} 
\caption{ The rescaled concurrence between two atoms in the Dicke mode. 
The concurrence  is rescaled in both for finite $N $ and in the thermodynamic 
limit.  The inset shows the finite size scaling. [From~\protect\cite{lambert04}]}
\label{rescal-conc}
\end{figure}

We first discuss the Tavis-Cummings model defined in Eq.(\ref{spin-singleboson}).
In this model the spin $S$ is proportional to the number of atoms, 
all interacting with a single mode radiation field. The  pairwise entanglement 
between two different atoms undergoing the  super-radiant  quantum phase 
transition~\cite{lambert04,lambert05,reslen} can be investigated through the 
rescaled concurrence $C_N=NC$, see Fig.\ref{rescal-conc}, 
similarly to what has been discussed above for the LMG models. 
In the thermodynamic limit the spin-boson model can be mapped onto a 
quadratic bosonic system  trough an Holstein-Primakoff transformation~\cite{emary03}. 
Many of the properties of the Tavis-Cummings model bear similarities with the 
ferromagnetic LMG model. In the thermodynamic limit the concurrence  
reaches a maximum value $1-\sqrt{2}/2$ at  the super radiant quantum phase 
transition with a square root singularity (see also~\onlinecite{schneider02}). 
The relationship between the squeezing of the state and entanglement was highlighted  
in~\cite{Sorensen} and analyzed in more details in~\cite{stockton03} where it was 
also suggested how to deal with 
entanglement between arbitrary splits of symmetric Hilbert spaces 
(like the Dicke states span). 

Entanglement between qubits and a single mode and between two spins 
with an Heisenberg interaction of the $XXZ$ type, additionally coupled to a 
single bosonic field was considered in~\cite{liberti,liberti1} and in~\cite{He06} respectively.

\subsubsection{Entanglement versus correlations in spin systems.}

From all the results summarized above it is clear that the anomalies 
characterizing the quantum critical points are reflected in the two-site 
entanglement. At a qualitative level this arises because of  
the formal relation between the correlation functions and  
the entanglement. 
A way to put this observation on a quantitative ground is provided
by a generalized Hohenberg-Kohn theorem~\cite{Wu06}. 
Accordingly, the ground state energy can be considered as a unique function 
of the expectation values of certain observables. 
These, in turn, can be related to
(the various derivatives of) a given 
entanglement measure~\cite{WuLidar04,Campos-Venuti06}.  
 
Specifically, for an Hamiltonian of the form $H=H_0+\sum_l \lambda_l A_l$
with control parameters $\lambda_l$ associated with  operators $A_l$, it
can be shown that the ground state reduced operators of the system 
are well behaved functions of $\langle A_l \rangle$. Then, any entanglement 
measure related to reduced density operators, $M=M(\rho)$ is a function of  
$\langle A_l \rangle$ (in absense of ground state degeneracy)
by the Hellmann-Feynman theorem: 
${\partial E}/{\partial \lambda_l}=
\langle{\partial E}/{\partial \lambda_l}\rangle=  \langle A_l \rangle$. 
Therefore it can be proved that 
\begin{equation}
M(\langle A_l \rangle )=M(\frac{\partial E}{\partial \lambda_l})
\label{ent-dft}
\end{equation}
where $E$ is the ground state energy.
From this relation it emerges how the critical behavior of the system is 
reflected in the anomalies of the entanglement. In particular, first order 
phase transition are associated to the anomalies of $M$ while second order 
phase transitions correspond to a singular behavior of the 
derivatives of $M$.
Other singularities like those  noticed in the concurrence for models with 
three-spin interactions~\cite{Yang05}, are due to the non-analyticity 
intrinsic in the definition of the concurrence as a maximum of two
analytic functions and the constant zero.

The relation given in Eq.(\ref{ent-dft}) was constructed  explicitly for the 
quantum Ising, $XXZ$, and LMG models~\cite{Wu06}.  
For the Ising model: $\sum_l \lambda A_l=h\sum_lS_l^z$; the divergence 
of the first derivative of the concurrence is then determined by the non analytical 
behavior of $\langle S^x S^x\rangle$\cite{WuLidar04}.  
For the XXZ model: $\sum_l \lambda_l A_l= \Delta\sum_l S^z_l S^z_{l+1}$. 
At the transition point $\Delta=1$ both
the purity and the concurrence display a maximum. 
It was proved that such a maximum is reflected also 
in a stationary point of the ground state energy as a function 
of $\langle S^z_i S^z_{i+1}\rangle$; the concurrence is continuous since the 
Berezinskii-Kosterlitz-Thouless transition is of infinite order.
A relevant caveat to Eq.~(\ref{ent-dft}) is constituted by  the uniaxial-LMG  
model in a transverse field (with $h_y=0$ and $\gamma =0$) that displays a first 
order QPT for $h_x=0$. The concurrence is continuous at the transition since 
it does not depend on the discontinuous elements of the reduced density 
matrix~\cite{Vidal-Palacios03}. 

The relation between entanglement and criticality was also studied in  the 
spin-$1$ $XXZ$ with single ion anisotropy. It was established that 
the critical anomalies  in the entropy experienced at the Haldane-large- $D$
(if  an axial 
anisotropy  $D \sum_i (S^z_i)^2$ is added to the Hamiltonian in Eq.(\ref{haldane-phase}))
transition fans out from the singularity of the local order parameter 
$\langle (S^z)^2\rangle$~\cite{Campos-Venuti06}.

A way to study the general relation between entanglement and critical 
phenomena was also pursued in ~\cite{Haselgrove03}.
It was argued how for systems with finite range interaction
a vanishing energy gap in the thermodynamic limit 
is an essential condition for the ground state to have 
non-local quantum correlations between distant subsystems.
    
\subsubsection{Spin models with defects}
\label{spin-def}

The problem to characterize entanglement in chains with defects was addressed
first for the quantum $XY$ models with a single defect
in the exchange interaction term of the Hamiltonian~\cite{Osenda03}. 
It was found that the effect of the impurity is to pin the entanglement.
Moreover the defect can induce a pairwise entanglement
on the homogeneous part of the system that were disentangled in the pure system. 
Even at the quantum critical point 
the finite size scaling of the critical anomaly of the concurrence is 
affected by the distance from the impurity.
This basic phenomenology was observed in a variety of different situation that we
review below.

The presence of two defects has been analyzed in the $XXZ$ chain. 
It turns out~\cite{Santos03} that various type of entangled states  
can be created in the chain by  spin flip excitations located at 
the  defects positions. The entanglement oscillates between 
the defects with a period that depend on their distance. Also the anisotropy 
$\Delta$ of the chain is a relevant parameter controlling the entanglement 
between the defects. Small anisotropies can suppress the 
entanglement~\cite{Santos-Rigolin}. The way in which this kind of localization 
can be exploited for quantum algorithms  was studied in~\cite{Santos-et.al}.
The entanglement  was also studied in systems with defects  
in the presence of an external magnetic field~\cite{Apollaro}.  
It was demonstrated that such defect
can lead to a entanglement localization within a typical lenght which coincides 
with the localization lenght. 

A possible way to mimick a defect is to change the boundary conditions.
The concurrence was studied for the ferromagnetic spin $1/2$ $XXZ$ chain with antiparallel 
boundary magnetic field which give rise to a term in the Hamiltonian of the form  
$H_{boundary}=h (S_1-S_N)$~\cite{Alcaraz04}. The boundary
field  triggers the presence a domain walls in the system that induces a first order 
phase transition between ferromagnetic and kink-type phases at 
$h_c=\sqrt{\Delta^2-1}$.
In the ferromagnetic  phase the pairwise entanglement vanishes. 
In the kink-type phase the concurrence acquires a finite value (for $h=0$ 
the ground state is factorized). For  a finite chain of  lenght $L$ and it is
enhanced at the center of the chain. 
In the gapless and in the antiferromagnetic regimes oscillation in   
nearest neighbor entanglement are established in the system resulting from  the tendency 
to reach the antiferromagnetic order. The oscillations are more pronounced in the gapped phase.
Finally a critical  inflection point was noticed in the measure of 
Meyer and Wallach~\cite{Wallach} for the global entanglement at the transition point. 
The spin $1/2$ $XXX$ antiferromagnetic chain with open boundary conditions 
with single defect was also studied in~\cite{Wang04}. It was proved that 
it exists a threshold value of the coupling between the impurity and the rest of the system 
at which the concurrence between them is switched on; for smaller values the entanglement is
dimerized in such a way that the monogamy property prevent the impurity to be 
entangled with the rest of the chain.

The case of many defects was also studied. 
For the quantum Ising model it was found~\cite{Huang04} 
that the disorder can shift the point at which the concurrence is maximum, eventually 
washing  out the critical behavior (strong disorder). The concurrence 
tends to be suppressed at the lattice site corresponding to the 
center of the gaussian; such effect is more robust for near  critical chains.
Quantum $XY$ and $XXZ$ chains with  
a gaussian disorder in the exchange interaction,  have been also studied to 
investigate how the quantum criticality of the 
concurrence is robust by the insertion of the inhomogeneities in the chain~\cite{Cai06,hoyos}.

\subsection{Two and three dimensional systems}
\label{ddim}
In higher dimensions nearly all the results were obtained by means of numerical 
simulations. 
The concurrence and localizable 
entanglement in two dimensional  quantum $XY$ and $XXZ$ models~\cite{Syl2} were considered. 
The calculations are based on Quantum Monte Carlo (QMC) simulations  and the use of stochastic series 
expansion for spin systems~\cite{QMC-sandvik1,QMC-Syljuasen}.
Although the concurrence for the 2d models results to be qualitatively very similar 
to the one-dimensional case, it is much smaller in magnitude.
It is the monogamy that limits the entanglement shared among the number of neighbor 
sites (which is larger in two dimensions as compared with chains).
Finally, it is observed that the  maximum in the concurrence occurs at a position very  closer to the critical point than in the $1D$ case.     

By studying appropriate bounds (concurrence of assistance and the largest 
singular value of the connected correlation functions), it was proved for the $XXZ$ model 
that the localizable entanglement is long ranged in the $XY$ region up to the 
isotropic antiferromagnetic point. Similarly to the case 
of the quantum $XY$ chain, the  bounds for the localizable entanglement are very tight 
in this case.

The pairwise entanglement in the $d$-dimensional XXZ model was 
studied in~\cite{Gu-d-dim}. The concurrence reaches its  highest
 value  at the antiferromagnetic quantum critical point $\Delta=-1$. A spin-wave analysis, 
corroborated by numerical exact diagonalization indicates that the concurrence 
develops a cusp in the thermodynamic limit, only for $d\ge 2$. Such  
behavior can be explained by noting that the  level crossing between the 
ground and the first excited states occurring at the antiferromagnetic 
point, causes  a non-analyticity in the   ground state energy. 
The enhanced  pairwise entanglement at the 
antiferromagnetic point  together with its non analyticity support 
the conjectured existence of long-range order for two dimensional 
antiferromagnets. Further support to this conjecture is the strong size dependence 
of the Von Neumann entropy that 
becomes singular in thermodynamic limit~\cite{GuTianLin05}.

The ground state entanglement in two dimensional $XYZ$ model were analyzed 
in~\cite{Firenze05} by means of quantum Monte Carlo simulations. The divergence 
of the derivative of the concurrence at the continuous phase transition, observed in 
$d=1$, was confirmed; also in this case the range of the pairwise 
entanglement extends only to few lattice sites. By studying 
the one- and the two tangle of the system, it was proved that 
the QPT is characterized by a cusp-minimum in the 
entanglement ratio $\tau_1/\tau_2$. The cusp is ultimately due to the 
discontinuity of the derivative of $\tau_1$. The minimum in the ratio $\tau_1/\tau_2$
signals that the enhanced role of 
the multipartite entanglement in the mechanism driving the phase transition.  
Moreover by looking at the entanglement it was found that 
the ground state can be factorized at certain value of the 
magnetic field. The existence of the factorizing field in $d=2$ was  
proved rigorously for  any $2d-XYZ$ model in a bipartite lattice. 
Unexpectedly enough the relation implying the factorization is very 
similar  to that one found in $d=1$.

Multiple spin exchange is believed to play an important role in the physics 
of several bidimensional magnets~\cite{Schollwock}. Entanglement in the ground state of 
a two leg ladder with four  spin ring exchange was evaluated by means of the 
concurrence~\cite{Song06}.

\subsection{Pairwise entanglement in fermionic models}
\label{pairfermion}

\subsubsection{Non interacting Fermions}
\label{sec-noninteracting}

The site-based entanglement of spin degrees of freedom
through the Jordan-Wigner transformation has been exploited for 
calculating the concurrence of nearest neighbor sites and the single site 
von Neumann entropy (see Section \ref{indistparts}) for the one-dimensional 
tight-binding model in presence of a chemical potential
for spinless fermions in ~\cite{ZanardiPRA02}. 
This model is related to the isotropic $XX$ model in a transverse 
magnetic field.
In this specific case, no double occupancy can occur and the concept of
entanglement coincides with that for spins 1/2.
It was found that the nearest neighbor concurrence of the ground state at 
$T=0$ assumes its maximum
at  half filled chain.
Due to particle-hole symmetry, the concurrence results symmetric 
respect to half-filling.
At finite temperatures it was  found that the threshold temperature 
for vanishing concurrence is independent of the chemical potential.
Raising the chemical potential leads from a monotonically
decreasing concurrence with raising temperature at low filling fraction
to the formation of a maximum at a certain temperature for high
filling fractions.

The continuous limit of the tight-binding fermion model is 
the ideal Fermi gas. In this system, the spin entanglement
between two distant particles has been studied in~\cite{Vedral03}.
There, depending on the dimensionality, the pairwise spin-entanglement
of two fermions has been found to decrease with their distance
with a finite range $R$ of the concurrence.
The two spin reduced density matrix is
\beq
\rho_{12}=\bigfrac{1}{4-2f^2}
\Matrix{cccc}{1-f^2&0&0&0\\0&1&-f^2&0\\0&-f^2&1&0\\0&0&0&1-f^2} 
\eeq
where $f(x)=d \frac{J_1(x)}{x}$ with $d\in\{2,3\}$ being the space dimension 
and $J_1$ the (spherical for $d$=3) Bessel function of the 
first kind~\cite{Vedral03,Oh04}.
This density matrix is entangled for $f^2\geq 1/2$.
As a consequence, 
there is spin entanglement for two fermions closer than 
$d_0\approx 0.65\frac{\pi}{k_f}$ for $d=3$
and $d_0\approx 0.55\frac{\pi}{k_f}$ for $d=2$
($k_f$ is the Fermi momentum).
A finite temperature tends to diminish slightly the range of 
pairwise spin entanglement~\cite{Oh04}.

It should not be surprising that non-interacting particles
are spin-entangled up to some finite distance.
It is true that the ground state and even an arbitrary thermal state
of non-interacting fermions has vanishing entanglement among the 
particles (which should not be confused with the
non-vanishing {\em entanglement of particles}~\cite{Dowling06}), 
since the corresponding states are 
(convex combinations of) antisymmetrized product states.
However, disentanglement in momentum space typically leads
to entanglement in coordinate space. A monochromatic plane wave
of a single particle for example corresponds to a $W$ state,
which contains exclusively pairwise 
entanglement in coordinate space for an arbitrary distance of the sites.
Furthermore does a momentum 
cut-off at $k_f$ correspond to a length scale of the order $k_f^{-1}$.

It is interesting that a {\em fuzzy} detection of the particles
in coordinate space increases the entanglement detected by the  
measurement apparatus. 
To this end~\cite{Cavalcanti05b}
calculated the two-position reduced density matrix defined by
$
\rho^{(2)}_{ss',tt'}=\expect{\Psi_{t'}(r')^\dagger\Psi_{t}(r)^\dagger
\Psi_{s'}(r')^{}\Psi_{s}(r)^{}}
$
with blurred field operators
$
\Psi_{s}(r)^{}:=\int\,\d r' \,\d p \psi_s(p) D(r-r')\e^{ipr'}
$
where $D(r-r')=\frac{1}{\sqrt{2\pi}\sigma}\exp{-\frac{|r-r'|}{2\sigma^2}}$
is a Gaussian distribution describing the inaccuracy of the 
position measurement.
This could be understood from the blurred field operators being coherent
sums of local field operators; the entanglement measured by the apparatus 
as described above then is the bipartite entanglement between the two regions
of width $\sigma$ around $r$ and $r'$. This entanglement is larger than
the average of all pairwise contributions out of it due to
the  super-additivity of the entropy/negativity.
An analysis in~\cite{Vedral04} for the three fermion spin density 
matrix revealed that
the state carries entanglement within the W-class~\cite{Duer00}, provided
the three particles are in a region with radius of the order of the 
inverse Fermi momentum; a similar reasoning applies to $n$ fermions
in such a region~\cite{Vedral04,lukens}.

\subsubsection{Pairing models} 
\label{supercond}
Itinerant systems, where the focus of interest 
is the entanglement of degrees of freedom forming
a representation of $su(2)$ in terms of the fermionic operators
have been also subject of intense investigation.  
This line has been followed in~\cite{ZanardiPRA02,Shi04}
for analyzing a connection to BCS superconductivity
and also to the phenomenon of $\eta$-pairing, a possible scenario for 
high $T_c$ superconductivity  (see also~\onlinecite{VedralNJP04,fan04,Vedral04}). 
Such states appear as eigenstates
of the Hubbard model with 
off diagonal long range order (see \ref{models-second}).
A simplified model of BCS-like pairing for spin-less fermions
has been studied in~\cite{ZanardiPRA02}.
The concurrence of the two qubits represented by the modes $k$ and $-k$
has been found to be a monotonically increasing function
of the order parameter; it drops to zero significantly before the 
critical temperature is reached, though.
For electrons with spin, a connection between the BCS order parameter
and the local von Neumann entropy in the particle number projected
BCS ground state has been proposed by~\onlinecite{Shi04} (see also~\cite{gedik}).

States with off diagonal long range order by virtue of $\eta$-pairing are defined in Section\ref{models-second}.
These  are symmetric states
and consequently, their concurrence vanishes in the thermodynamic limit
due to the sharing property of pairwise entanglement of $su(2)$ degrees 
of freedom.
Consequently, a connection to the order parameter of off diagonal long range order
$
{\cal O}_{\eta} = \Expect{\Psi}{\eta^\dagger_j\eta^{}_k}{\Psi}
= N(L-N)/L(L-1) {\longrightarrow} n(1-n) \;
$
(with $N,L\longrightarrow \infty$ and fixed filling fraction $n$)
can not be established,
not even for the rescaled concurrence, since
$
C \longrightarrow 1/L
$ 
(see also the analysis for the LMG model in 
Section~\ref{ground-critical}).
Nevertheless, the state is entangled, as can be seen from the
entropy of entanglement and the geometric measure of entanglement~\cite{wei03}.
The latter is tightly connected to the relative entropy~\cite{wei04}.
Both have been calculated in~\cite{VedralNJP04} and clearly indicate
the presence of multipartite entanglement.

\subsubsection{Kondo models}
\label{kondopairwise}

The Kondo models are paradigms to explore the  
quantum impurity problems. They identify a special class of 
physical systems whose macroscopic properties are  
dramatically influenced by the presence of few impurities with 
quantum internal degrees of freedom~\cite{Hewson}. In its simplest formulation, 
the effective Hamiltonian  describes a single impurity spin interacting with a  
band of free electrons. The many-body screening 
of the impurity spin provided by the electrons as a collective effect, 
leads  the system from weak coupling to strong coupling regimes~\cite{Andrei}; 
the lenght scale of the screening cloud is 
$\xi_K=v/T_K$, where $v$ is the speed of the low lying excitations and $T_K$ 
is  the Kondo temperature.
 
In the first studies of the entanglement the charge degrees of 
freedom of the electrons were considered frozen. 
The pairwise entanglement of spin degrees of freedom  
in the isotropic Kondo model was analyzed~\cite{Oh06} 
within the variational formalism of Yosida where the Kondo singlet 
is described as 
$
|\Psi_s\rangle=\frac{1}{\sqrt{2}} \left (
|\phi_\downarrow\rangle |\chi_\uparrow\rangle- |\phi_\uparrow\rangle 
|\chi_\downarrow\rangle\right )
$. 
In the previous equation $|\chi_\sigma\rangle$ denote the impurity spin states;
$|\phi_\sigma\rangle$ represent the electronic states  with 
an unbalanced spin $\sigma$~\cite{Hewson}.  
In agreement with
the common wisdom, the reduced density operator of the impurity is found to be  
maximally  mixed, meaning that the Fermi sea and the impurity 
spin are  in a maximally entangled state (the Kondo singlet). 
The impurity spin and a single electron are  
in a Werner state made of a superposition of the back ground and 
the Kondo singlets. 
Due to the 
entanglement monogamy (the electrons cooperatively  
form a singlet with the spin) two electrons cannot be
entangled with each other within the Kondo cloud and the 
single-electron spin entanglement  vanishes in the
thermodynamical limit. 
Pairs of electronic spin can be nevertheless entangled in a finite system through the  
scattering with the spin impurity; this effect might be used 
to manipulate the electron-electron entanglement by performing 
a projective measure on the  impurity spin~\cite{Yang-kondo}. 
This suggests that some amount of electron-electron entanglement  
might be extracted even in the thermodynamical limit where it was 
demonstrated that the Kondo resonance is washed out by the 
measurement~\cite{Katsnelson}--effectively removing the constraint of the 
entanglement sharing.

The two impurity Kondo model was studied as well. 
The new feature here is the  
Ruderman-Kittel-Kasuya-Yosida (RKKY) effective interaction between 
the impurity  spins ${\bf S_1}$ and ${\bf S_2}$, that competes with 
the Kondo mechanism  (favouring non magnetic states); 
it is ferromagnetic or antiferromagnetic depending  on the distance 
between the impurities. Because of such interplay a 
quantum critical point emerges in the phase diagram separating 
the spin-spin interaction regime from the phase where the two spin are 
completely screened~\cite{Jones88,Affleck92}.

As for the single impurity, the two impurity spins  
are in a Werner state, for which the  concurrence is characterized by a single 
parameter $p_s$, exhibiting a singlet type of entanglement between the two 
spin impurities.  The concurrence is found to vanish at the critical point.

For ferromagnetic RKKY interaction the concurrence between the impurity spin 
vanishes identically as the result of a $S=1$ Kondo screening.
It turns out that~\cite{Cho06} the impurity spins can be entangled 
(with a finite  concurrence) by the RKKY interaction only when 
certain amount of antiferromagnetic correlations 
$f_s=\langle {\bf S_1} {\bf S_2}\rangle$ is 
established in the system; such value of the 
correlation function  is that one reached at the quantum critical point. 
The entanglement between the conduction electrons and the Kondo impurities is 
quantified by a combined analysis of the  Von Neumann entropies  
of the two impurities and of the single impurity (tracing out both the 
electronic spins 
and the remaining impurity). The latter quantity is maximized 
independently on $f_s$, meaning that the  impurity spin is completely screened 
either by the Kondo cloud or by the other impurity spin.  In the regime where 
the Kondo mechanism dominates, the concurrence cannot be finite 
because of the entanglement sharing.   
  
Entanglement in the Kondo physics of double quantum dots in an external 
magnetic field was studied in~\cite{Ramsak06}.   
The main phenomenology 
results to be consistent with the scenario depicted in~\cite{Cho06}  
especially if the dots are arranged in series (each dot is coupled to the 
leads exclusively, resembling the configuration of the Kondo spins embedded
 in the electrons). The concurrence switches to finite values for a certain 
threshold of the inter-dot coupling (for which the 
assumptions  of negligible charge fluctuations results still valid).
The temperature weakens the entanglement  between the qubits also at $T>T_K$.
For the side- and parallel-coupled dots a more intense coupling amoung the 
qubits is required to entangle them. For the side-coupled dots this results 
because one has to win on the enhanced Kondo effect on the dot coupled to 
the leads ('two stage Kondo' effect); therefore the critical  inter-dots coupling is 
$\sim T_K$. For the parallel-coupled arrangement 
the concurrence is zero because the effective RKKY interaction turns out 
to be ferromagnetic up to a certain value that is the threshold to 
entangle the electrons. 

The RKKY interaction controls in an effective way 
the entanglement  amoung the qubits also 
in the case of many impurity spins arranged 
as in the   Kondo necklace model~\cite{Saguia03}.     
The Hamiltonian describes a Kondo lattice where the 
localized impurity spins, displaced in every  
lattice site, interact with the (pseudo) spins of the 
electrons(see \onlinecite{kondo-lattice} for a review on Kondo 
lattice models).  It results that the 
additional on-site spin-spin interaction 
impose a 'selective' monogamy of the entanglement, depending on whether 
the Heisenberg interactions is ferromagnetic or antiferromagnetic. 
The effects of finite temperature and magnetic field were  
considered also in anisotropic models. 
It emerges that a critical field exists separating different patterns in 
the thermal entanglement between the eigenstates of the model.

The effect of the fluctuations of the charge degrees of freedom of the 
electrons (frozen in the references cited above) in the RKKY 
mechanism was discussed 
quantifying the entanglement of particles in a small cluster described by the
 periodic Anderson model~\cite{Samuelson06}.  It was evidenced that the ground 
state of the system is characterized by a double occupancy of the electronic 
levels, whose entanglement can  be only partially captured by assuming them as simple qubits. 

\subsection{Entanglement in itinerant bosonic systems}

In contrast to the free Fermi gas, in bosonic systems
the phenomenon of Bose-Einstein condensation (BEC) takes place
at sufficiently low temperatures. Then, a macroscopic portion
of the bosons is found in the single particle ground state
of the system. This state
clearly is a symmetrization of a product state in momentum space,
and in fact all eigenstates of the ideal Bose gas are of that structure.
Nevertheless, in principle entanglement could  be present  when
going in the coordinate basis. 
It results, however, this not the case. Neither do
two distant bosons carry spin entanglement~\cite{Vedral03}
nor are two distant groups of $n$ and $m$ particles 
entangled~\cite{Dowling06} when superselection rules for the 
particle number is applied in both regions~\cite{Wiseman03}.
In Ref.~\cite{Simon02} a very different notion of entanglement has
been employed: the state of either part of a certain bipartition
of the BEC has been viewed at as a {\em qudit}, or more precisely,
a $N$-level system. $N$ is the number of bosons in the condensate
and the different states in both regions are labelled by its 
occupation number. The entropy of entanglement for a spatial bipartition
of the BEC is then non-zero.
A proposal for entangling internal atomic degrees of freedom
in a weakly interacting BEC has been put forward in 
Ref.~\cite{Soerensen00,Helmerson01}.

The disentangled modes in a BEC naturally become entangled by means of 
interactions/scattering between these modes.
This has been exemplified in~\cite{Vedral03} (see also \onlinecite{Shi04})
for the case where the scattering strength is independent 
of the momentum transfer $q$: 
$
H=\sum_p \eps(p)a^\dagger_p a^{}_p+
V\sum_{p,p',q}a^\dagger_{p+q}a^\dagger_{p'-q} a^{}_{p'} a^{}_p
$
The Hamiltonian can be  diagonalized with for $q=0$ and $p'=-p$ by means of  a Bogoliubov 
transformation $ a^{}_p=:u_p b^{}_p+v_p b^\dagger_{-p} $. We observe that such 
transformation entangle the two modes
$\pm p$. 
The corresponding entropy of entanglement is~\cite{Vedral03}
$
S_{p,-p}=-(\frac{v_p}{u_p})^2\ln (\frac{v_p}{u_p})^2
         -(1-(\frac{v_p}{u_p})^2)\ln(1-(\frac{v_p}{u_p})^2)
$
If $u_p=v_p$, the reduced state for mode $p$ is maximally mixed,
and hence the modes $p$ and $-p$ are maximally entangled.
The entanglement entropy for a bipartition in positive and negative
modes is then given by $\sum_{p>0}S_{p,-p}$.
This constitutes a simple example on how mode mixing generates entanglement.
Such a scenario is rather generic; a curious example being the entanglement
of the accelerated vacuum and from the viewpoint of relatively accelerated 
observers due to the Unruh effect 
(see e.g.~\onlinecite{Vedral03,Alsing03,Benatti04,Fuentes05}).

Further studies of entanglement in bosonic system include the analysis in 
two-mode condensates~\cite{hines}, in optical 
lattices~\cite{ng} and in two-species spinor Bose condensates~\cite{shiniu}.

\subsection{Entanglement of particles}

Studies which use measures for indistinguishable particle
entanglement (see Section~\ref{indistparts}) in the area of many-body systems are 
is still only few, particularly regarding  the use of 
the fermionic concurrence, giving account 
for the possibility of double occupancy (with spin degree of freedom).
The entanglement of particles and its difference with the usual spin entanglement, 
is discussed in~\cite{Dowling06}, starting with
very small systems as two spinless fermions on four lattice sites and the 
Hubbard dimer, and then for the tight binding model in one 
spatial dimension, in order to compare with existing results for the 
spin entanglement elaborated in~\cite{Vedral03}. 
For the Hubbard dimer (a two-site Hubbard model), 
the authors compare with their results with those for
the entanglement measured by the 
local von Neumann entropy without superselection rule for the local
particle numbers~\cite{ZanardiPRA02}.
Whereas the latter signals decreasing entanglement in the ground state
with increasing $U/t$, the entanglement of particles 
increases~\cite{Dowling06}.
This demonstrates that imposing superselection rules
may lead to qualitatively different behaviour of the entanglement.  
Interestingly, an increase with $U/t$ is observed also for the
entanglement of modes without imposing 
superselection rules~\cite{DengGu04}.

We would like to finish this section with the notice of a recent 
proposal of an experiment in order to decide whether even entanglement
merely due to the statistics of the indistinguishable particles
can be useful for quantum information processing~\cite{Cavalcanti06}.

\section{Entanglement entropy}
\label{block-entropy}

\begin{figure}
	\vspace*{2mm}
	\includegraphics[width=.77\linewidth]{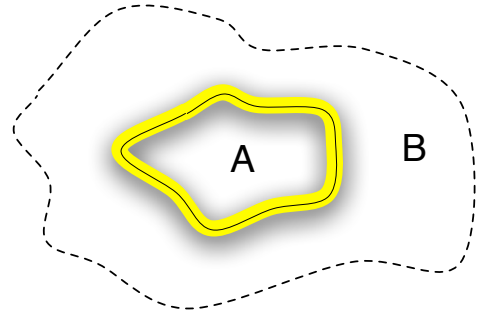}
	\caption{The block entropy is evaluated after partitioning the system 
	in regions $A$ and $B$. For finite range correlations, it  
is intuitive  that the wave function of the system is factorized  
$|\Psi_A\rangle |\Psi_B\rangle$ by removing the  region at the boundary 
(yellow). Accordingly the reduced entropy would vanish} 
\label{block}
\end{figure}

An important class of works analyzing the entanglement in many body system 
considered a bipartition of the system dividing it in two distinct regions 
$A$ and $B$ as shown in Fig.\ref{block}.
If the total system is in a pure state than a measure of the entanglement 
between $A$ and $B$ is given by the von Neumann entropy $S$ associated to the 
reduced density matrix of one of the two blocks ($\rho _{A/B}$)

Motivated by the pioneering work of Fiola {\em et al.}~\cite{fiola} and 
Holzhey, Larsen and Wilczek~\cite{holzhey94}
born in the context of black hole physics, the problem was first reanalyzed in 
the framework of quantum information by Vidal and coworkers for 
quantum spin chains~\cite{GVidal03}  and by Audenaert and 
coworkers for harmonic lattices~\cite{Eisert02}. 

In studying the properties  of block entropy it is important to understand 
its dependence on the properties (topology, dimensions, $\cdots$) of the two regions 
$A$ and $B$. A key property which is intensively explored to understand its range 
of validity is what is known as the {\em area law}~\cite{srednicki93}.  When it 
holds the reduced entropy $S$  would depend only on the surface of 
separation between the two regions $A$ and $B$. In d-dimensional system this means 
that $S \sim \ell^{d-1}$ where $\ell$ is of the order of the size of one of the block 
(see~\onlinecite{riera} for a recent discussion of various aspects of the area law).
In the rest of the Section we discuss several different physical systems in 
one- and higher dimensional lattices and see when the area law hold. Here we 
consider only many-body systems in their ground state,the thermal effect and the dynamical properties of  the entropy 
      will be discussed in Sections~\ref{th} and~\ref{dyn} respectively.

The entanglement entropy is not a mere theoretical concept but it might be measured. 
Following the procedure proposed in~\cite{silva} the measure of $S$ can be related to 
the measurement of the distribution of suitably chosen observables.

\subsection{One-dimensional spin systems}

We start our review on the properties of the block entropy by analyzing the case of 
one-dimensional spin systems to which a large body of work has been devoted.  By 
means of the Jordan-Wigner transformation it is possible 
to map the models onto a lattice fermions, hence the results discussed here are 
applicable to fermionic models (after the appropriate mapping) as well. 
A particular important case which is amenable of an exact solution is 
the $XY$ model (see Section~\ref{models-spin-short}) which can be mapped onto a 
free fermion model. For this case we discuss in more details the method to calculate 
the block entropy. In this section we consider only chains with short-range 
interaction.

\subsubsection{Spin chains}

In one dimension the surface separating the two regions is constituted by two points, 
therefore the area's law would imply that the reduced 
entropy is independent on the block size. This is indeed the case when the system 
is gapped and hence the correlation lenght $\xi$ is finite (see \onlinecite{Hastings03} 
for a rigorous proof). In the gapless case, 
$\xi = \infty$, logarithmic corrections appear and  the prefactor of the block 
entropy is {\em universal}, related to the central charge of the underlying 
conformal field theory.
%
Holzey {\em et al.}~\cite{holzhey94}, benefiting from an earlier work of 
Cardy and Peschel~\cite{cardy88}, analyzed the block entropy of a $1+1$ dimensional
massless bosonic field. Vidal {\em et al.} studied numerically one-dimensional Ising 
and Heisenberg chains~\cite{GVidal03,latorre04} and conjectured that the block entropy 
would saturate for non critical chain while would diverge logarythmically with a 
prefactor related to the central charge of the underlying conformal theory~\cite{holzhey94}.
Such violation of the area law in critical systems reflects how the 
mixedness of the state increases by the partial tracing operation, regardless of the 
spatial extension of the spin block. 

A calculation of the block entropy by means of conformal field theory, generalizing 
in several respect the the results of Holzey {\em et al.} by including 
the case of free and periodic boundary conditions, different partitions, non critical  
systems and finite temperature has been performed by Calabrese and Cardy~\cite{calabrese04}.
Starting from the work on the $XX$ model of Jin and Korepin~\cite{jin04b}, 
important explicit analytic calculations for a number of one-dimensional $XY$ spin 
(free fermion) models have been carried out in~\cite{Its,its06,keating05,Peschel,
popkov05,franchini1,franchini2,jin04b,korepin04,weston,Eisler05,peschel05b}. 
Numerical calculations on the $XX$ and $XXZ$ models were also performed 
in~\cite{Laflorencie05,dechiara06a}.
The study of entanglement entropy is of great interest in a more general context 
[see~\onlinecite{holzhey94,fiola,casini05,casini05b,ryu,ryu2}
and references therein], Casini and coworkers, for example, evaluated the entanglement 
entropy both in the massive scalar field theory~\cite{casini05} and for Dirac 
fields~\cite{casini05b}).  
Very recently the area law for gapped one dimensional 
systems was proved by~\cite{hastings07,hastings07a}.

The main features of the reduced entropy in one-dimensional spin/fermi systems 
can be summarized as follows (for clarity we discuss only the long distance 
behaviour as dictated by the underlying conformal field theory):

\begin{itemize}

\item
At criticality a one dimensional system has a block entropy which diverges 
logarithmically with the block size. If the block is of lenght $\ell$ and the system 
is $L$ long with periodic boundary condition then $S_{\ell}$ is given by 
\begin{equation}
	S_{\ell} =\frac{c}{3} \log_2 \left[\frac{L}{\pi a} 
	\sin\left(\frac{\pi}{L} \ell\right)\right] +A
\label{Sell}
\end{equation}
where $c$ is the central charge of the underlying conformal field theory and $a$ is an 
ultraviolet regularization cut-off ( for example the lattice spacing in spin systems).
$A$ is a non-universal constant. For the Ising model $c=1/2$ while for the Heisenberg 
model $c= 1$ (see Fig.7).  

\item
Slightly away from criticality, in the case in which the system has a large but 
finite correlation lenght $\xi\gg a$ and the Hamiltonian 
is short-ranged, the block entropy saturates to a finite value (see Fig.7)
\begin{equation} 
\displaystyle{
	S_{\ell} \sim \frac{c}{3} \log_2 \frac{\xi}{a} 
                 \;\;\;\;\;\; \mbox{for} \;\;\;\;\;\;
                 \ell \to \infty }
\end{equation}  
\item
An extension to finite temperature in the critical case has been obtained 
by means of conformal field theory~\cite{calabrese04} and by conformal 
mapping together with  the second law of thermodynamics~\cite{korepin04} with the result
\begin{equation}
	S_{\ell} =\frac{c}{3} \log_2 \left[\frac{\beta}{\pi a} 
	\sinh \left(\frac{\pi}{\beta} \ell\right)\right] +A \,\,.
\label{Sbeta}
\end{equation}
where $A$ is a constant and $\beta$ the inverse temperature.
In the finite temperature case however, the block entropy is not a measure of
the entanglement between the two partitions as the state to start with is mixed. A comparison of numerical data with the CFT 
                         predictions is shown in Fig. 7 for the $XXZ$
 model.
\end{itemize}
 \begin{figure}
	\includegraphics[width=.67\linewidth]{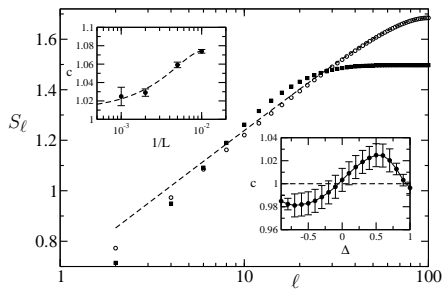}
\caption{The Block entropy $S_\ell$ for $L=200$ in the case of an $XXZ$ Heisenberg chain 
	for a critical value $\Delta=0.0$ 
	(circles) and non-critical value $\Delta=1.8$ (squares). The critical 
	data compared with the conformal field theory prediction (dashed line). 
	Lower inset: central charge extrapolated by fitting the numerical data 
	$S_\ell$ with Eq.(\ref{Sell}) (with a factor 1/2 as  in the numerical 
	calculation the block is taken at the boundary) for different values of $\Delta$. 
	The data are for $L=1000$. Upper inset: scaling of $c$ extrapolated as 
	a function of $1/L$ for the worst case $\Delta=0.5$ and compared to a quadratic fit 
	(dashed line).[From \cite{dechiara06a}]} 
\label{blockentropy}
\end{figure}

By now it is clear that the various measures of entanglement are sensitive to 
the presence of quantum phase transitions, the scaling of the entropy
gives excellent signatures as well. Recent works tried to construct 
efficient ways to detect quantum phase transitions by analyzing the reduced entropy 
for small clusters. One-site entropy has been 
considered in~\cite{chen06,gu04}. Chen {\em et al} analysed  
the entanglement of the ground states in $XXZ$ and dimerized Heisenberg spin chains as 
well as in a two-leg spin ladder suggesting that the phase boundaries 
might be identified based on the analysis on the local extreme of the 
entanglement entropy~\cite{chen06}. Legeza and coworkers~\cite{Legeza06,legezapre} 
pointed out that in the biquadratic spin-1 Heisenberg chain, see Eq.(\ref{haldane-phase}), 
the two-site entropy is ideal to highlight the presence of a 
dimerized phase. They also considered the two-site entropy also for the 
ionic Hubbard model~\cite{hubbard81}. 

\begin{figure}
\includegraphics[scale=0.3,angle=-90]{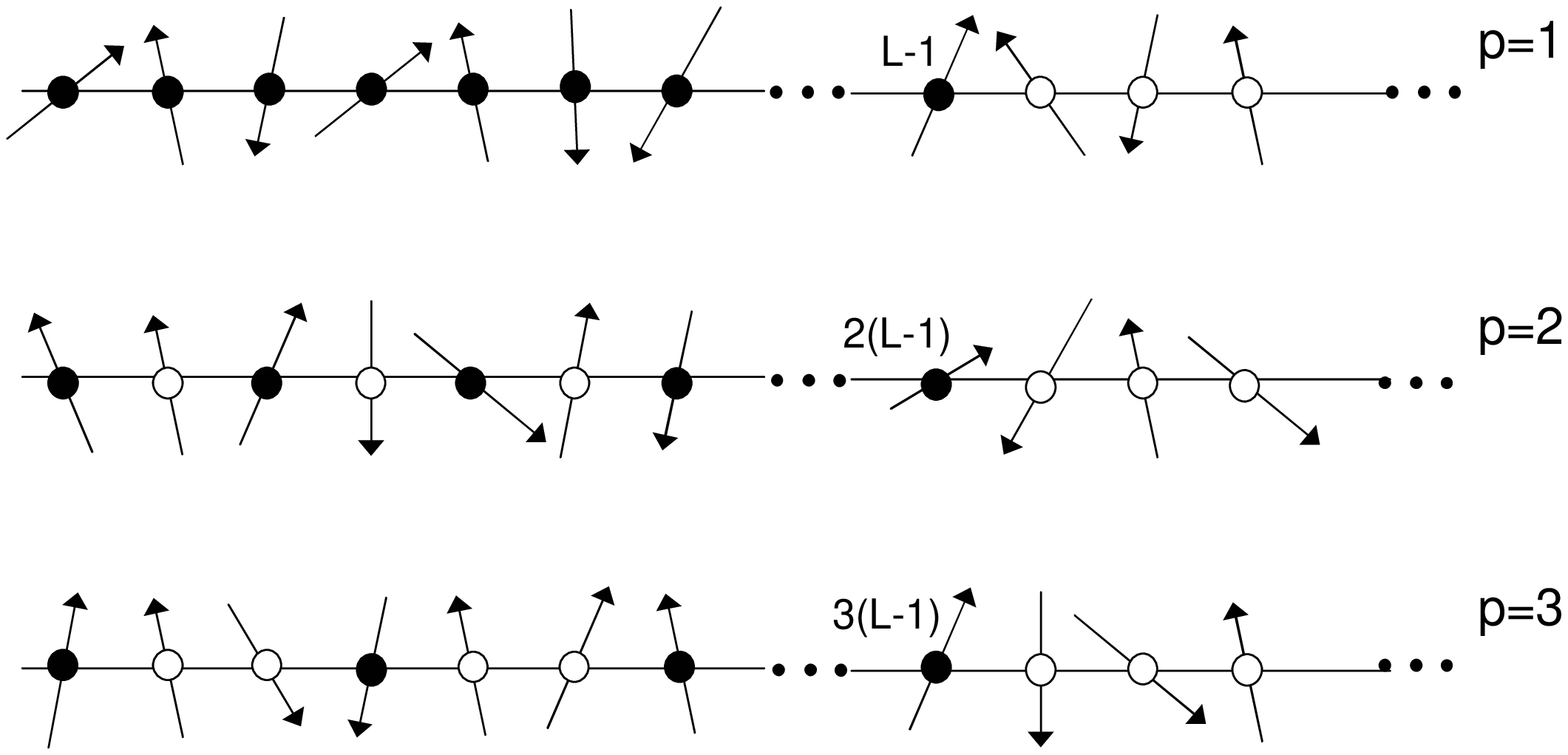}
\caption{The idea of the comb-partition suggested by Keating Mezzadri and Novaes is  
	illustrated for three different values of
	the spacing $p$. In all the case the subsystem $A$, denoted by black
	circles) contains $\ell$ spins while the other (denoted by empty
	circles) contains the rest of the chain. The case, $p=1$, corresponds 
	to the well known
	`block' division.[From~\cite{keating06}]}
\label{comben}
\end{figure}

The idea of partitioning the system in a more elaborated way in order to 
analyze additional properties of entanglement lead to the introduction of the 
concept of comb entanglement~\cite{keating06}. This is 
illustrated in Fig.\ref{comben}. The two blocks $A$ and $B$ are not chosen 
contiguous  but  $A$ consists of
$\ell$ equally spaced spins, such that the spacing between the spins in
this subsystem is  $p$ sites on the chain while $B$
contains the remaining spins. For this choice of the partition, the 
"surface" separating the two blocks grows with the system size (differently
from the case $p=1$ where it is composed by two links).
As a consequence non-local properties of entanglement between the 
two blocks can be investigated. For $p > 1$ the leading contribution to the 
entropy scales linearly with the block size 
$
S_{\ell}(p) = {\cal E}_1(p) \ell +{\cal E}_2(p) \ln \ell
$
In the limit $p \to \infty$ the coefficient ${\cal E}_1(p)$ is, to leading order, 
sum from single spin contributions. The unexpected result is that the corrections vanish
slowly, as $1/p$, differently from other measures like concurrence where 
these long-range corrections are not present.

We conclude this section by briefly discussing the single copy entanglement $E_1$ 
introduced by Eisert and Cramer~\cite{eisert05} and studied in details for 
one-dimensional spin systems~\cite{eisert05,orus05,peschel05}. Single copy 
entanglement $S_{\mbox{sc},\ell}$ quantify the amount of entanglement that can be distilled from 
a single specimen of a quantum systems. For spin chains it can be shown~\cite{eisert05}
that single copy entanglement asymptotically is half of the entanglement entropy
$	
\lim_{L \to \infty} S_{\mbox{sc},\ell}/S_{\ell} = 1/2
$. 
This result was later generalized to conformally invariant models~\cite{peschel05,orus05}.

\subsubsection{$XY$ chains and free fermion models }

As we saw in Section~\ref{models-qIsing} the $XY$ chain ($\Delta =0$ in 
Eq.(\ref{general-spin})) can be mapped onto a 
model of free fermions. As a result an analytical (albeit non-trivial) approach 
is possible for the calculation of the block entropy. An analytic proof of the logarithmic 
dependence of the block entropy in the isotropic $XY$ model was obtained 
in~\cite{jin04b,Its,franchini1,franchini2}. The relation between the entanglement
entropy of this model and the corner transfer matrices of the triangular Ising lattice
has been derived in~\cite{Peschel}. Keating and Mezzadri considered 
a more general free fermion Hamiltonian in which the matrices ${\bf A}$ and
${\bf B}$ (defined in Eq.(\ref{eq:quadratic})) do not have  the tridiagonal structure which appear in the case of 
the $XY$ model~\cite{keating05}. They showed that under certain conditions the entropy
can be expressed in terms of averages over ensembles of random matrices.
In this section we recall the main steps of the derivation leading to the 
evaluation of the entanglement entropy , more details can 
be found in the review of Latorre, Rico and Vidal~\cite{latorre04} and in the 
above mentioned papers~\cite{jin04b,Its,franchini1,franchini2,keating05}.

The reduced density matrix of a block of $\ell$ spins can be expressed 
in terms of averages of strings of $\ell$ spin operators with weights given by the 
averages of these strings on the ground states. By means of a mapping of the spin 
operators in terms of the Majorana fermions, 
$
a_{2l-1} = \left( \prod _{m<l}\sigma^z_m \right) \sigma^x_l 
$,
$
a_{2l} =  \left( \prod _{m<l}\sigma^z_m \right) \sigma^y_l
$
and given the fact that the resulting 
fermionic Hamiltonian is  quadratic (Wick theorems holds), it is possible to express 
the block entropy in terms of the elements of the correlation matrix $B_{\ell}$
\begin{equation}
	B_{\ell} = \left[
 	\begin{array}{cccc}
	\Pi_0  & \Pi_1   & \cdots & \Pi_{\ell\!-\!1}   \\
	-\Pi_1 & \Pi_0   &  & \vdots\\
	\vdots & & \ddots & \vdots\\
	 -\Pi_{\ell\!-\!1}   & \cdots &\cdots & \Pi_0 
	\end{array}
	\right].
\label{corrmatrix}
\end{equation}   
where
$
	\Pi_l = \left[\begin{array}{cc}
	0 & g_l \\
	-g_{-l} & 0
	\end{array}
	\right]
$
with real coefficients $g_l$ given as, for $L \rightarrow \infty$, by
\begin{equation}
	g_l = \frac{1}{2\pi}\int_{0}^{2\pi} d\phi e^{-il\phi}
	\frac{ \cos \phi - \lambda - i \gamma \sin \phi}{| \cos 
	\phi - \lambda - i \gamma \sin \phi|}.
\end{equation}
It is crucial to notice that the matrix $B_{\ell}$ are {\it block Toeplitz matrices}, 
that can be thought as usual Toeplitz matrices but with non commuting entries.
After the transformation of $B_{\ell}$ into a canonical form $\oplus^L_{m=1}=\nu_m i \sigma_y$,
the system is described by a set of $\ell$ independent two-level systems. Therefore 
the entanglement entropy is given by
\begin{equation}
	S_{\ell} = - \sum_{m=1}^{\ell} \left[ \frac{1+\nu_m}{2}\log \frac{1+\nu_m}{2} + 
	\frac{1-\nu_m }{2}\log \frac{1-\nu_m}{2}\right]
\label{blockxy}
\end{equation} 
Numerical and analytical analysis of (\ref{blockxy}) has been performed 
leading to the behaviour described in the preceding subsection.
In order to obtain the analytical formula for the asymptotics of the entropy (\ref{blockxy}), 
the first step is to recast it, by the Cauchy formula,  into a
contour  integral \cite{jin04b,Its} 
\begin{equation}\label{integ}
S=\lim_{\epsilon\rightarrow0^+}\lim_{\delta\rightarrow0^+}\frac{1}{2\pi
i}\oint_{c(\epsilon,\delta)}e(1+\epsilon,\lambda)\frac{d \ln
D_A(\lambda)}{d\lambda}d\lambda,
\end{equation} 
where
$
e(2x,2y)=-\sum_{\pm}(x \pm y)\log_2\left(x \pm y\right)
$.
The contour of integration $c(\epsilon,\delta)$ approaches the interval $[-1,1]$
as $\epsilon$ and $\delta$ tend to zero without enclosing the
branch points of $e(1+\epsilon,\lambda)$.
The matrix in $D_A(\lambda)={\rm
det}(\lambda I-B_{\ell})$ is again of the block Toeplitz type. The asymptotics of the 
entropy can then be 
obtained from the asymptotics of  $D_A(\lambda)$. This, in turn can be done 
resorting to the Riemann-Hilbert approach to  the theory of Fredholm integral 
equations~\cite{Korepin-book}. This 
allows to generalize the standard Szeg\"o  theorem for scalar Toeplitz matrices  to 
obtain the leading formula for the determinant 
of  the block Toeplitz matrix  $D_A(\lambda)$. This leads to the asymptotics,
$\ell \to \infty$, for the entropy 
\begin{equation}
S={{1}\over{2}} \int_1^\infty \ln \left ( \frac{\theta_3[\beta(x) +{{\sigma \tau}\over{2}}]
\theta_3[\beta(x) -{{\sigma \tau}\over{2}}]}{\theta_3({{\sigma\tau}\over{2}})} \right )dx 
\label{entropy-asymp}
\end{equation}
where $\theta_3(s,\tau)$ is one of the Jacoby elliptic function, 
$\beta (x) = (1/2 \pi i) [\ln (x+1) - \ln (x-1)$, $\sigma =1,0$ for $h_z <,> h_c$ 
and $\tau \sim [\log(|1-\lambda^{-1}|)]$.
The critical behaviour of Eq.(\ref{entropy-asymp}) can be obtained 
by the asymptotic properties of $\theta_3(s,\tau)$ for small $\tau$, and the  leading 
term of the critical 
entropy results
$
S = -(1/6) 
\log \left(|1-\lambda^{-1}|/(4\gamma)^2\right) + {\cal O} \left 
[ |1-\lambda^{-1}|(\log |1-\lambda^{-1}|)^2\right ] \;.  
$
The previous expression  can be obtained in a more direct way 
resorting a duality relation connecting the quantum Ising chain with the Ising model on a 
square or triangular lattice~\cite{peschelkaulke,Peschel,calabrese04}. 
In particular the reduced density matrix can 
be written as the trace of the corresponding (corner) transfer matrix.  In this way, 
however the expression for the critical entropy can be obtained only for $\gamma^2+h_z^2>1$.

In the isotropic case $\gamma=0$, $c=1$  and then  the prefactor of the log 
divergence is $1/3$. 
In this case the critical entropy can be obtained as an average,
by realizing that the block Toeplitz matrices (for $\gamma=0$) are
unitary. Then the contour integral can be recast into an integral in the ensemble of 
unitary  matrices. 
 This remarkable observation~\cite{keating05} allows to relate the spectral statistics 
of the model with the entanglement encoded in the ground state, following a reasoning 
that has many analogies in spirit with random matrix 
theory. The entanglement entropy was obtained explicitly for  
matrices  {\bf B}, in Eq.(\ref{eq:quadratic}), being elements of classical groups. 
It is interestingly that only {\bf B} affects the  the prefactor of the logarithm  
in entropy. It is proportional to  $ 2^{w_G}$ 
where $w_G$ is a universal quantity related solely to the classical group establishing the 
symmetries, the constant of proportionality is Hamiltonian dependent.
 
\subsubsection{Disordered chains}
Conformal invariance implies universal properties for the 
entanglement entropy. What happens when conformal invariance is lost as in the case of 
certain one-dimensional disordered spin systems? Refael and Moore~\cite{refael04} were 
the first to look at this question by computing the block entropy for  the Heisenberg, 
XX, and quantum Ising chains with random nearest-neighbor coupling. Their approach 
was based on a real space renormalization group developed earlier~\cite{ma79,fisher94} 
for random spin chains where disorder is relevant  and drives the system at low 
energies in the so-called random singlet phase which can be thought as a collection 
of singlet bonds of arbitrary length. Consequently the entropy of a given segment of the 
chain is just $\ln 2$ times the number of singlets crossing the boundary between the 
two regions in which the systems is partitioned.  Refael and Moore showed that the 
entropy, as in the case of clean critical chain, grows as the logarithm of the block size
$
  S_{\ell} \sim \tilde{c} \; \log \ell
$
with a ``renormalized central charge'' $\tilde{c} = c \ln 2$.

\begin{figure}[!ht]
\begin{center}
\includegraphics[width=6.5cm]
{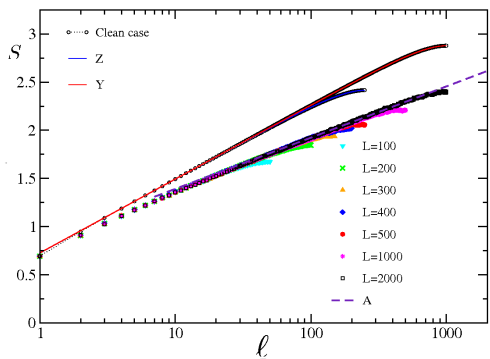}
\caption{(color online) Entanglement entropy of a subsystem of size
  $\ell$ embedded in a closed ring of size $L$, shown vs $\ell$ in a
  log-linear plot. Numerical results obtained by exact
  diagonalization performed at the XX point. 
For clean non random systems with $L=500$ and $L=2000$ (open circles), 
$S_{\ell}$ is in agreement with Eq.~(\ref{Sell} (red and blue curves).
$A=0.8595 +\frac{\ln {2}}{3}\ln \ell$,
$Y=0.72602 +\frac{1}{3} \ln (\frac{2000}{\pi} \sin \frac{\pi \ell}{2000})$,
$Z=0.72567 +\frac{1}{3} \ln (\frac{500}{\pi} \sin \frac{\pi \ell}{500})$,
[from \cite{Laflorencie05}]}
\label{laflo}
\end{center}
\end{figure}

A numerical test of this prediction was performed both for the $XX$~\cite{Laflorencie05}
and for the Heisenberg models~\cite{dechiara06a}. In Fig.\ref{laflo} we report the data of 
Laflorencie, the two curves represent the results for the clean and disordered case and 
fully confirm the prediction of Refael and Moore.
The result that the ratio between the random and pure values of the prefactor of 
the block entropy is the same for all the different chains studied in~\cite{refael04}
might suggest that this value might be determined by the flow from the pure to the 
random fixed point.
This conjecture was recently questioned by analyzing the 
entanglement entropy for a family of models which includes the N-states random Potts 
chain and the $Z_N$ clock model. In this case it was shown that the ratio
between the entanglement entropy in the pure and in the disordered system is model 
dependent~\cite{santachiara06}.

\subsubsection{Boundary effects}

Boundaries or impurities may alter in a significant way the the entanglement 
entropy.

The result given in Eq.(\ref{Sell}) was obtained for periodic boundary conditions.
If the block is at the boundary of  the chain then the prefactor is modified and 
the block entropy is 
one half of the one given in Eq.(\ref{Sell})~\cite{calabrese04}. 
\begin{equation}
	S_{\ell} =\frac{c}{6} \log_2 \left[\frac{2L}{\pi a} 
	\sin\left(\frac{\pi}{L} \ell\right)\right] + g + \frac{A}{2}
\label{Sboun}
\end{equation}
where $A$ is the non-universal constant given in Eq.(\ref{Sell}) 
and $g$ is the boundary entropy~\cite{affleck91}.
The case of open boundary conditions in critical $XXZ$ chains was also recently reconsidered
in~\cite{laflorencie06}. In addition to the log divergence 
there is a parity effect depending on the number of spins of the block being 
even or odd. The amplitude of the resulting oscillating term decays as a power law with the 
distance from  the boundary. The origin of this oscillating term is easy to understand
qualitatively as an alternation of strong and weak bonds along the chain. The boundary spin 
has a strong tendency to form a singlet pair with its nearest neighbour on the right-hand 
side; due to the monogamy of the entanglement this last spin will be consequently less 
entangled with its partner on the third site of the chain. Furthermore it was also 
shown that the alternating contribution to the entanglement entropy is 
proportional to a similar term in the energy density (the constant of proportionality 
being related to the lattice constant and to the velocity of the excitations).
The effect of open boundary conditions on the entanglement entropy of a resonant 
valence bond solid was studied as well~\cite{fan06}. In this case however the corrections
due to the open ends decay exponentially.

Different type of boundaries can appear in the AKLT quantum spin chain, 
with  bulk spin-1 and two spin-1/2 at the ends. The entanglement entropy 
has been studied in~\cite{fan04}. They showed
that  the block entropy approaches to a constant 
value exponentially fast with $\ell$

The entanglement entropy of one-dimensional systems is affected by the presence of 
impurities in the bulk~\cite{levine04,peschel05b,zhao06} or aperiodic couplings~\cite{Igloi}. 
In these cases the entanglement entropy has the same form as in Eq.(\ref{Sell}) but 
with an effective value which depends on the strength of the defect. 
The entanglement properties of anisotropic open spin one-half Heisenberg chains with 
a modified central bond were considered in~\cite{zhao06} where the entanglement entropy 
between the two half-chains was calculated using  the DMRG
approach. They find a logarithmic behavior with an effective central charge varying 
with the length of the system. 
The numerical simulations of~~\cite{zhao06} show that 
by going from the antiferromagnetic to the ferromagnetic case the effective central 
charge grows from zero to one in agreement with~\cite{levine04}.
The combined presence of interaction between the excitation and a 
local impurity modifies in an important way the properties of a one-dimensional system.
Starting from the work of Kane and Fisher~\cite{kane92} is by now understood that at 
low energies the scattering with the impurity is enhanced or suppressed depending on the 
interaction being repulsive or attractive. 
It is therefore expected that the entanglement entropy is affected as well.
Levine, by means of bosonization, studied the entanglement entropy 
in a Luttinger liquid interrupted by an impurity and found that there is a correction 
to $S$, due to the impurity which scales as $\delta S_{imp} \sim - V_B \log (\ell/a)$ where 
$V_B$ is the renormalized backscattering constant~\cite{levine04}. In the repulsive case the 
backscattering flows to large values suggesting that the total entropy would vanish 
(the correction is negative). In the opposite case of attractive interactions, 
the impurity potential is shielded at large distances and the entropy would approach 
the value of the homogeneous liquid. 

The single copy entanglement in the presence of boundaries has been considered as 
well~\cite{zhou05}. Differently from the bulk contribution here the boundary contribution 
to the von Neumann entropy equals that of the single copy entanglement.

Some of these results provided the fertile ground  to study 
the entanglement encoded in the Kondo 
cloud. 
Specifically, the block entropy $S_{imp}$ of a spin cloud of radius $r$ around the impurity
with the rest of the system is analyzed~\cite{Sorensen06}. 
By using a combination of 
Bethe ansatz results, conformal field theory and DMRG methods, the authors 
demonstrated that $S_{imp}$  is a universal scaling function of $r/\xi_K$. 

\subsection{Harmonic chains}
\label{harmblocks}
Static systems of harmonic chains have been first
analyzed in~\cite{Eisert02}, where periodic arrangements of harmonic oscillator
modes have been considered.
The oscillators have been coupled in the standard way
via their coordinate variables and the Hamiltonian has been chosen to
be translational invariant. The entanglement in these systems has been analyzed for
both the ground state and of thermal states; both belong to the class of
Gaussian states. 
Here we review the results obtained in~\onlinecite{Eisert02}. 
For work on higher dimensional
lattices and emphasize on the entropy area law we refer
to Section~\ref{blockharm}. 

Using the covariance matrix defined in Eq.(\ref{covarianceThS}), the 
logarithmic negativity can be expressed directly in terms of the interaction
matrix $\mathbb{U}$~\cite{Eisert02}
$
E_{N}=\trace \log_2 \mathbb{U}^{-1/2}P\mathbb{U}^{1/2}P
$
where $P$ is a diagonal matrix, with non-zero entries $P_{jj}=-1$ where the partial 
transposition is performed and $P_{jj}=1$ elsewhere.
This entanglement monotone has been analyzed
for bipartitions of a ring containing an even number of oscillators.
It is convenient to define
$
\mathbb{U}=\Matrix{cc}{U'&U''\\U''&U'}
$
For the symmetric bisection into equally large connected parts,
a lower bound for the logarithmic negativity has been obtained as
\beq\label{lowerbound}
E_{ N}\geq\frac{1}{2}|\trace F_n\log_2 \mathbb{U}|=
\frac{1}{2}\log_2(1+4\sum_{m=0} \alpha_{2m+1})
\eeq
where the coefficients $\alpha_n$ have been defined in Sections~\ref{harmodels}.
$F_n$ is the $n\times n$ flip matrix with ``$1$'' in the cross-diagonal and
``$0$'' elsewhere.
Equality holds if $F_{\frac{n}{2}}U''$ is semi-definite, which is
the case for nearest neighbor interaction. For this case one obtains
$
E_{N}^{n.n.}=\frac{1}{2}\log_2(1+4\alpha_1)\; .
$
Remarkably, this result is independent of the size of the ring.
This also tells us that the negativity of the symmetric bisection
for a model including couplings $\alpha_d$ of arbitrary range  
is higher than that of the corresponding
chain with only nearest neighbor coupling and coupling strength 
$\sum_{d}\alpha_d$. It is interesting to anticipate here that
for critical systems, the lowest eigenvalue of $\mathbb{U}$ tends to zero
with growing system size.
This leads to a symplectic eigenvalue of $V$ that diverges with the system size
with a consequent divergence of the negativity.

\begin{figure}[ht]
\begin{center}
\includegraphics[width=0.35\textwidth]{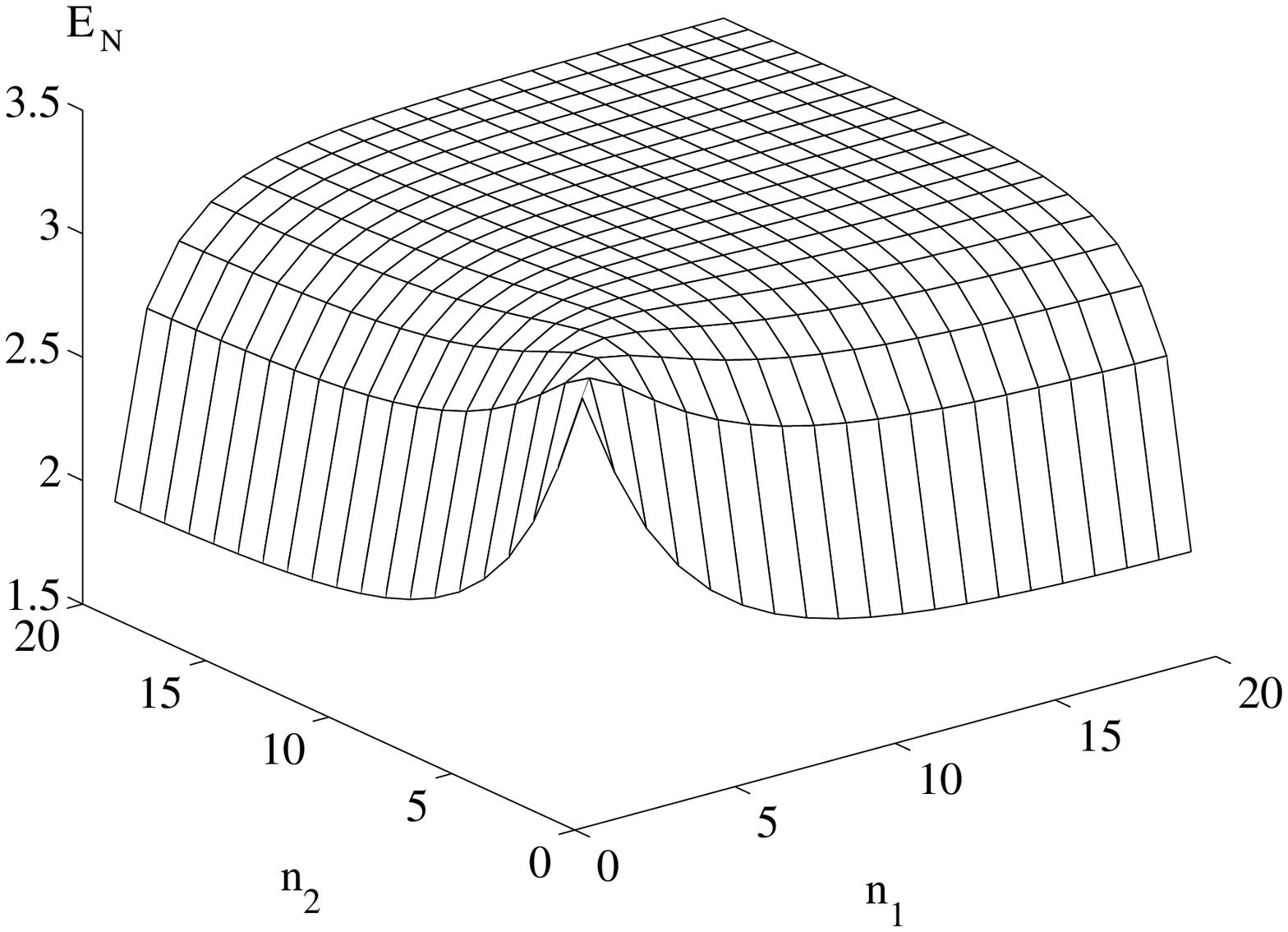}
\hspace*{7mm}
\includegraphics[width=0.35\textwidth]{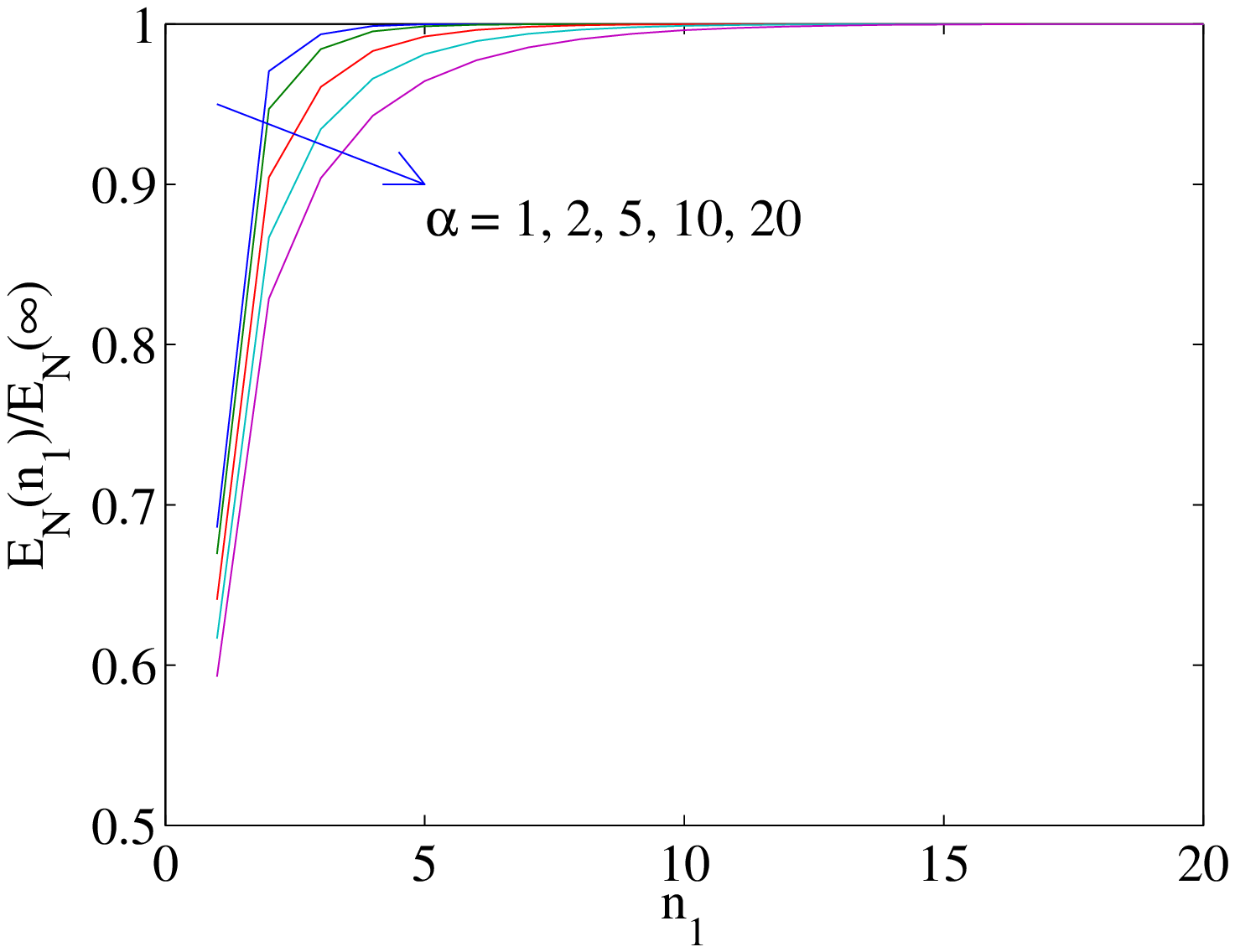}
\caption{
\label{AEPW5-6}
 Logarithmic negativity for harmonic chains.
{\em Top panel:}
The logarithmic negativity for coupling strength $\alpha=20$ and a bipartition
in $n_1$ and $n_2$ oscillators (total number of oscillators $N=n_1+n_2$).
For sufficiently large parts, a plateau is reached. For one part consisted
of only few oscillators, the negativity decreases as a function of the 
system size. 
{\em Bottom panel:} The logarithmic negativity relative to its plateau value
$E_N(\infty)$ for $n_2=20$ as a function of $n_1$ and varying coupling strength
$\alpha$. For small coupling the plateau limit is reached faster; since the 
plateau limit is connected to a local quantity (the average energy per
oscillator), this can be this can be explained by a correlation length
                that grows with $\alpha$.
[From~\cite{Eisert02}]}
\end{center}
\end{figure}
The analysis for general bisections revealed that for nearest neighbor 
couplings the negativity of a single oscillator with the rest of
the chain monotonically decreases with the size of the chain.
This single-oscillator negativity turned out to establish also a lower 
bound for the negativity of any connected set of two or more oscillators 
with the rest of the same chain
(see top panel of Fig.~\ref{AEPW5-6}).
In all cases the maximum negativity has been observed for the 
symmetric bisection.
Both features are expected to be generic to coupled ensembles
of harmonic oscillators~\cite{Eisert02}. 
In particular should the infinite size limit of the symmetric bipartition
negativity establish an upper bound:
$\lim_{m\to\infty} E_N(m,m)\geq E_N(n_1,n_2)$.
This upper bound appears as a plateau in the top panel of Fig.~\ref{AEPW5-6},
which as a function of $n_1$ and $n_2$ is reached 
already for not too small $n_1$ and $n_2$. The plateau value 
is essentially proportional to the average energy per oscillator.
With increasing nearest neighbor coupling strength, a more shallow 
approaching of the plateau value is observed (see bottom panel of Fig.~\ref{AEPW5-6}).

The situation changes when the negativity of two disconnected parts
of the chain is considered. The particular limiting case
of an alternating bipartition, consisting in all the oscillators located 
at odd sites and the complement being all the oscillators 
at even sites, has been analyzed in the presence of nearest neighbor 
couplings only~\cite{Eisert02}.
In this setting the logarithmic negativity is no longer limited by the
average energy per oscillator but instead it grows linearly with
the system size $n$.
The conclusion  is that one part of the negativity
can be related to the energy; the second part can be seen as a surface
term, proportional to the area of the boundary forming the contact
between both bipartitions.
In one spatial dimension this ``area'' is the number of contacts between
both parts, which in the periodic setting described above equals $n$.
This interpretation finds further support in the result for the logarithmic 
negativity of a symmetric bisection in an open chain of oscillators, which is then
roughly half the logarithmic negativity of the corresponding chain 
with periodic closure (see discussion in ~\onlinecite{Eisert02}).

\begin{figure}[ht]
\begin{center}
\includegraphics[width=0.35\textwidth]{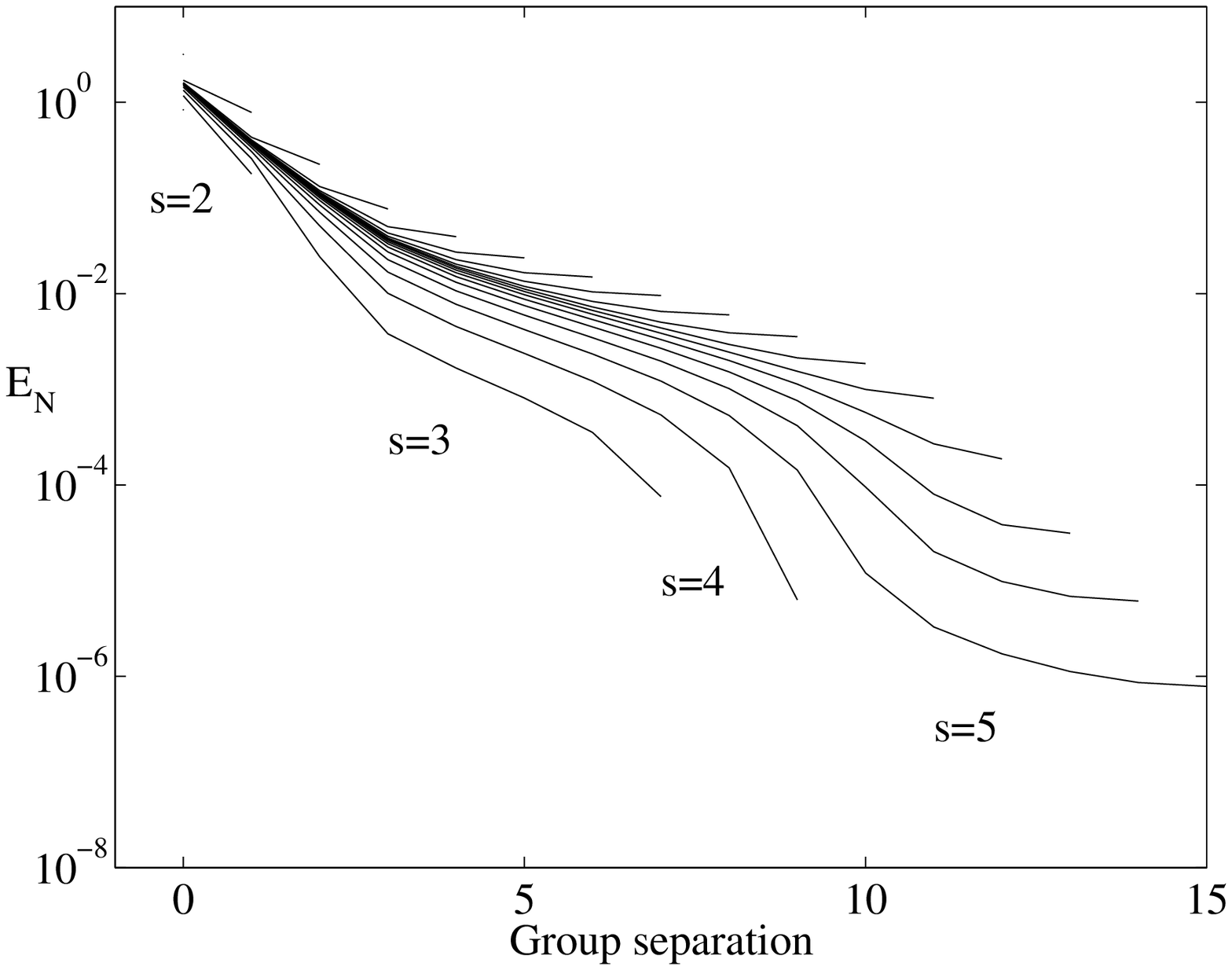}
\hspace*{7mm}
\includegraphics[width=0.35\textwidth]{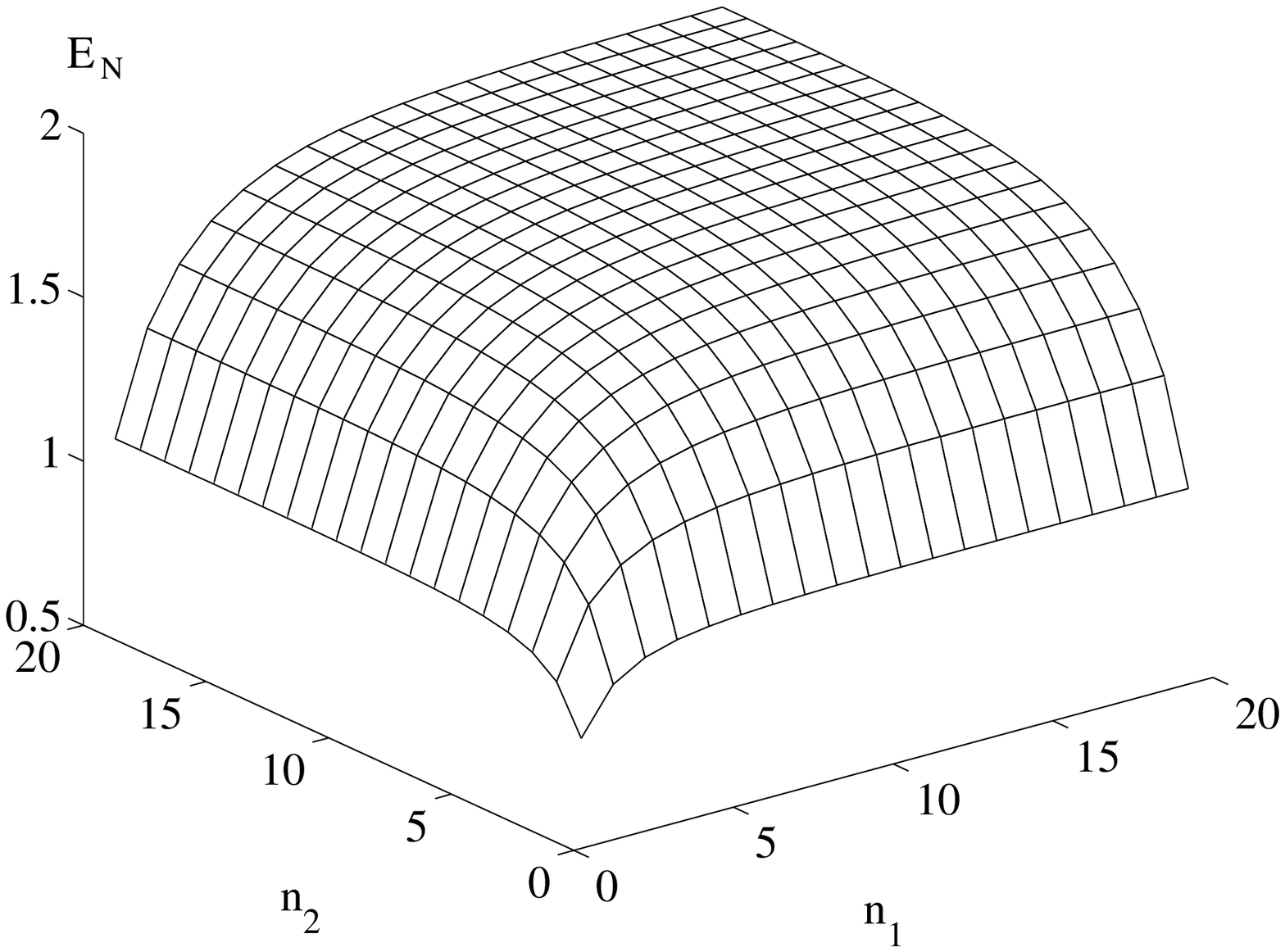}
\caption{
\label{AEPW10-14}
Logarithmic negativity for harmonic chains.
{\em Top panel:}
The logarithmic negativity of two connected groups of $s$ oscillators
as a function of the distance between both groups. It is seen that the range of 
the negativity is finite but grows with the group size $s$. For $s=1$, its 
range is zero, meaning that no two single oscillators share pairwise 
entanglement unless they are neighbors.
{\em Bottom panel:} Logarithmic negativity as in the top panel of 
fig.~\ref{AEPW5-6} but for an open chain. The value of the negativity is 
roughly half as large as for the periodic chain since the number of connection
points of the parts is halved. In contrast to the periodic chain, the 
negativity of one small part with the rest of the open chain grows 
with system size  [From~\cite{Eisert02}]}  
\end{center}
\end{figure}
An interesting puzzle is given by the analysis of the negativity of
two connected parts of the chain that are separated from each other
by a finite number of sites. Besides an expected all-over exponential
decay of the logarithmic negativity with the distance between the two
equally large groups, the negativity is also of limited
range (see top panel of Fig.~\ref{AEPW10-14}). 
This range increases with the size of the two parts.
In particular is there no pairwise negativity between two single
oscillators unless they are neighbors. This implies that the 
entanglement of distant groups of oscillators cannot be due to
``free'' pairwise entanglement of single oscillators 
(as opposed to ``bound'' entanglement not detected by
the negativity). So either ``bound'' pairwise entanglement
is responsible for the entanglement present in distant groups
or multipartite entanglement might play an important part.

It is instructive to mention that pairwise correlations between single
oscillators do exist notwithstanding a vanishing pairwise 
negativity~\cite{Eisert02}.
In any case the presence of correlations is necessary for 
quantum entanglement of the parts. 
It is worth noticing that both the plateau exhibited by the negativity
for not too small size of the parts, and the decrease of the single 
oscillators' negativity with the system size
find a plausible explanation merely in terms of the correlation length.
The same plausibility arguments predict the single oscillators'
negativity to increase with growing system size when open chains are 
considered; in fact this is what the authors observed (see the discussion
in ~\onlinecite{Eisert02}).
The observed short range of the negativity in particular
for small connected sets of oscillators overstretches this simple
reasoning and demonstrates that the connection between entanglement
and correlations is indeed more subtle.

\subsection{Systems in $d >1 $ and the validity of the area law}
\label{area-fermi}

The scaling of the entanglement entropy in systems of higher dimensions has been 
subject of intense investigation in various fields of research. In the context of 
quantum information the understanding of the scaling of the entropy as a function 
of the block size has important consequences on the simulability of a quantum 
system by a classical algorithm and therefore is attracting a lot of interest. 
The picture that emerged from the analysis of the one-dimensional case, i.e. the 
violation of the area law when the system is critical, does not seem to hold in 
higher dimensions. The situations appears more complex. 
The higher dimensional generalization of matrix product states, the projected 
entangled pair states, satisfy the area law~\cite{peps1,peps2} despite a divergent 
correlation length. Another example is ground-state of antiferromagnetic Ising-spin 
networks embedded on planar cubic lattices~\cite{wellard} where the area law is fullfilled 
also at the critical point. 
In the following of this section we concentrate on the ground state of some many-body 
Hamiltonian for which the block entropy has been recently computed. We first consider 
the case of hopping Hamiltonian of (free) fermions and bosons  and then we discuss the 
case of harmonic lattices (note that, hystorically the bosonic case was considered first).

\subsubsection{Fermi systems}
In one dimension, by virtue of the Jordan-Wigner transformation, 
the block entropy of a system of interacting spins is tightly 
connected to that of a (free) Fermi gas. 
It is of great interest to understand what are the 
properties of the block entropy for free fermions in $d$ dimensions.
This question has been studied in~\cite{Gioev06,Wolf06} where it 
was shown that logarithmic corrections persist also in higher 
dimensions
\begin{equation}
   S_{\ell} \sim \ell^{d-1}\log_2 \ell \;\; .
\end{equation} 
The expression of the constant of proportionality in the equation above has been 
obtained by~\cite{Gioev06} resorting the Widom conjecture. \onlinecite{Wolf06} 
exploited the quadratic lower bound of~\cite{Fannes03}.  
The corrections to the area law are a Fermi surface effect. 
In the case of fractal dimension of either the Fermi
or the block surface, the scaling is modified into 
$
   S^{fr}_{\ell} \sim \ell^{d-\beta}\log_2 \ell
$ 
where $1-\beta$ is the maximum fractal enhancement of dimension of
either the Fermi or the block boundary. An interesting case where the system 
undergoes a Lifhitz phase transition has been considered in~\onlinecite{CramerEisert}.
Indeed as it was pointed out by Cramer {\em et al} these transitions related to 
a change in the Fermi surface manifest in a non-analytic behaviour of the 
prefactor of the leading order term entanglement entropy.

For regular block and Fermi surface, numerical analysis has confirmed
the modified area-law for critical 
two-dimensional~\cite{BarthelChung06,LiDing06}
and three-dimensional~\cite{LiDing06} models.
\onlinecite{BarthelChung06} study the tight binding model as an 
example for a two-dimensional 
model with a connected Fermi surface as well as the model
$H=-\sum_{x,y}[(1+(-1)^y) c^\dagger_{x,y}c^{}_{x,y+1}
+c^\dagger_{x,y}c^{}_{x+1,y+1}
+c^\dagger_{x,y}c^{}_{x-1,y+1}]
$
with a disconnected Fermi surface and
$H=-\sum_{x,y}\left[h c^\dagger_{x,y}c^{}_{x+1,y}+
(1+(-1)^{x+y}) c^\dagger_{x,y}c^{}_{x,y+1}\right]
$
with a zero dimensional Fermi surface, as for the one-dimensional case.
Whereas in the first two cases, the entropy is found to obey the modified area law,
this is no longer true for the third model 
with zero-dimensional Fermi surface.
There, the corrections to the standard area law $S\sim \ell^{d-1}$ are sub-logarithmic.

The same feature has been observed by~\cite{LiDing06} studying
the spin-less fermionic in two and three spatial dimensions. 
The authors conjecture an interesting
connection between the modified area law to be observed and
the density of states at the Fermi energy. They formulate this in terms
of the {\em co-dimension} at the Fermi energy: i.e. the dimension of momentum space
minus the dimension of the degeneracy at the Fermi energy in momentum space, providing a measure 
of the relative portion of the gapped excitations in the low lying spectrum of the model.   
In agreement with the findings for the model zero-dimensional Fermi surface,
the authors observe only sub-logarithmic corrections to the area law if
the co-dimension at the Fermi energy is $2$. 
The authors conjectured from this that in two spatial dimensions a
co-dimension less or equal to $1$ is necessary for the modified
area law $S \sim \ell \log_2 \ell$ to apply. They do not mention implications
of fractal co-dimension due to a fractal Fermi or block surface.
This finding would be worth further investigation in direction 
to higher dimensions
in order to fix the  connection between 
area law and co-dimension at the degeneracy point. 

We finally mention the  interesting connection between the 
block entropy and the Berry phase in lattice models of fermions recently 
discussed in~\cite{hatsugai}.

\subsubsection{Harmonic systems}
\label{blockharm}

Harmonic systems have been also investigated to understand the validity of the 
area law. They provide one of the few physical systems for which exact 
analytical treatements
are avaliable (see~\onlinecite{Cramer06} and references therein).

We first consider a system of non-critical harmonic oscillators 
with nearest neighbor interaction and periodic boundary conditions. 
Non-criticality implies that the lowest eigenvalue of the interaction matrix
$\mathbb{U}$, $\lambda_{min}(\mathbb{U})$, is well separated from zero.
Further peculiar characteristics of
the covariance matrix, in particular its symmetric and circulant form,
allowed to give estimates for upper and lower bounds of the block 
entropy of some compact $d$-dimensional hyper-cubic region with edge 
length $\ell$ and surface proportional to $\ell^{d-1}$~\cite{Plenio05}.

The upper bound has been established directly from the 
logarithmic negativity (see~\onlinecite{Eisert02}), 
whereas for the lower bound several estimates for the dominant 
eigenvalue of the reduced density matrix have been employed.

Key ingredients to the problem are the largest 
eigenvalue of the covariance matrix
and the uncertainty relation which constrains all eigenvalues of the 
covariance matrix to lie above $\hbar /2$.
The result is that both bounds go proportional to ${\ell}^{d-1}$, hence
the entanglement entropy is indeed proportional to the surface of the block.
An extension to general block shapes has been formulated 
as well for Gaussian states~\cite{Cramer06}. For pure states,
lower and upper bounds are given as before, which both scale linearly with the
surface area of the block. For a given finite range of interaction
beyond nearest neighbors, the area law could be stated only in terms
of an upper bound (as in \cite{Plenio05}, this upper bound has been 
obtained from the logarithmic negativity).
Therefore~\onlinecite{Cramer06} could not  exclude the block entropy
to scale with lower dimensionality than the block-area.
The area-law can even be present in disordered systems; the crucial 
requirement for the area law to hold in this case
is that $\kappa:=\frac{\lambda_{max}(\mathbb{U})}{\lambda_{min}(\mathbb{U})}$
is bounded for all admissible disorder realizations.
The presence of a finite temperature $T$ enters only in the proportionality
factor, when thermal equilibrium states
(which are still Gaussian) are considered.

It is interesting to notice that an algebraically diverging correlation
length does not automatically imply a violation of the 
area law
\onlinecite{Cramer06} find that for non-critical systems of harmonic 
oscillators and algebraic decay $r^{-\alpha}$ of the correlations
with the distance $r$ still leads to the area-law as long as $\alpha>2d$.
Essential for this conclusion is that  
$\kappa$
is bounded.
The authors conjectured $d$ as a tighter bound at least for 
hyper-cubic blocks (When compared to the one-dimensional
Ising model, already at $\alpha=d=1$ logarithmic corrections appear) .

For critical systems the situation is different: criticality leads to
eigenvalues of $\mathbb{U}$ arbitrarily close to zero and hence unbound $\kappa$.
In~\cite{Unanyan05-b}, a one-to-one connection between criticality
and logarithmic corrections to the area law has been reported for
one-dimensional systems with finite range interaction.
The authors also report on evidence for this connection to hold also
in higher dimensional setups. However the absence of logarithmic corrections for critical 
           two-dimensional arrays of harmonic oscillators and nearest neighbor 
           interaction has been evidenced numerically in~\cite{BarthelChung06}.
           The analytical calculations of~\cite{CramerEisert} conclude the area law to apply
           for arbitrary number of dimensions, opposed to the conslusions of~\cite{Unanyan05-b}.
As support for their claim, \onlinecite{Unanyan05-b} quote
a factoring interaction matrix, which however corresponds to a noninteracting
array of one-dimensional harmonic chains; this can be seen as the limiting
case of an anisotropic interaction with finite range and does not give support
to the original claim. It is indeed fundamentally different from harmonic $d$-dimensional 
lattices with isotropic finite range interaction, 
as discussed in~\cite{BarthelChung06,CramerEisert}. 
Nevertheless it raises the question for a critical anisotropy for the coupling
of the harmonic oscillators, which on the background of 
the findings in~\cite{BarthelChung06,CramerEisert} could be phrased as:
``Does a finite critical anisotropy exist beyond which the harmonic lattice
is quasi one-dimensional?'' 

Logarithmic corrections are also being observed, when infinite range
interactions are considered, which drive the system 
towards criticality~\cite{Unanyan06}. 
To this end, the authors consider
a two-dimensional array of harmonic oscillators, with an interaction
of finite range in $x$-direction and an infinite range in the $y$-direction.
This is a very instructive example in that it leads to the logarithmic 
correction $\sim {l_x}\ln l_y$. Interestingly, the logarithm contains 
the length $l_y$
	     of the block, where the interaction has infinite range. 
 The prefactor of the logarithm is half the length
of the block in $x$ direction (with finite range interaction).

\subsection{LMG Model}
The logarithmic divergence with the block size of the entanglement entropy is 
not exclusive of one-dimensional systems. The block entropy of the LMG 
model was studied both in the ferromagnetic~\cite{stockton03,latorre05b,barthel06,dus} and 
antiferromagnetic~\cite{unanyan05} case. In the LMG model each 
spin is interacting with all the other spin in the network therefore the idea 
of a block as depicted in Fig.\ref{block} does not fit very well. Nevertheless
it is perfectly legitimate to define the reduced entropy of $\ell$ spins once the 
other $N-\ell$ ($N$ is the total number of spins) have been traced out. Evidently 
the entropy is independent on which spins have been selected to be part of the block.
 \begin{figure}
	\includegraphics[angle=0,width=0.75\linewidth]{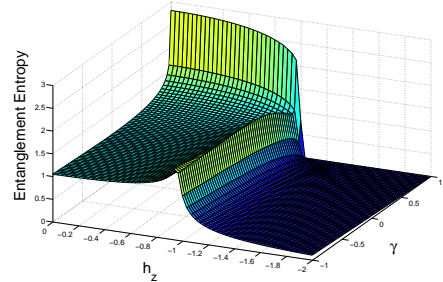} 
\caption{Entanglement entropy for $N=500$ and $\ell=125$ as a function of the 
	parameters $h_z$ and $\gamma$ of the ferromagnetic LMG model.
	For $\gamma \ne 1$, there is an anomaly at the critical point $h_z=1$, 
	whereas the entropy goes to zero at large $h_z$ since the ground state is 
	a fully polarized state in the field direction. In the zero field
	limit, the entropy saturates when the size of the system increases. 
	For $\gamma = 1$, the entropy increases with the size of the system in 
	the region $0 \le h_z < 1$ and jumps to zero at $h_z=1$.
	[From \cite{latorre05b}]} 
\label{lmgen}
\end{figure}
In Fig.\ref{lmgen} the representative behaviour of the entanglement entropy as a 
function of the various regions of the phase diagram is shown for a fixed value
of the block and system sizes. Below we summarize the main findings related to the LMG model. 
In the ferromagnetic case and in the case of $\gamma =1$ (isotropic model),
the entropy diverges logarithmically with the block size 
 $S \sim (1/2) \log \ell$ while
at fixed $\ell$ and $N$ diverges when the external magnetic field approaches 1 from 
below. Also in the antiferromagnetic case~\cite{unanyan05}, when no transition as a 
function of the field is present, the entropy grows logarithmically with the size of 
the block in the isotropic limit. 
Differently from the one-dimensional case where the prefactor is universal and 
related to the central charge, here the origin of the prefactor of the logarithmic 
divergence is related to the presence of the Golstone modes and to the number of 
vanishing gaps. The recent work of Vidal {\em et al} clarifies this issue by studying 
a number of collective spin model by means of $1/N$-expansion and scaling 
analysis~\cite{dus}.

\subsection{Spin-boson systems}
\label{entropy-spinboson}
The entropy in models  of spins interacting with harmonic oscillators have 
been analyzed as well. Here the separation between spin and bosonic degrees of freedom is 
natural and the partition leads to study the reduced entropy of one subsystem (say the spin). 
The entanglement entropy was 
studied for the Jaynes-Cummings~\cite{boseJC}, Tavis-Cummings model~\cite{lambert04,lambert05} 
and for the spin-boson model~\cite{costi03,stauber06,jordan04,kopp06}.
Lambert and coworkers analyzed how the  super-radiant quantum phase transition 
manifests in the entanglement between the atomic ensemble and the field mode.
They compute the von Neumann entropy numerically at finite 
$N$ and analytically in the thermodynamical limit. They found that the entropy 
diverges at the phase transition as (see Fig. 13)
\begin{equation}
  S \sim - ({1}/{4}) \log \mid \lambda - \lambda _c \mid
\end{equation}
where $\lambda$ is the coupling between the spins and the boson field 
and $\lambda _c$ is the value at which the transition takes place (see Fig.13).
\begin{figure}
\centerline{
\includegraphics[clip=false,width=6.5cm]{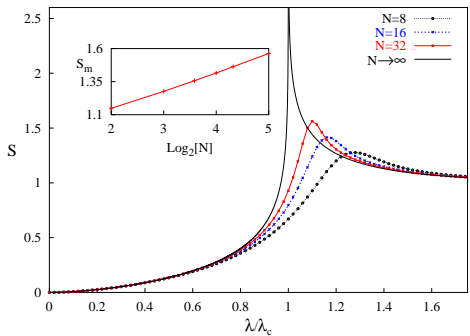}
} 
\caption{ Entropy $S$ between two-level atom and cavity field as 
           a function of the atom-radiation coupling for both
 $N \rightarrow \infty$ 
  and finite $N$. In the inset the scaling of the value of the entanglement maximum as a
  function of $\log N$. [From~\cite{lambert04}]}
\label{dicken}
\end{figure}
The entropy of the spin-boson model (see Eq.(\ref{spin-boson})) was studiedby numerical renormalization group in~\cite{costi03}. An analytic 
calculation, including other dissipative models, has been presented in~\cite{stauber04} 
and more recently in~\cite{kopp06,stauber06,Stauber-errata}.  
In the broken-symmetry state has an effective 
classical description and the corresponding von Neumann entropy is zero.  
In the symmetric phase the entropy can be easily expressed as a function 
of the ground state energy.
and $\delta$ (defined in Eq.(\ref{spin-boson})). The previous expression allows
At the transition point the entropy is discontinuous with a jump given by~\cite{kopp06}
$\Delta S = \ln 2 + \delta /4 \delta_c \ln (\delta / \delta_c)$ ($\delta _c$ is a high 
energy cutoff). A systematic analisys  of the entropy in the spin-boson 
model for 
different coupling regimes was pursued recently in~\cite{Kopp07,LeHur}.

We finally mention the interesting connection between entanglement and energy 
fluctuations introduced by~\cite{jordan04} and exploited in details both for 
a spin and for an harmonic oscillators coupled to a bath. This connection 
might be useful in the light of possible experimental measure of entanglement
(see also~\onlinecite{silva}).
For example, as pointed by~\cite{jordan04}, in certain mesoscopic realization of 
qubits as metallic rings of superconducting nano circuits, a measurement of 
persistent current can be directly related to a  measurement of 
the entropy.

\subsection{Local entropy in Hubbard-type models}
\label{hubbardsent}
A very important class of interacting fermion models is that
of Hubbard type models (see \ref{models-second}). First studies of 
entanglement in the one-dimensional case have appeared in~\cite{korepin04} 
and in~\cite{gu04}. Most of the studies in this type os systems analyzed the 
properties of the local entropy.

Gu {\em et al} analyzed the local entropy for the one-dimensional extended 
Hubbard model for fermions with spin 1/2.
Due to the conservation of particle number and $z$-projection of the spin,
the local density matrix of the system takes the simple form
\beq
\rho^{(1)}_j=z\proj{0}+u^+\proj{\up}+u^-\proj{\down}+w\proj{\up\down} \nonumber
\eeq
independent of the site number $j$ because of translational symmetry.
\begin{figure}[ht]
\begin{center}
\includegraphics[width=0.4\textwidth]{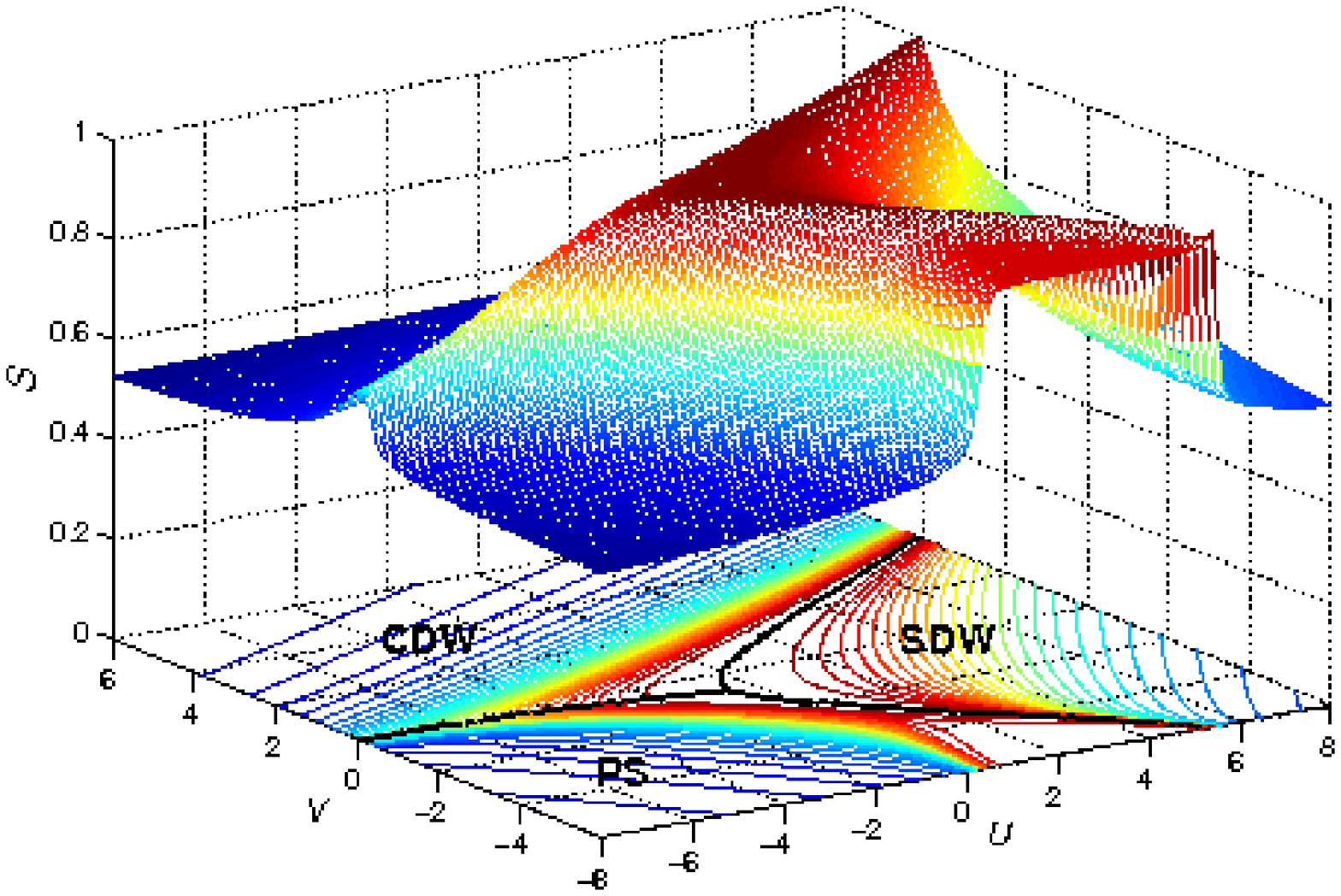}
\hspace*{7mm}
\includegraphics[width=0.36\textwidth]{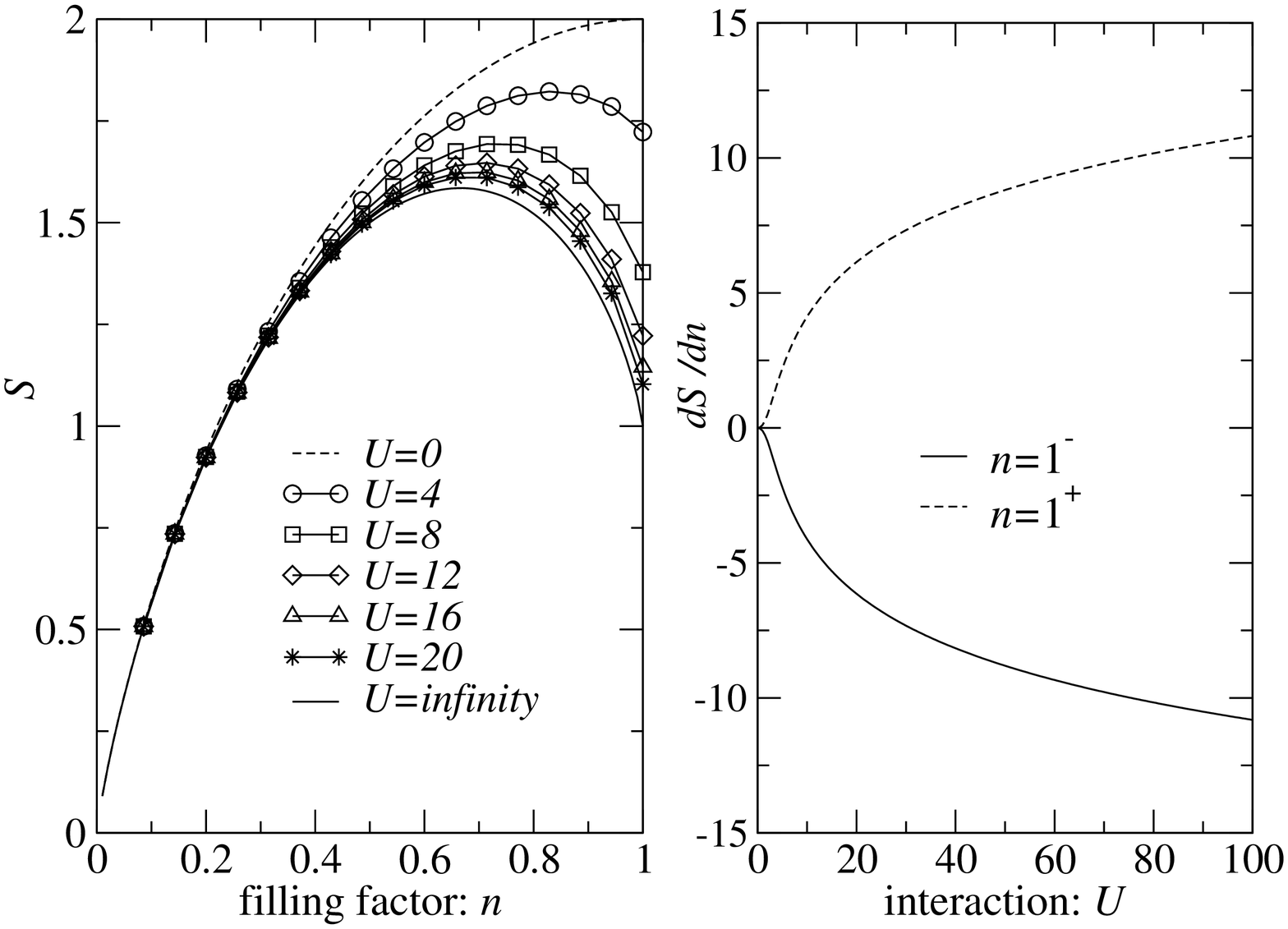}
\vspace*{1mm}
\caption{
\label{GuDeng1-4}
{\em Upper panel:}
The local entropy as a function of the on-site coupling $u$ and 
nearest neighbor coupling $v$. The contour plot below includes
the known phase diagram of the model (full black line). Except the 
superconducting phase, the phase diagram is nicely reproduced.
{\em Lower panel:} 
The local entropy for the standard Hubbard model and different on-site
couplings $u$ as a function of the filling ($1$ corresponds to half filling).
Except for $U/t=0$, where a maximum appears, this quantity shows a dip
at half filling, i.e. at the metal-insulator transition.
[From~\cite{gu04}]}
\end{center}
\end{figure}
The broken translational invariance 
      in the charge density wave phase has not been 
      taken into account in this work.
 This does not affect the 
central result but might affect the entropy
within the charge density wave phase.
Except the superconducting phase, the phase diagram at half filling (for $\mu=0$) 
 of this model has been nicely reproduced by 
the contour plot of the local entropy (see top panel of Fig.~\ref{GuDeng1-4}),
where the phase transition coincides with its crest. 
This turns out to be a general feature of local entropies 
- also
for spin models - as opposed
to entanglement class specific measures, as e.g. the concurrence 
for pairwise entanglement whose maxima in general appear at a certain distance 
to quantum critical points 
and hence are not associated to the quantum phase transition. 
In view of the monogamy of entanglement 
this is interpreted as evidence of dominant multipartite entanglement 
in the vicinity of quantum phase transitions.\\
For the Hubbard model Eq.(\ref{Hubbardmod}) and fixed $U/t$, the local entropy 
as a function of $n$  shows
a dip at the critical filling fraction $n_c=1$, where a metal-insulator
transition occurs (for $U>0$) (see bottom panel of Fig.~\ref{GuDeng1-4}).
For the two limiting cases $U=0,\infty$ the maximum instead is 
located at filling fractions, where the ground state is a singlet
of the largest symmetry group. Gu {\em et al} conjecture
that this was true for general $U>0$ and then the presence
of an unknown phase transition at these maxima.

This analysis clearly points out that the local entropy indicates
different phase transitions in different ways, essentially depending on
whether the quantity is sensitive to its order parameter or not.
Due to the $u(1)$ symmetry of the model, the single site reduced 
density matrix is a functional of occupation numbers only.
These operators cannot, however, describe order parameters of
superconductivity or some order parameter of the metal-insulator 
transition.
Indeed, the superconducting  phase can be predicted if the entropy of 
entanglement is calculated for a block of spins, instead of for just 
a single site~\cite{DengGu05}. A reduced density matrix of at least
two sites is necessary for being sensitive to superconducting correlations
[see also Ref.~\onlinecite{Legeza06} for a similar result obtained for the 
ionic Hubbard model] to be seen.

Another model studied with this respect is the so called
bond-charge extended Hubbard model (see \ref{models-second}). 
In phases II and III of fig. 15 there are  superconducting correlations
which are due to $\eta$-pairing and hence indicate the presence
of multipartite entanglement, as discussed before~\cite{VedralNJP04}.
At the bond-charge coupling corresponding to $x=1$,
the entanglement of the model has been analyzed in Ref.~\cite{AnfossiGiorda05}
and for general $x$ and $n=1$ in Ref.~\cite{AnfossiBoschi05,Anfossi06}. Besides the local entropy
of entanglement $S_i$, in \cite{AnfossiGiorda05},  also the negativity~\cite{Vidal02}
and the quantum mutual information~\cite{Groisman05}
have been used. 
\begin{figure}[ht]
\includegraphics[width=0.28\textwidth]{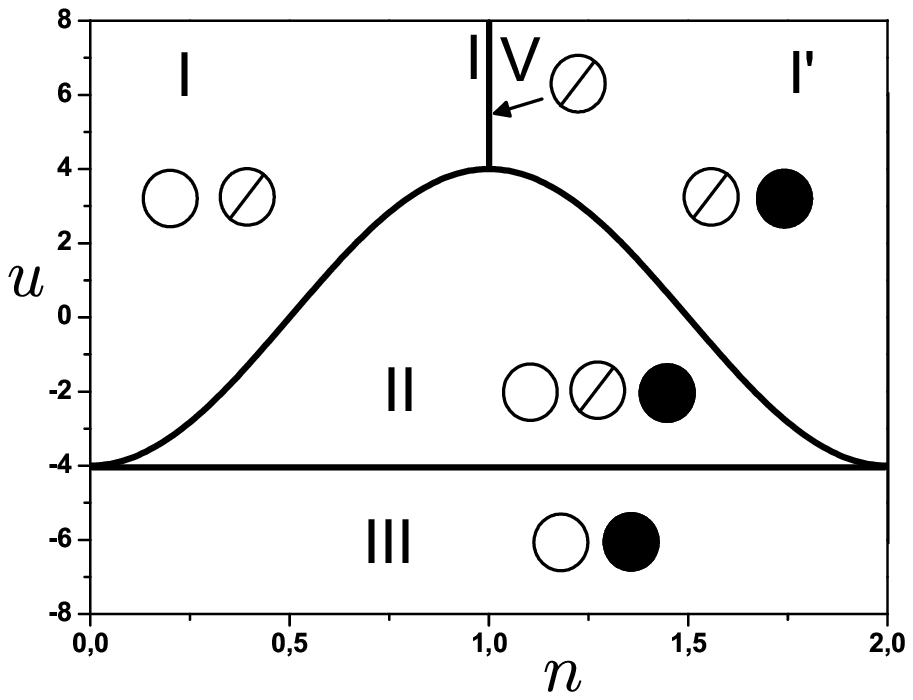}
\includegraphics[width=0.28\textwidth, viewport= 60 250 580 825,clip
]{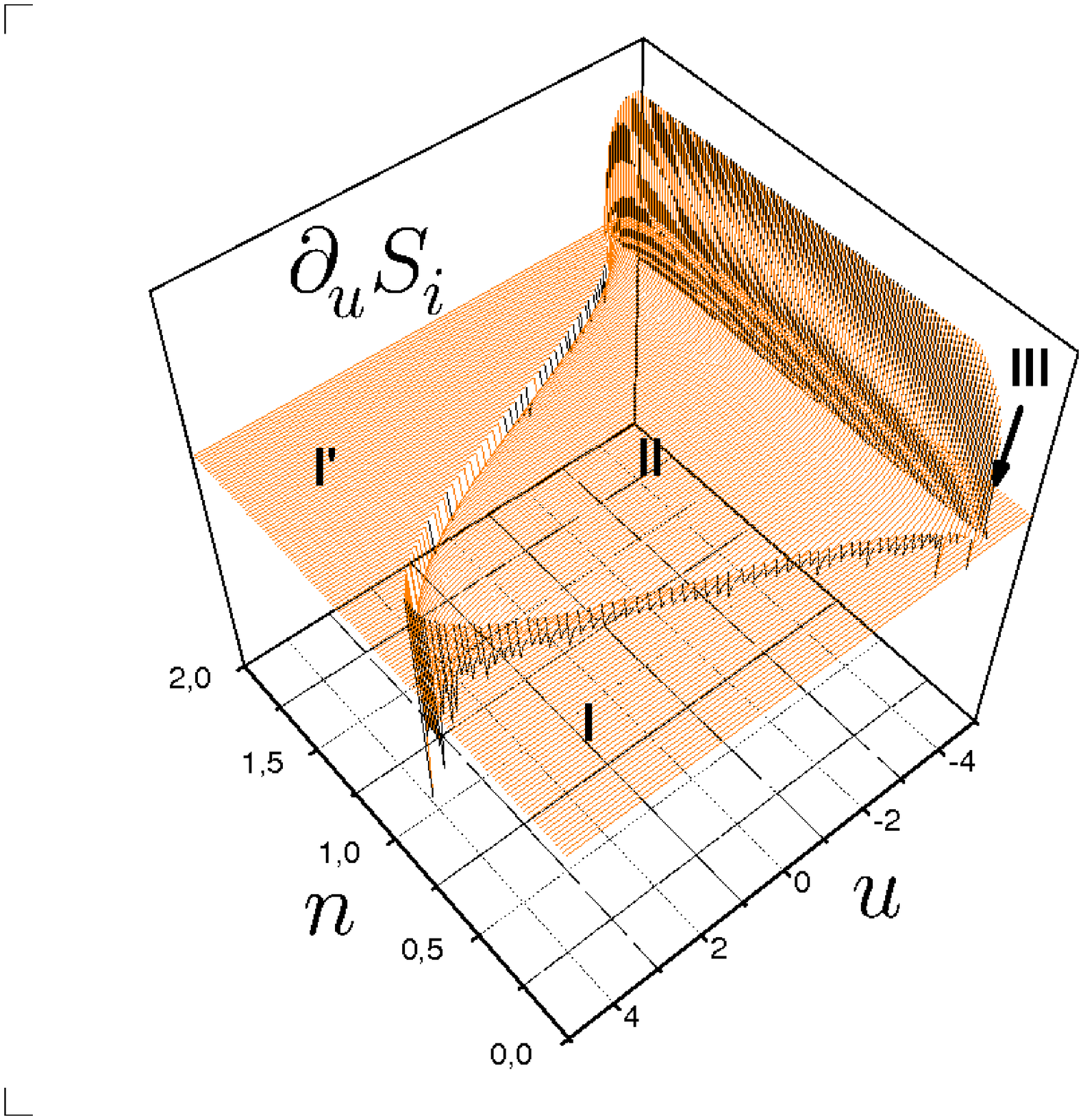}
\caption{{\em Upper panel:} The ground state phase diagram of the
Hirsch model at $x=1$. Empty, slashed and full circles 
indicate the presence of empty, singly and doubly occupied sites,
respectively. 
{\em Bottom panel:} Except the insulating line IV, 
the phase diagram is nicely reproduced by 
$\partial_u\mathcal{S}_{i}$. [From~\cite{AnfossiBoschi05}]}
 \label{Anfossi1-2}
\end{figure}
While $S_i$ measures all (pairwise and multipartite) quantum correlations
involving this specific site, the negativity offer a 
lower bound for the quantum correlation of two specific sites, 
and the mutual information accounts for pairwise quantum and classical correlations.
Therefore, this combination of correlation measures opens the possibility
to decide, what type of correlation is relevant at a
quantum phase transition. The results are shown in the upper panel
Fig.~\ref{Anfossi1-2}.
The different phases are shown  of Fig.\ref{Anfossi1-2}: they are  discriminated by local occupation numbers 
as described  in the top panel;
consequently, the entropy $S_i$ bears 
the information on all the phase diagram except the insulating line IV.
This is seen from the plot of $\partial_u S_i$ (with $u\doteq U/t$) as a function of the 
on-site Coulomb coupling $u$ and the filling fraction $n$.
A comparison of first derivatives respect to $y=n,u$
(depending on the phase transition) of all three correlation measures 
reveals common singularities for $\partial_y S_i$ and
$\partial_y {\cal I}_{ij}$ only for the transitions II-III and II-IV; 
furthermore, for  both of them it was proven that the  range of the concurrence $R$ diverges\cite{Anfossi06}. 
These  facts allow to characterize  the transitions II-III and II-IV (at $n=1$ and arbitrary $x$) as governed 
by pairwise entanglement, that is the more spread along the chain the closer the transition is got. 
For the  transitions II-I  and II-I' instead the  multipartite entanglement is relevant, with a finite range 
of the concurrence. A similar behaviour was encountered for {\it non-critical} spin models where the divergence of $R$  
is accompanied by the emergence of a fully factorized ground state (see Sec.~\ref{mod} and~\ref{concu.spin}). 
Here $R$ diverges close to QPT; it was  also noticed by Anfossi {\it et al.} that, while the ground state in  IV is indeed factorized,
the phase III is made of a superposition of doubly occupaied and empty states.  Observing that the pairwise entanglement is 
vanishing in  both the phases III and IV, the authors conjectured that the divergence of $R$ marks an 'entanglement transition' 
in solely in the pairwise entanglement.    
 
In order to detect the transition II-IV at $n=1$ and any $x$, $\partial_x S_i$ has been
calculated by means of DMRG~\cite{AnfossiBoschi05}. 
Its singularities allow to accurately determine the charge gap as a function
of the bond-charge coupling $x$.

We now proceed with the Hubbard model in a 
magnetic field (see \ref{models-second})
Also here, the local entropy $S_i$ has been looked at in order to analyze 
its entanglement. 
As in the examples before,
$S_i$ indicates the second order phase transitions in terms of
divergences of its derivatives $\partial _h S_i$ and
$\partial_\mu S_i$, respectively. Indeed, it has been demonstrated that
$\partial_h S_i$ and $\partial_\mu S_i$ can be expressed in terms of 
spin and charge susceptibilities~\cite{Larsson05}, 
hence bridging explicitly between the standard method
in condensed matter physics for studying phase transitions and
the approach from quantum information theory.

The local entropy for thebosonic version of the Hubbard model, the Bose-Hubbard model, was 
computed in~\cite{Giorda04,Buonsante07} for different graph topologies. The attention in this work was 
on the dependence of the entanglement on the hopping amplitude. The authors showed that for 
certain classes of graphs the local entropy is a non-monotonic function of the hopping. Also 
for the bosonic case the local entropy is a good indicator for the presence of a 
(superfluid-insulator) quantum phase transition.

Summarizing, the body of work developed  suggest  to conclude that 
that local entropies can detect QPTs in systems of itinerant fermions, particularly 
if the transition itself is well predicted by a mean field approach for local observables 
of the model, [see also~\cite{Larsson06}]. 
Furthermore, translational invariance is necessary for the prediction
to being independent of the site, the local entropy is calculated for.
If this symmetry is absent, it might prove useful to average over the
sites; the resulting measure is then equivalent to the 
$Q$-measure~\cite{Wallach}. 

\subsection{Topological entanglement entropy}
We close this section by summarizing the ongoing research activity
studying the subleading corrections to the block entropy
in the two-dimensional systems.
Most of the results were demonstrated for quantum two-dimensional lattices
(though generalizations two higher
dimension is straightforward).

Fradkin and Moore considered  quantum critical points in two
spatial dimensions with scale invariant ground state wave functions,
characterizing for example the scaling limit of quantum eight
vertex models and non abelian gauge theories (see \onlinecite{Ardonne04}
for a recent reference).
The main result is that  a universal logarithmically divergent correction,
determined by the geometry of the partition of the system, emerges  in
addition to
the area's law term in the entanglement block entropy~\cite{fradkin06}.

Such work benefit of the earlier seminal contribution of Kitaev and
Preskill, and Levin and Wen~\cite{kitaevpreskill,levinwen}
demonstrating that the correction to the area law is of  topological origin.
Namely the entanglement entropy was demonstrated to scale as
\begin{equation}
S=\alpha {\ell} -\gamma + {\cal O} ({\ell}^{-1})
\end{equation}
The coefficient $\alpha$ is non universal and ultraviolet divergent.
In contrast, the quantity $\gamma$ can be expressed as $\log D$, where $D$
is known as the {\it total quantum dimension},  is universal, and related
to the topological winding number of the
theory.
The calculations were pursued with methods of topological field theory,
giving an explicit expression for $\gamma$.

We remark that  the result of  Kitaev and Preskill, and Levin and Wen could
provide an alternative avenue to lattice gauge theory methods\cite{Wen-book},
detecting a genuine topological order in the system (when descriptions
based on local order parameters fails) by  direct inspection of the wave
function.

When a topological order is present, the ground state of the system
acquires a peculiar degeneracy when its lattice has a 
non trivial genus $g$.
Hamma {\em et al.} studied the entropy of the Kitaev model on a two
dimensional lattice with generic genus $g$.
The Kitaev model~\cite{Kitaev} is a two dimensional exactly solvable lattice
model  with double periodic boundary conditions,
whose Hamiltonian can be realized with a set
of spins in a square lattice with ring exchange  and vertex interactions.
The ground state of the Kitaev model is characterized by the presence of a
topological order \footnote{The Kitaev model  was suggested to provide a
realization of the so caaled 'toric code', namely a topological quantum
computers made by
a physical system with anionic excitations
(see also \onlinecite{Rasetti})}.
For such systems Hamma {\em et al.} related the degeneracy of the ground
state (that is $4^g$), to
the block entropy~\cite{hamma1,hamma2}.

As a step toward models with more generic topological orders, the
topological entropy was studied numerically
for the quantum dimer model in a triangular lattice~\cite{Furukawa} and
for fractional
quantum Hall states~\cite{Haque}.

We finally mention that the topological term in the entanglement entropy
in the context
of quantum gravity was evidenced in~\cite{ryu,Fursaev06}.

\subsection{Entanglement along renormalization group flow}

One of the original motivations put forward by Preskill to investigate entanglement in 
many-body systems~\cite{preskill00} was the idea that quantum information could elucidate
some features of the renormalization group which is a cornerstone method in modern physics.
It is natural to think that the procedure of tracing out high energy modes in a renormalization 
group step has some kind of irreversibility built in. Quantum information 
concepts could prove to be useful in elucidating issues related to this irreversibility and 
possible help could come from relating the celebrated $c-$theorem~\cite{zamolodchikov86} 
to the loss of information.  

Entanglement loss along a renormalization group trajectory was studied in spin 
chains~\cite{latorre05}. More recently a number of relations relating renormalization 
group, conformal field invariance and entanglement loss were derived in~\cite{orus05b}. 
According to~\cite{latorre05} entanglement loss it can be characterized at three 
different levels:\\
\noindent
\underline{- Global entanglement loss -}   
  By using the block entropy as a measure of entanglement, for 
  which we know the result of Eq.(\ref{Sell}), and an inequality on the central charges which 
  derives from the $c-$theorem, it follows that    $S^{UV} \ge S^{IR}$.
  The block entropy at the ultraviolet fixed point cannot be smaller than that 
  at the infrared fixed point.\\
\noindent
 \underline{- Monotonous entanglement loss -} It is also possible to follow the entanglement 
  along the whole transformation. Also in this case the entropy is a non-decreasing function along 
  the flow. As a simple example one can consider the block entropy of and Ising chain close to the critical 
  point which goes as $S \sim \ln |\lambda -1|$, from which monotonicity follows.\\
\noindent
 \underline{- Fine-grained entanglement loss -} The monotonicity of the entanglement seems to 
  be present at a deeper level in the structure of the density matrix. It is possible to 
  show~\cite{latorre05,orus05}
  through majorization relations that the spectrum of the reduced density matrix gets more ordered 
  along the flow. By denoting with $r_m$ the eigenvalues of the reduced density matrix $\rho$, majorization 
  relation between the two set of spectra (corresponding to two different parameters) means 
  that there is a set of relations  for which 
  $\sum_{i=1,n} r_i \ge \sum_{i=1,n} r_i'$ for $n= 1, \; \cdots , \; d$ 
(d is the dimension of $\rho$).


Motivated by ultraviolet divergencies of the entropy of entanglement
in quantum field theory,
Casini and Huerta introduced a quantity $F(A,B)$ related to the entropy
measuring the degree of entanglement between the two regions $A$ and $B$. 
The function $F$ is defined as 
$
F(A,B) = S(A) + S(B) - S(A \cap B) - S(A \cup B)
$
which coincides with the mutual information Eq.(\ref{mutual}) in the case of non-intersecting 
regions. In two dimensions, it is a finite positive function with the property
$F(A,B)\leq F(A,C)$ for $B\subset C$ if $A\cap C=\emptyset$.
Then, for sets with a single (path-connected) component
in two dimensional conformal field theories they showed that 
it allows to prove an alternative entropic version of the c-theorem~\cite{casinien}.

\section{Localizable entanglement}
\label{local}
\subsection{Localizable entanglement and quantum criticality}
\label{ground-localizable}
The study of localizable entanglement (see Section~\ref{locsection}) in spin chains allows to find 
a tighter connection 
between the scales over which entanglement and correlations decay (we saw in the 
previous sections that the two spin entanglement, expressed by the concurrence, 
does not decay on a the same range of correlations)~\cite{verstraete04,popp05,popp06}.
One expects that the procedure of entangling distant sites by a set 
of local measurements will be less effective as the distance between the two particles 
increases thus leading to 
a definition of entanglement length $\xi_E$. For a translational invariant system $\xi_E$  can be defined  
in analogy of the standard correlation lenght

\begin{equation}
        \xi_E^{-1} = - \lim_{|i-j| \to \infty} \log \frac{E_{loc}(|i-j|)}{|i-j|} \;.
\label{elenght}
\end{equation} 
By definition the entanglement length cannot be smaller than the correlation length, 
$
        \xi_E \ge \xi
$, 
therefore at a second order phase transition the localizable entanglement length diverges. 
In addition there may also appear "transition points" associated solely to a divergence  
in $\xi_E$. In order to avoid misinterpretations, it must be stressed that
the {\em localizable} ``classical'' two-point correlations then diverge as well. 
Thus, the essence of the phenomenon is that
correlations can be localized between arbitrarily distant sites 
by means of suitable local operations and classical communication 
despite a finite correlation length; necessary for this is the presence of global
entanglement~\cite{Popescu92}.  

For the Ising model in a transverse field it can be shown that~\cite{VerstraetePC04}
\begin{equation}
         \max_{\alpha =x,y,z} |Q_{\alpha}^{ij} | \leq
         E_{loc}(i-j)\leq\frac{1}{2}\sum_{\pm}\sqrt{s_{\pm}^{ij}}
\label{loc.ent.bounds}
\end{equation}
where
$
s_{\pm}^{ij}= \left( 1 \pm \langle S_i^z S_j^z \rangle \right)^2- \left(\langle
S^z_i \rangle \pm \langle S^z_j \rangle \right)^2 \,\, 
$
and 
$
Q_{\alpha}^{ij} = \langle S_i^{\alpha}S_j^{\alpha}\rangle - 
\langle S_i^{\alpha}\rangle \langle S_j^{\alpha}\rangle \;\;\; .
$ 
 
In this case, the lower bound in Eq. (\ref{loc.ent.bounds}) is determined by 
the two-point correlation function in the x-direction.
In the disordered phase ($\lambda < 1$) the ground state possesses a small 
degree of entanglement and consequently its  entanglement length is finite. 
The situation changes at the other side of the critical point. 
Here, although the correlation length is finite, the entanglement length is 
infinite as asymptotically the correlation tends to a finite values. 
The divergence  of $\xi_E$ indicates that the ground state is a globally entangled state,
supporting the general idea that multipartite entanglement is most 
relevant at the critical point~\cite{Osborne02,Firenze04}. 

The properties of localizable entanglement were further investigated for a 
spin-1/2 $XXZ$-chain in~\cite{jin04,popp05} as a function of the anisotropy 
parameter $\Delta$ and of an externally applied magnetic field $h$. 
The authors used exact results for  correlation functions relying on the 
integrability of the  models to find the bounds in Eq.(\ref{loc.ent.bounds}).
For the antiferromagnetic $XXX$-case they provided the following lower bound
$
	E_{loc}(i-j) \ge \frac{2}{\pi^{2/3}}\frac{\ln |i-j|}{|i-j|} \;\; .
$
The presence of the anisotropy increases the lower bound of the localizable 
entanglement. At the Berezinskii-Kosterlitz-Thouless critical point ($\Delta = 1$) 
the lower bound of the nearest-neighbour localizable 
entanglement shows a kink~\cite{popp05}. As pointed out by the authors this might have  
implications in the general understanding of the Berezinskii-Kosterlitz-Thouless 
phase transitions where the ground state energy
and its derivatives are continuous as well as the 
concurrence (see \ref{ground-critical} and Fig.\ref{xxz-gs}).

The localizable entanglement in two-dimensional $XXZ$ model was discussed as 
well~\cite{Syl} by means of quantum Monte Carlo simulations. 
A lower bound has been determined by studying the maximum 
correlation function which for $\Delta > -1$ is $Q_{x}$, the long-range 
(power law) decay of the correlation implying a long ranged localizable entanglement.

The definition of localizable entanglement has an interesting connection with the 
concept of quantum repeaters introduced in~\cite{briegel98}. Quantum repeaters have 
been designed to enhance the transmission  of entanglement through noisy channels. The 
idea is to distribute along the channels a number of intermediate sites where a certain 
number of local operations are allowed in order to maximize the entanglement between the 
transmitter and a receiver. This is the very definition of 
localizable entanglement. 

Localizable entanglement has been defined in Sec.\ref{locsection} as an average 
over all possible measuring processes it is of interest to understand also 
the statistical fluctuations around this average value. To this end~\cite{popp05} 
analyzed also the variance associated to the entanglement fluctuations:
$ 
       \delta E_{loc}^2 = \sum_{s}p_s E^2(|\psi_s \rangle ) - E^2_{loc} \,  
$, 
where $E$ is a measure of pairwise entanglement.
The fluctuations of the entanglement increase in the vicinity of a critical point. 
This was checked explicitly for the one-dimensional Ising model.  

As detailed below, additional interesting results were obtained for spin-1 systems 
where a true transition in the entanglement (with a diverging $\xi_E$ but finite 
correlation length) has been found.

\subsection{Localizable entanglement in valence bond ground states}
\label{gapped}
For half-integer spins, gapped non degenerate ground states are characteristic for systems in 
a disordered phase (consider paramagnets for example). A finite gap in the 
excitation spectrum of the system in the thermodynamic limit makes the correlations 
decaying exponentially. This is the Lieb-Schultz-Mattis theorem, 
establishing that, under general hypothesis, the ground state of a spin system is 
either unique and gapless or gapped and degenerate~\cite{LIEB} (see \onlinecite{Hastings03} for recent results). It was a surprise, 
when Haldane discovered that systems of integer spins can violate
this theorem~\cite{HALDANE-CONJ1,HALDANE-CONJ2}. 
This suggests to investigate whether the entanglement in the ground state 
might play some role in establishing the hidden order characteristic for the Haldane phases. 
An aspect that might be relevant to this aim was recently addressed by studying 
the localizable entanglement in  AKLT models~\cite{verstraete04}. 
The ground state of this class of models is of the valence bond type as 
discussed in~\ref{models-spin1}
\begin{equation}
|gs\rangle_{AKLT}=\left (\otimes_k A_{k,\bar{k}}\right ) 
|I\rangle_{\bar{0}1} |I\rangle_{\bar{1}2} \dots |I\rangle_{\bar{N}N+1} \;. 
\label{valence-AKLT}
\end{equation}
$|I\rangle$ are singlets and $A$ are $3\times 4$ operators projecting the 
Hilbert space of two combined spins on its symmetric part, at the given site. 
This is nothing else than a matrix product state (see Section~\ref{models-spin1}).
For this state it was demonstrated that a singlet state made of two 
spins-$1/2$ located at the ends of the chain can be always realized 
(see Fig.\ref{rvb}). This implies that the localizable entanglement is 
long ranged despite the exponentially decaying correlation~\cite{verstraete04}. 

The localizable entanglement can be  related to the string 
order parameter $O_{string}^\alpha$ defined in Eq.(\ref{string}).
The key to make explicit this relation is to observe that the
localizable entanglement can be calculated as an expectation value~\cite {Campos-Venuti05} as
$
\displaystyle{L(|\psi\rangle)=\langle \psi| \sigma_0 G_s(\psi) \sigma_N |\psi \rangle }
$
where $G_s =\sum_s|s\rangle \langle \bar{s}| 
\mbox{sign} (\langle \psi| \sigma^y\otimes \sigma^y|\psi\rangle)$,
and
$|s\rangle$ is the optimal basis  which maximize the entanglement of assistance.
For the AKLT model, the expression above reads
\begin{equation}
L(|gs\rangle_{AKLT})=\langle \prod_{i=1}^N e^{i\pi S^y_i} \rangle
\end{equation}
In this case, both the localizable entanglement and
all the three components of the string order parameter saturate.
Perturbing the AKLT ground state, namely making the resonating valence bonds with 
non-maximally entangled states $(\ket{10} - \displaystyle{\e^{-i2\phi}}\ket{01})/\sqrt{2}$, 
the relation between the hidden order and the localizable entanglement
is weakned  (as compared to the AKLT model):
it can be demonstrated that the string order parameters for $\alpha=x,z$
are finite, while the localizable entanglement vanishes exponentially 
with the deformation $\phi$~\cite{popp05}, but 
a tight connection of the localizable entanglement 
and $O_{string}^y$ is observed to persist also
for the $\phi$ deformed ground state~\cite{Campos-Venuti05}.

The valence-bond-solid phase order was further studied by looking at the hidden 
order in chains with more complicated topology.  
The von Neumann entropy was studied in spin-$1$ $XXZ$ model with biquadratic 
interaction and single ion anisotropy  in~\cite{GuTianLin05,Wang-spin1} and in~\cite{Campos-Venuti06}.
Some of the features of the corresponding phase diagram are captured.
The Haldane transitions exhibited in the phase diagrams are marked by 
anomalies  in the Von Neumann entropy; its  maximum at the 
isotropic point is not related to any critical phenomenon 
(the system is gapped around such a point), but it is due to the 
equi-probability of the three spin-$1$ states occurring at that point~\cite{Campos-Venuti06}. 
Since the  Berezinskii-Kosterlitz-Thouless 
transition separating the $XY$ from the Haldane or large-$D$ phases connects
a gapless with a gapped regime, it was speculated that an anomaly in the entanglement 
should highlight such transition~\cite{GuTianLin05}.

\section{Thermal entanglement }
\label{th}

Though the  very nature of entanglement is purely quantum 
mechanical, we saw that it can persist for macroscopic systems 
and will survive even in the thermodynamical limit.  
Entanglement survives also at finite temperatures. 
This temperature could be 
as high as $100$ Kelvin in high-temperature 
superconductors~\cite{VedralNJP04} (see also~\onlinecite{Fanllo04}).
In this section we review the 
properties of entanglement in many-body systems at finite temperatures 
(see also~\onlinecite{andersrev}).  
We will see that the analysis could shed new light on the interplay between 
the quantum nature of the system and its thermodynamics.
Moreover, somewhat surprisingly, macroscopic state variables can be 
used to detect entanglement. Thermodynamics describes large scale systems 
with macroscopic properties, its state variables  $T, N, V, p$, its external 
fields $h$ and its response functions the susceptibility and heat capacity, respectively 
$\chi, C, ...$.  Addressing entanglement as a thermodynamical property 
raised significant amount of interest in various communities. 
One wants to know, for example, under what conditions  can we detect 
and extract entanglement. Can we see 
entanglement itself as a state variable, just like pressure is for a collection 
of atoms in a gas? What could be the corresponding thermodynamical potential? 
Is entanglement extensive? Since entanglement is closely related to entropy,
we would expect the answer to the last question to be ``yes''. 

The states describing a system in thermal equilibrium states, are determined 
by the Hamiltonian  and the inverse temperature 
$\beta=(1/T)$. The density matrix is $\rho=Z^{-1}e^{-\beta \hat H}$.
where $Z = tr[e^{-\beta \hat H}]$
is the partition function of the system. 
The thermal states expanded in the energy eigenbasis 
$|e_i \rangle, i= 0, 1, ...$ are then
\begin{equation}
	\rho = {e^{-\beta E_0} \over \sum_i e^{-\beta E_i} } |e_0 \rangle\langle e_0| 
	+ {e^{-\beta E_1} \over \sum_i e^{-\beta E_i} } |e_1 \rangle\langle e_1|
 	+ ...
\label{thstate}
 \end{equation}
Any separable state, or classical state, with respect to this split into subsystems 
$A, B, C, D, ..$ (for example the sites of a spin system) can then be written as convex 
mixture of tensor products of states of the respective subsystems $A, B, C, D, ...$ with 
probabilities $p_i$ 
$        
\rho = \sum_i p_i \rho_i^A \otimes \rho_i^B \otimes \rho_i^C \otimes \rho_i^D \otimes ...\
$.
If the state in Eq.(\ref{thstate}) {\em cannot} be written in the form given above then it is 
entangled. In this section we will discuss the properties of this {\em thermal entanglement}.

\subsection{Thermal pairwise entanglement}

Extensive efforts have been made to understand how to quantify 
thermal entanglement in many-body systems starting from the initial 
papers~\cite{nielsenphd,Arnesen01,Gunlycke01}.
Several models of interacting spins in arrays were discussed. Entanglement as 
measured by concurrence was shown to exist  at nonzero temperatures in the 
transverse Ising~\cite{Osborne02}, Heisenberg~\cite{wangth,wangzan,tribedi,Asoudeh04}, 
$XXZ$~\cite{canosa05,canosa06}, $XYZ$ models~\cite{rigolinth,zhangzhu},
ferrimagnetic transition~\cite{wangwang06} and spin-one chains~\cite{zhangli}.
Sevaral non-trivial aspects of the behaviour of the pairwise entanglement at finite 
temperatures can be illustrated by considering the simple case of 
two-sites systems. 

\paragraph{$XXX$ Model}
We start our discussion on thermal entanglement by considering 
the $XXX$ antiferromagnet (see 
Eq.(\ref{general-spin})). In this case the thermal state of this system can 
be written as a Boltzmann mixture of the singlet and the triplet states:
\begin{eqnarray}
\rho_T & = & \frac{1}{Z} e^{3\beta J}|\psi^-\rangle\langle \psi^-| + 
e^{-\beta (J+2h)}|00\rangle\langle 00|\nonumber\\
& + & e^{-\beta (J-2h)}|11\rangle\langle 11| + e^{-\beta
J}|\psi^+\rangle\langle \psi^+|
\label{thstatexxx}
\end{eqnarray}
where $Z= e^{3\beta J}+e^{-\beta (J+2h)}+e^{-\beta
(J-2h)}+e^{-\beta J}$ is the partition function of the system and 
$|\psi^{\pm} \rangle = |01\rangle \pm |10\rangle$.
For the sake of simplicity we focus our attention only on two regimes

The first regime is when the coupling $J$ is large compared to the external field $h$.
The ground state is then the singlet
state, and at low temperature the system is therefore entangled. At
higher temperatures the triplet becomes mixed into the singlet, and
when (roughly) $T>J/k$, the entanglement completely disappears (when
the external field is zero). Therefore, in order to have  high-temperature entanglement 
in dimers we need a large value of the
coupling constant $J$.

\begin{figure}
\centering
\includegraphics[width=7.5cm]{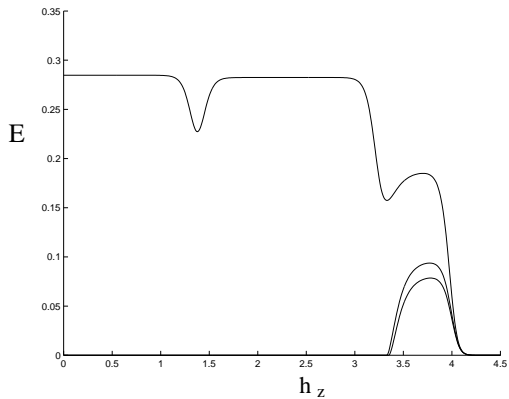}
\caption{
Entanglement between two qubits
interacting according to the antiferromagnetic Heisenberg model as a function
of the external field $h_z$ and temperature (multiplied by the
Boltzmann's constant) $T$ with coupling $J=1$.
The topmost plot shows the variation of nearest-neighbor
entanglement with the magnetic field.
The middle and the bottommost plot show the same for next
nearest and next to next nearest neighbors, respectively. 
From~\protect\onlinecite{Arnesen01}} 
\label{arnes}
\centering
\end{figure}

When $J$ is fixed, the second regime occurs when we can vary the
value of the external field $h$. When $h$ is large (greater than
$2J$), the ground state is $|11\rangle$ and at zero temperature the
dimers are therefore not entangled. The point where the singlet
state becomes replaced by $|11\rangle$ as the ground state
corresponds to the quantum phase transition (occurring in  the 
thermodynamical limit). 
However, if we start to
increase the temperature, the singlet state---which is the first
excited state under these circumstances---starts to become populated.
{\em Entanglement can then be generated by increasing the
temperature.} The behaviour of entanglement as a function of the 
magnetic field is shown in Fig.\ref{arnes}.

\paragraph{Ising model}
Another interesting case is that of a Ising coupling.
At zero temperature only the lowest energy level is
populated. In the case $N=2$ the tangle of this pure state can easily be calculated
from the density matrix, for $h>0$,
$
  \tau_1 =J^2/(J^2+h_z^2) 
$.
It is clear that the entanglement is
highest for nearly vanishing magnetic fields and decreases
with increasing field amplitude
(this expression however is not valid for strictly $h=0$, where no entanglement is present).
Let us now turn our attention to the  case of non-zero
temperatures. For a general pure state only one of the
eigenvalues of the Hamiltonian weight Eq.(\ref{thstate}) is non-zero and
therefore equal to the tangle.  For low temperature and
magnetic field, i.e. $h,T\ll J$, it is a good
approximation to assume that only the two lowest energy levels
are populated.  In this case, the combination of the two lowest states also
combines their concurrences in the following way~\cite{Gunlycke01}:
\begin{equation}
  \label{eq:weightconcurrence2}
  C = \max\{|w_0C_0 - w_1C_1|,0\},
\end{equation}
where  the index $0$ refers to the ground state, while $1$ refers
to the excited state and $w_0$ and $w_1$ are the thermal weights of the
ground and excited states respectively, see Eq.(\ref{thstate}).  
This phenomenon has been named as {\em concurrence mixing}. In this case, the first 
excited state is the Bell state, $|\Psi^{-}\rangle = (|01\rangle-|10\rangle)/\sqrt{2}$, 
and Eq. (\ref{eq:weightconcurrence2}) reduces to
$
 C =|w_0\frac{J}{\sqrt{J^2+h_z^2}} - w_1|
$.
In general, the first term in the above equation is larger than
the second, and in this case the concurrence decreases with
temperature as $w_0$ decreases and $w_1$ increases. Moreover it is also 
possible to see that, for a given temperature, the entanglement can 
be increased by adjusting the magnetic field and is generally largest 
for some intermediate value of the magnetic field. This effect can be 
understood by noting that $w_0$ increases with increasing $h$ as the
energy separation between the levels increase, but $J/\sqrt{J^2+h_z^2}$ decreases. 
As a result the combined function reaches a peak as we vary $h$ and 
decreases subsequently, inducing analogous behavior for the concurrence. 

\subsection{Pairwise entanglement in the $T \ne 0$ critical region}

At finite temperatures but close to a quantum critical 
points, quantum fluctuations are essential to describe  
the properties of the systems~\cite{Halperin89,Sachdev99}. 
In order to study the interplay between the 
thermal entanglement and the quantum fluctuations caused by the critical point at $T=0$   
the  analysis on small clusters is no longer sufficient.  
For the presentation we consider the one-dimensional quantum $XY$ models. Such systems 
cannot exhibits any phase transitions at finite temperature, but the very existence of 
the quantum critical point is reflected in the crossover behaviour at $T\ne 0$.
The renormalized--classical crosses-over the quantum disordered regimes through the so called
{\em quantum critical region}~\cite{Sachdev99}.  In
the $T-h$ plane a $V$-shaped phase diagram emerges, characterized by
the cross-over temperature customarily defined as $T_{cross}\doteq |\lambda^{-1}-\lambda_c^{-1}|$.
For $T\ll T_{cross}$ the thermal De Broglie length is much smaller than the average spacing
of the excitations; therefore the correlation functions
factorize in two contributions coming from quantum and thermal
fluctuations separately.  The quantum critical region is characterized
by $T \gg T_{cross}$.  Here we are in the first regime and the correlation functions 
do not factorize.  In this regime the interplay between quantum and thermal effects is the
dominant phenomenon affecting the physical behaviour of the system.

Thermal entanglement close to the critical point of the quantum $XY$ models was recently 
studied in~\cite{Amico05}.
In analogy with the zero temperature case they demonstrated that the  entanglement 
sensitivity to thermal and to quantum fluctuations obeys universal $T\neq 0$--scaling 
laws. The  crossover  to the quantum disordered and renormalized classical 
regimes in the entanglement has been analyzed through the study of derivatives of  
the concurrence $\partial_{\lambda} C$ and  $\partial_T C$. The thermal entanglement 
results to be very rigid when the quantum critical regime is accessed from the 
renormalized classical and quantum disordered regions of the phase diagram; such 
a 'stiffness' is reflected in a maximum in $\partial_T C$ at  $T\sim T_{cross}$. 
The maximum in the derivatives of the concurrence seems a general feature of the 
entanglement in the crossover regime. In this respect we mention that also 
the concurrence 
between two Kondo impurity spins  
discussed in~\cite{stauber04,Stauber-errata} experiences the largest 
variation again in the crossover phenomenon.

\begin{figure}[ht]\centering
\includegraphics[width=7.5cm]{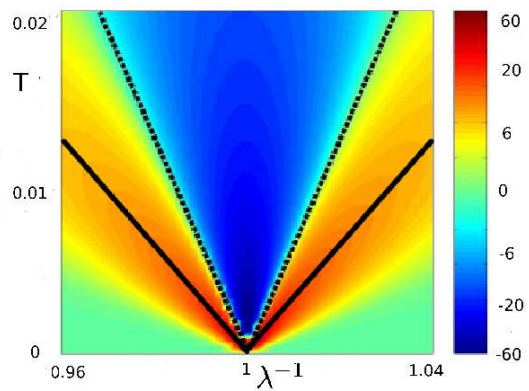}
\caption{The effect of temperature on the anomalies originated from 
the critical divergence of the field-derivative of $C(R)$ can be measured by 
$\partial_T[\partial_{a} C(R)]$. The density plot  corresponds to $\gamma=1$ and 
$R=1$. $T=T^*$ and $T=T_M$  are drawn as dashed and thick lines respectively. 
Maxima below $T^*$ are found  at $T_M =\beta  T_{cross}$ 
with $\beta\sim 0.290 \pm 0.005$ and they are independent of $\gamma$ and $R$; 
the crossover behaviour is enclosed in 
between  the two flexes of  $\partial_T[\partial_{a} C(R)]$ at  
 $T_{c1}$ $T_{c2}$; such values are fixed to: $T_{c1}=(0.170  \pm 0.005) T_{cross}$ and 
$T_{c2}=(0.442  \pm 0.005)T_{cross}$ and 
found to be independent of $\gamma$ and $R$. For $T\lesssim T_{c1}$ 
$\partial_T[\partial_{a} C(R)]\simeq 0$.
Scaling properties are inherited in $\partial_T[\partial_{a} C(R)]$  
from $\partial_{a} C(R)]$  [From \protect\onlinecite{Amico05}].}
\label{mixedcross}
\end{figure}

\begin{figure}
\includegraphics[width=60mm]{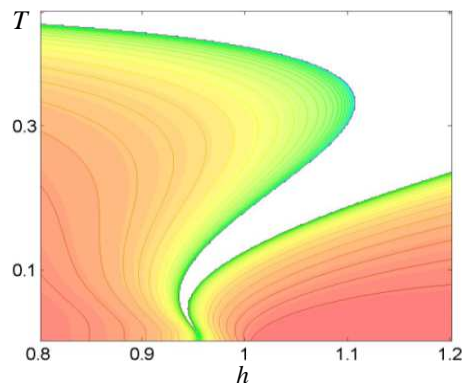} 
\caption{Contour plot of $\tau_2$ in the $h-T$ plane, for $\gamma=0.3$ 
(i.e. $h_f=0.9539...$). The white area indicates where 
$\tau_2=0$.[From \onlinecite{Amico06}]}  
\label{f.fig2} 
\end{figure}

Due to the vanishing of the gap at the quantum critical point,  
in the region $T\gg T_{cross}$ an arbitrarily small temperature 
is immediately effective  in the system (see Fig.~\ref{mixedcross}).  
From  analysis  of the quantum mutual information (see Eq.~\ref{mutual})) it emerges 
that the contribution of the 
classical correlations is negligible in the crossover, thus providing 
the indication that such a phenomenon  is driven solely by the thermal entanglement. 
It is interesting to study how the  exsistence of the factorizing field $h_f$ affects 
the thermal pairwise entanglement (vanishing at zero temperature). 
It is found that  the two tangle $\tau_2$ still vanishes 
in a region of the $h-T$ plane, fanning out from just $h_f$, (white region 
in Fig.~\ref{f.fig2})\cite{Amico06}. In there entanglement if present, 
is shared between three or more parties.
It is  further observed that, in contrast to the analysis of the ground state, at
finite temperature we cannot characterize the two separate phases of
parallel and antiparallel entanglement by. In fact, the two types of
entanglement (though well defined also for mixed states) can swap by varying
$T$ and/or $r$. The exchange between parallel
and antiparallel entanglement occurs in a non trivial way, that
ultimately produces the reentrance of $\tau_2$ seen in
Fig.\ref{f.fig2}.  

The common feature in all cases for which the existence of
entanglement could here be proved is that both high temperatures
and high values of magnetic field move the thermal states away
from the region with non-zero entanglement. This is understandable
because high values of magnetic field tend to order all spins
parallel to the field which corresponds to an overall state being
a product of the individual spin states. There is upper limit 
of this phenomenon, since an increase in the temperature  is disruptive for entanglement 
due to thermal fluctuations.

\subsection{Thermal entanglement witnesses}
\label{thermal-witness}

At first sight it may be very surprising that thermodynamical variables can witness 
entanglement since the  only need to obtain them are the system's eigenenergies, 
and no eigenstate information is required. Since entanglement resides in the fact 
that the state is inseparable (and is not related to the value of its energy) it 
would appear that the partition function is not enough to characterize entanglement. 
This logic, although seemingly simple, is not entirely 
correct. The reason is subtle and lies in the fact that the whole Hamiltonian is 
used for constructing the partition function, so in a roundabout way we do have 
the information about the states as well.  Entanglement witnesses in spin systems
have been considered in~\cite{Toth05a,bruknervedral,wulid05,hide07} (see also the next Section 
on multipartite entanglement)

We will now illustrate how and why we can derive entanglement witnesses for the 
partition function. Suppose that the system is described by an antiferromagnetic 
Heisenberg model. First we have the following identity coming from the Hellman-Feynman theorem
$
\sum_{i=1}^{N} \langle S_{i}^{x} S_{i+1}^{x} + 
S_{i}^{y} S_{i+1}^{y} + S_{i}^{z}
S_{i+1}^{z}\rangle \sim \frac{d}{dJ} (\ln Z)
$
Now this means that the two point correlations function summed over all nearest 
neighbors can be derived from the partition function. This is also a quantity 
that can be accessed experimentally as is usually performed very frequently 
in the solid state experiments. Most importantly, this average has in general 
different values for separable and entangled states. It can, therefore, serve 
as an entanglement witness as will be seen shortly. 
The amazing fact that will emerge is that, in order to say if a state is 
entangled,  we really do 
not need to have the analytically calculated form of the eigenstates
in order to tell if the resulting mixture is entangled. One price to pay for this is that
we will only be able to derive a sufficient condition for entanglement that is typical
of entanglement witnesses. Namely, we will be able to tell if for some conditions the resulting
thermal state is entangled, but we will not be able to say  with certainty 
that the state is not entangled if these conditions are violated. 

Using $U\!=\!\langle H
\rangle$ and $M^z\!=\! \sum_{j=1}^{N} \langle S^z_j \rangle$ we
obtain
\begin{equation} \frac{U+hM^z}{NJ}=-\frac{1}{N} \sum_{i=1}^{N}
(\langle S_{i}^{x} S_{i+1}^{x} \rangle + \langle
S_{i}^{y} S_{i+1}^{y} \rangle + \langle S_{i}^{z}
S_{i+1}^{z}\rangle) 
\label{okako}
\end{equation}
The right-hand of Eq.(\ref{okako})
is an entanglement witness as shown in~\cite{Toth05a}: for any
separable state, that is, for any classical mixture of the
products states: $\rho=\sum_k w_k \rho^1_k \otimes \rho^2_k
\otimes ... \otimes \rho^N_k$, one has
\begin{equation}
\frac{1}{N} |\sum_{i=1}^{N} (\langle S_{i}^{x}
S_{i+1}^{x} \rangle + \langle S_{i}^{y} S_{i+1}^{y}
\rangle + \langle S_{i}^{z} S_{i+1}^{z}\rangle)| \leq \frac{1}{4}.
\label{vudubluz}
\end{equation}
This is also valid for any
convex sum of product states (separable states). The upper bound
was found by using the Cauchy-Schwarz inequality and knowing that
for any state $\langle S^{x} \rangle^2 + \langle S^{y}
\rangle^2 + \langle S^{z} \rangle^2\leq 1/4$. It is important
to note that the same proof can also be applied if one considers
$XX$ model. 
We now give our thermodynamical entanglement witness: if, in the
isotropic $XXX$ or $XX$ Heisenberg model, one has
\begin{equation}
  |U+hM^z| > N|J|/4, 
\label{niko}
\end{equation}
then the solid state system is in an entangled state. The
entanglement witness is physically equivalent to the exchange
interaction energy or, equivalently, to the difference between the
total (internal) energy $U$ and the magnetic energy $-hM$. From
the tracelessness of the Pauli operators one can easily see that
$\lim_{T\rightarrow \infty} U \rightarrow 0$. This means that the
value of the internal energy as given by Eq.(\ref{niko}) should be
defined relatively to the referent value of zero energy in the
limit of high temperatures.
In order to complete the proof we need to give an explicit example of a 
state that violates Eqs.~(\ref{vudubluz}) (or the corresponding inequality for 
the $XX$ model). This 
implies that that Eq.(\ref{niko}) is indeed an entanglement
witness and not just a bound that is trivially satisfied by any
quantum state.

As an example of such a state we take the ground
state of the antiferromagnetic isotropic XXX Heisenberg model with
zero magnetic field. The energy of this state was found to
be~\cite{hulthen}: $|E_0/JN|\!=\! 0.44325 \!>\! 1/4$.
Furthermore, due to the symmetry of the $XXX$ Heisenberg Hamiltonian
one has $E_0/(3NJ)\!=\!\langle S_{i}^{x} S_{i+1}^{x}
\rangle_0 \!=\! \langle S_{i}^{y} S_{i+1}^{y} \rangle_0
\!=\! \langle S_{i}^{z} S_{i+1}^{z}\rangle_0 \!=\!
-1.773/12$ for every $i$. This implies that $1/N|\sum_{i=1}^N
(\langle S_{i}^{x} S_{i+1}^{x} \rangle_0 + \langle
S_{i}^{y} S_{i+1}^{y} \rangle_0)|=$ $ 0.295 > $ $ 1/4$.
Therefore, Eq.~(\ref{niko}) is an entanglement witness for the
solid state systems described by $XXX$ or $XX$ Heisenberg interaction.

We will now discuss various concrete models of spin interaction of
which some are exactly solvable and for which dependence of
internal energy $U$ and magnetization $M$ on temperature $T$ and
magnetic field $h$ are known. This will help us to determine the
parameter regions of $T$ and $h$ within which one has entanglement
in the solids.

We first consider $XXX$ Heisenberg model with no magnetic
field, in this case the magnetization vanishes
and thermodynamical witness, Eq.(\ref{niko}) reduces to
$
|U| > N|J|/4
$.
It was shown that concurrence $C(1)$  is zero at any
temperature in the ferromagnetic case and that it is given by
$C=\frac{1}{2} \text{ max }\left[0,|U|/(NJ)-1/4\right]$ in the
antiferromagnetic case~\cite{wangzan}. Thus $C$ is nonzero if and only if
$|U|/(NJ)>1/4$. This shows that the thermodynamical entanglement
witness can detect entire bipartite entanglement as measured by
concurrence. Furthermore, the fact that the value of the
entanglement witness for the ground state is well above the limit
of $1/4$ suggests that entanglement could exist
and be detected by the thermodynamical witness at nonzero
temperatures as well.
In the presence of a finite magnetic field the low temperature 
partition function is given by $Z\!=\!e^{\beta(J+h_z)/4}(1+e^{-\beta
h_z/2}N/\sqrt{2\pi\beta J})$. Using this we obtain $|U+h_zM|/(NJ)\!=\!1/4$
and thus no entanglement can be detected in agreement with~\cite{Asoudeh04,pratt1}.

The $XX$ Heisenberg model with nonzero magnetic field is the most interesting as it is
exactly solvable, the partition function was found in~\cite{katsura}. 
Let us introduce the following dimensionless
quantities: $b=h_z/T$ and $K=J/T$ (note a difference of factor 2
in the definitions of $J$ and $K$ with respect to~\cite{katsura}) and the function
$
f(K,b,\omega)= \sqrt{2K^2+2K^2\cos{2\omega}-4bK\cos{\omega}+C^2}
$
for convenience. Then the internal energy is given
by~\cite{katsura}
\begin{equation}
\frac{U}{N} = -\frac{T}{4\pi} \int_{0}^{\pi} f(K,b,\omega)
\tanh{f(K,b,\omega)} d\omega, \label{energy}
\end{equation}
and the magnetization by~\cite{katsura}
\begin{equation}
\frac{M}{N} = -\frac{1}{2\pi} \int_{0}^{\pi}
\frac{4K^2\cos^2{\omega}}{f(K,b,\omega)} \tanh{f(K,b,\omega)}
d\omega 
\label{magnetization}
\end{equation}
both in ferromagnetic and antiferromagnetic case.

We use Eqs.~(\ref{energy}) and (\ref{magnetization}) to determine
the parameter regions of temperature $T$ and magnetic field  $h_z$
for which entanglement exists in the solid state system.
The critical values of $T$ and $h$ below
which entanglement can be detected is of the order of $J$, which
can be as high as $10$ Kelvin~\cite{experiment}.

It should be stressed that the analysis based on the entanglement witness 
could be applied to any
model for which we can successfully obtain the partition function. 
This feature is the main advantage of using thermodynamic 
witnesses approach to detecting entanglement.  
This method for determining entanglement in solids
within the models of Heisenberg interaction is useful in the cases
where other methods fail due to incomplete knowledge of the
system. This is the case when only the eigenvalues but not
eigenstates of the Hamiltonian are known (which is the most usual
case in solid state physics) and thus no measure of entanglement
can be computed. Furthermore, in the cases where we lack the complete
description of the systems one can approach the problem
experimentally and determine the value of the thermodynamical
entanglement witness by performing appropriate measurements. It is important to 
emphasize that any other thermodynamical function of state could be a 
suitable witness, such as the magnetic susceptibility or heat capacity~\cite{marcin}
(see next Section). 

The temperature as well as other thermodynamical state variables have been shown to 
behave  also  as entanglement witnesses of {\em spatial entanglement}~\cite{anders}. 
This general feature was explicitly worked out in the case of a non-interacting Bosonic gas.
It was found that entanglement can exist at arbitrarily high temperatures, provided that 
one can probe smaller and smaller regions of space. 

The methods outlined here are not only applicable to the models we considered.
There are several interesting questions and
possibilities for generalizations such as consideration of
Hamiltonians with higher spins, two- and three-dimensional systems,
non-nearest interactions, anisotropies, other thermodynamical
properties (e.g. heat capacity, magnetic susceptibility) and so
on. Similar analysis can be done for continuous thermal entanglement 
in a field. It has been shown that for non-interacting bosons, 
entanglement exists when their de Broglie thermal wavelength is 
smaller than their average separation, $a$~\cite{anders}. 
The precise condition is $kT < \hbar^2/2ma^2$, where $m$ is the mass 
of bosonic particles. We can now introduce the following correspondence 
between spin coupling $J$ and the continuous variables bosonic kinetic 
energy, $J=\hbar^2/2ma^2$.  This further implies that we can think of 
the thermal de Broglie wavelength for spins as $\lambda_{dB} = a \sqrt{J/T}$, 
where $a$ is the spin separation. The condition for entanglement that the 
wavelength is larger than the lattice spacing $a$ now leads us to the 
condition that $T<J$ which is exactly the result obtained from a more 
detailed analysis above.

\subsection{Experimental results}
\label{thermal-experiments}

The question of having macroscopic entanglement is not only
fascinating in its own right but it also has a fundamental
significance as it pushes the realm of quantum physics well into
the macroscopic world, thus opening the possibility to test
quantum theory versus alternative theories well beyond the scales
on which theirs predictions coincide. It also has important
practical implications for implementation of quantum information
processing. If the future quantum computer is supposed to reach
the stage of wide commercial application, it is likely that it
should share the same feature as the current (classical)
information technology and be based on solid state systems. It
will thus be important to derive the critical values of physical
parameters (e.g. the high-temperature limit) above which one
cannot harness quantum entanglement in solids as a resource for
quantum information processing.

Recently it was demonstrated experimentally that entanglement can 
affect macroscopic properties of solids, albeit at very low (critical) 
temperature (below 1 Kelvin)~\cite{ghosh}. This extraordinary result 
opens up an exciting possibility that purely quantum correlations 
between microscopic constituents of the solid may be detected by only 
a small number of macroscopic thermodynamical properties. 

\begin{figure}
\centering
\includegraphics[width=7.5cm]{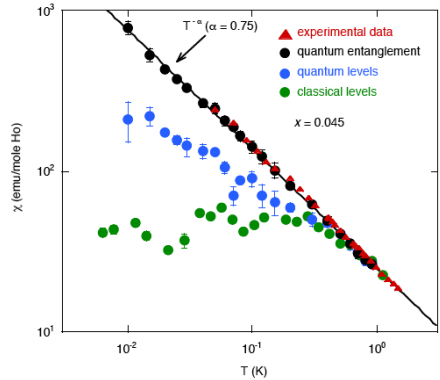}
\caption{
Magnetic susceptibility vs. temperature $T$ from simulations of the diluted,
dipolar-coupled Ising magnet,compared to experimental data (red
triangles).  The black circles
use quantum decimation as well as the correct quantum mechanical form of
susceptibility, utilizing the entanglement of the low-lying energy doublet
with the excited states. [From~\protect\onlinecite{ghosh}]} 
\label{gh}
\centering
\end{figure}

Ghosh {\em et al} made measurement on an insulating salt, the $\mbox{LiHoF}_4$.
At low temperatures the susceptibility deviates from a simple Curie-like law.
They find that the the temperature dependence is well fitted by a power law 
$\chi \sim T^{-\alpha}$ with $\alpha =0.75$.
The key observation made in~\cite{ghosh} is that the experimental data at low 
temperatures cannot be explained by simply resorting to a classical 
approximation. By itself this might not be enough. It is remarkable 
however that the authors are able to put in close connection the 
power law divergence of the susceptibility with the entanglement present in 
the low-lying excited states.

V\'ertesi and Bene studied the magnetic susceptibility of $\mbox{NaV}_3\mbox{O}_7$ and used 
macroscopic entanglement witnesses as discussed in the previous Section to estimate the critical 
temperature  below which thermal entanglement is present. The experimental value of 
this temperature is $365$K approximately three times higher than the critical 
temperature corresponding to the vanishing of bipartite entanglement~\cite{vertesi}.

We also mention the very recent experiment reporting on macroscopic magnetic measurements 
of the pyroborate $\mbox{MgMnB}_22\mbox{O}_5$ and warwickite $\mbox{MgTiOBO}_3$~\cite{rappoport}
(see also~\onlinecite{continentino})

Earlier experimental data witnessing entanglement in bulk properties of solids 
have been reanalyzed in~\cite{cav}. They discusses the experimental results of neutron
scattering measurement of CN obtained in 2000~\cite{broholm} and
show that they provided a direct experimental
demonstration of macroscopic entanglement in solids. The
experimental characterization of the dynamic spin correlations for
next neighboring sites enabled them to determine the
concurrence and show the existence of entanglement at moderately high
temperatures (as high as 5 Kelvin). 
In the same work they also showed that magnetic susceptibility 
at zero magnetic field is a macroscopic thermodynamical entanglement witness for the class
of solid states systems that can be modeled by strongly alternating
spin-1/2 antiferromagnet chain~\cite{cav}.The measured
values for magnetic susceptibility of CN in 1963~\cite{berger} imply presence of 
entanglement in the same temperature range (below 5 Kelvin).

\begin{figure}
\begin{center}
\includegraphics[clip=true, angle=0,width=8cm]{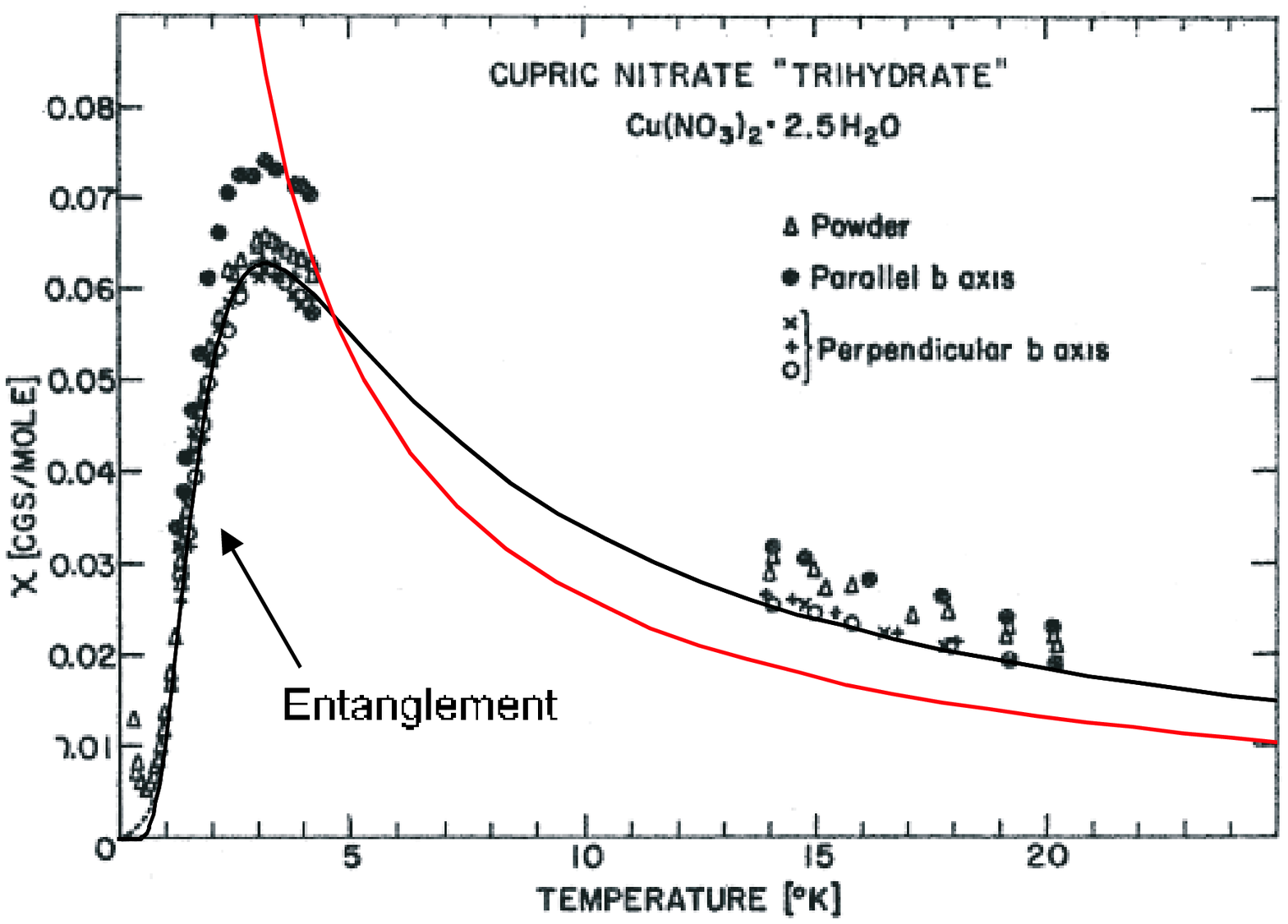}
\caption{The temperature dependence of magnetic susceptibility of
powder CN (triangles) and a single-crystal CN, measured at low field
parallel (open squares) and perpendicular (open circles, crosses,
filled circles) to the monoclinic $b$ axis. The data and the figure
are from~\cite{berger}.  The intersection point of this
curve and the experimental one defines the temperature range (left
from the intersection point) with entanglement in CN around 
$ \approx 5K$. [From \protect\onlinecite{cav}]} \label{figcav}
\end{center}
\end{figure}

An analysis of the experimental results of a magnetic susceptibility measurement of
CN~\cite{berger} showing that the values measured at low
temperatures cannot be explained without entanglement being present
was performed in~\cite{cav}. This was  based on the general proof 
that magnetic susceptibility of any strongly alternating antiferromagnetic
spin-1/2 chain is a macroscopic thermodynamical entanglement
witness. As discussed in~\cite{cav} the magnetic susceptibility for
separable states is bounded by the value
\begin{equation}
\chi_{\mbox{sep}} \geq \frac{g^2 \mu^2_B N}{T} \frac{1}{6}.
\label{macrowitness}
\end{equation}
The results of their analysis are reported in 
Fig.\ref{figcav}

\section{Multipartite entanglement }
\label{multip}

Most of the results reviewed in the previous section are for pairwise entanglement.  
Although much has been learned from the study of those 
quantities, the structure of entanglement in many-body systems is much richer and 
it is natural to expect that multipartite entanglement is present 
both in the ground state and at finite temperatures. 
Although multipartite entanglement in many-body systems is much less studied,
some important results have been already obtained. 

A number of groups showed that in certain limits the state of a spin chain 
can resemble that of known multipartite states. 
For small chains Wang first noted that the ground state tends to have 
multipartite entanglement~\cite{wang02}. This analysis was further pursued in 
Ising and Heisenberg rings where the ground state has GHZ-~\cite{stelmachovic04} and
W-like~\cite{bruss05} correlations, respectively.
Multipartite mixed states can be also realized in the case of ferromagnetic rings 
where the ground state is fully polarized 
along the direction of the field and the first excited state is a W-state. 
At finite but low temperatures the density matrix is approximately as
$
      \rho = p \mid 0 \cdots 0 \rangle \langle 0 \cdots 0   \mid + 
             (1-p) \mid W\rangle \langle W \mid
$
where $p$ is the Boltzman factor and $ \mid 0 \cdots 0 \rangle$ is the ferromagnetic 
ground state. For three qubits $\rho$ has been shown to contain 
tripartite entanglement in~\cite{bruss03}. 

These examples show that several models naturally have multipartite entangled 
ground states. 
At the same time it also shows that it is important to quantify multipartite 
entanglement in many-body systems. Analysis in this direction is reviewed below.

\subsection{Multipartite entanglement in spin systems}

\begin{figure}
\includegraphics[scale=0.25,angle=0]{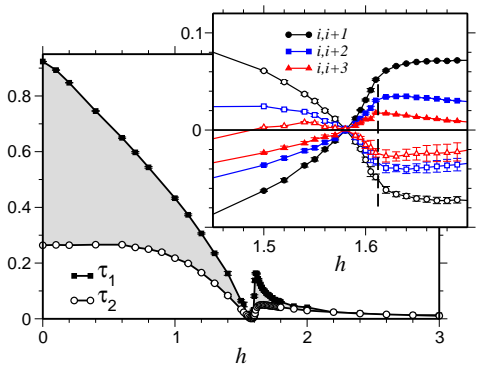}
\caption{One-tangle $\tau_1$ and the sum of squared concurrences 
$\tau_2$ as a function of the applied magnetic field in $z$-direction
for the $XYZ$ model with exchange along $y$: $J_y= 0.25$ (in unit of exhange along $z$). 
Inset: contributions to the concurrence between j-th neighbors; full
symbols stand for $C^{I}(j)$, open symbols for $C^{II}(j)$. The dashed
line marks the critical field $h_c$.  [From \protect\onlinecite{Firenze04}] } 
\label{tau2tau1}
\end{figure}

A first way to estimate multipartite entanglement in spin system is provided 
by the entanglement ratio $\tau_2/\tau_1$ as the amount of two spin  
relative to global entanglement. For $1d-XYZ$ models it was shown that 
a small value of such a ratio is generic in these systems 
with a cusp minimum at the quantum critical point~\cite{Firenze04}. 
This is was shown numerically for the XYZ  chain in a field 
(Fig.\ref{tau2tau1}).

In order to {\it quantify} the multipartite entanglement, 
Wei {\em et al.} calculated the geometric measure of entanglement~\cite{wei03}, 
see \ref{measure-multipartite}, for the transverse XY chain~\cite{wei05}. 
\begin{figure}
\includegraphics[scale=0.3,angle=0]{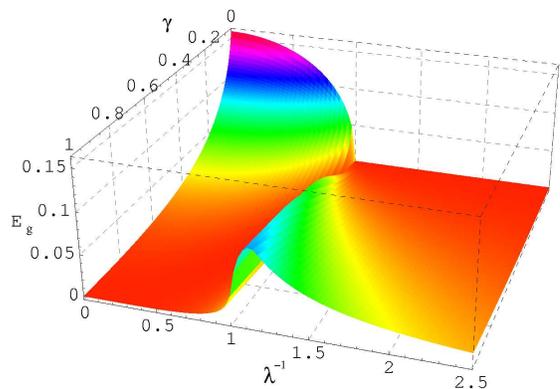}
\caption{
The geometric entanglement measure Eq.(\ref{weimeas}) per site
is plotted versus $\gamma$ and $h_z$ 
for the $XY$ model. There are three different phases: the ordered oscillatory phase 
for $\gamma^2+h_z^2<1$ and $\gamma \ne 0$; the ordered ferromagnetic phase between 
$\gamma^2+h)z^2>1$ and $h<1$; and the paramagnetic phase for $h_z>1$. 
There is a sharp rise in the entanglement
across the quantum phase transition line $\lambda^{-1}=h_z=1$. 
At $\gamma=0$ lies the XX model, which belongs to a different universality
class than the anisotropic XY model. [From \protect\onlinecite{wei05}] } 
\label{wei}
\end{figure}
The calculation can be done by a clever use of symmetries: 
translational invariance and periodic boundary conditions. 
In this case, the set of all possible separable 
states can be described by a global rotation around the $y$-axis of the 
fully polarized state.
The maximization is thus reduced to only one variable. 

As well as in the case of bipartite entanglement, also the multipartite measure of 
Wei {\em et al.} is very sensitive to the existence of QPTs. 
As a paradigmatic example the authors analyzed the phase diagram in the anisotropy-magnetic 
field plane. Their results are shown in Fig.\ref{wei}.  As expected there 
is no divergence in the measure itself but in its derivative. The new aspect 
here is that differently from the concurrence the multipartite entanglement measure in 
Eq.(\ref{weimeas}) can be expanded as a function of n-point correlators. 

The geometric entanglement cannot discriminate between different n-particle 
entanglement classes. 
A comprehensive classification in spin systems has been recently given 
by~\cite{guehne05} via the concept of  
k-producibility (see \ref{measure-multipartite}). 
The systems analyzed in detail are the one-dimensional $XY$ and Heisenberg models.
Different types of n-particle quantum correlated states lead to
distinct bounds for the internal energy (or the ground state energy at $T=0$).
A violation of these bounds then implies the presence of certain
k-party producable entanglement in the system.
As pointed out in~\cite{guehne05}, aiming at the thermodynamic
limit of an infinite number of spins, the notion of k-producibility is easier to 
handle than the n-separability (see \ref{measure-multipartite})
as its definition is independent of the number of sites in the system.

A systematic approach for deriving energy bounds for states 
without certain forms of multipartite entanglement has been developed 
in~\cite{guehne06}. 
The method  allowed to investigate also higher 
dimensional and frustrated systems.
As an example we report on the results for the Heisenberg model. 
In $d$ dimension, the energy bounds per bond for one-, two-, 
three-, and four-producible states are given by

\begin{center}
\begin{tabular}{c|r|r|r|r}
$-4\langle H \rangle/J$ &1p & 2p & 3p & 4p\\
\hline
$1d$    &  1         &   3/2  & 1.505  &  1.616\\
$2d$     & 1   &   13/12   & 1.108 & 1.168\\
$3d$    &  1   &   31/30   &   1.044   &  1.067
\end{tabular}
\end{center}
%
It is striking how relatively close the 2- and 3-producible bounds are
in all cases.
All the bounds given above are found to be violated in the ground state. 
In the previous expression the superscripts refer to the dimensionality of the model and 
the subscripts to the k-party entanglement for which the bound is obtained.
There is a factor 1/4 of difference with respect to the original paper because 
of the different notation used in this review.

\begin{figure}
\includegraphics[scale=0.6,angle=0]{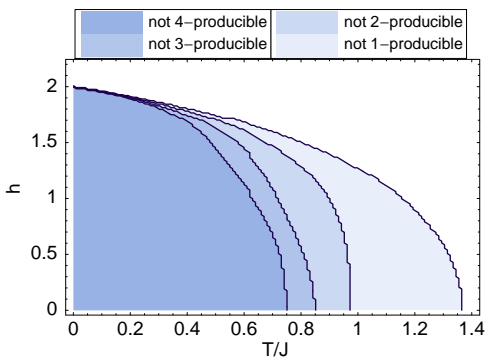}
\caption{
Entanglement in thermal states of the  $XX$-model in 
a transverse magnetic field. The different regions correspond to 
different types of multipartite entanglement contained in the 
equilibrium thermal state. [From \protect\onlinecite{guehne06}]}
\label{Tentangl}
\end{figure}

Corresponding to the energy scales fixed by the bounds 
there are different temperature scales at which the various n-party entanglement 
types disappear. As one would expect, these crossover temperatures 
are monotonic in k; i.e. $T_k \ge T_{k+1}$. 
The example given in Fig.\ref{Tentangl} clearly shows how 
higher order multipartite entanglement 
progressively disappears as the temperature increases.

\subsection{Global  entanglement}
\label{globalent}

Multipartite entanglement  close to quantum  phase transitions was quantified 
by the global-entanglement $E_{gl}$ measure of 
Meyer and Wallach (see \ref{measure-multipartite}) in ~\cite{oliveira,oliveira06b,Somma04}. 
Together with the Meyer-Wallach measure, de Oliveira {\em et al} also introduced 
slight extension of it as 
\begin{equation}
    E^{(2)}_{gl} = \frac{4}{3}\frac{1}{N-1}\sum_{l=1}^{N-1} \left[ 
    1 - \frac{1}{N-1} \sum_{j=1}^N \mbox{Tr} \rho_{j,j+l}^2 \right]
\label{E2}
\end{equation}
where $\rho_{j,j+l}$ is the reduced density matrix associated to the sites $j$ and 
$j+l$. Similarly one can consider also three-body reduced density matrices and 
construct the corresponding global entanglement measure. According to de Oliveira 
{\em et al.} the above hierarchy might provide a comprehensive description of 
entanglement in many-body systems. Explicit calculations for the  
one-dimensional Ising model~\cite{oliveira} showed that both 
$E_{gl}$ and $E^{(2)}_{gl}$ are 
maximal at the critical point (with $E_{gl} < E^{(2)}_{gl}$ for any value of the 
coupling constant) suggesting that the quantum critical point is characterized 
by the presence of multipartite entanglement. As in the case of concurrence the 
non-analyticity associated to the critical point is manifest in the derivative of 
the global entanglement measure. 
By extending an earlier approach developed in~\cite{WuLidar04}, 
de Oliveira {\em et al.} also showed how the non-analytic behavior of $E^{(n)}_{gl}$ 
is related to that of the ground state energy. Note that from Eq.(\ref{E2}) it is 
possible to define an entanglement length proportional to the correlation length $\xi$. 
This differs considerably from that one defined by the localizable entanglement 
(see Eq.(\ref{elenght})); the latter is always bounded from below by the 
correlation length and can even be divergent where $\xi$ is finite.

As discussed in~\cite{facchi06,facchi06a} the analysis of the average purity 
might not be sufficient and the analysis of the distribution of the purity for 
different partitions could give additional information. 
Rather than measuring multipartite entanglement in terms of a single number, one 
characterizes it by using a whole function. One studies the distribution function 
of the purity (or other measures of entanglement) over all bipartitions of the
system. If the distribution is sufficiently regular, its average and
variance will constitute characteristic features of the global entanglement: the
average will determine the ``amount'' of global entanglement in the
system, while the variance will measure how such entanglement
is distributed. A smaller variance will correspond to a larger
insensitivity to the choice of the bipartition and, therefore, will
be characteristic for different types of multipartite entanglement. 
The application of this 
technique  to the one-dimensional quantum Ising model in a transverse field shows that
the distribution function is well behaved and its average and second
moment are good indicators of the quantum phase transition~\cite{facchi06b}. This is
in agreement with previously obtained results. At the onset of the
QPT both the average and the standard deviation exhibit a peak that
becomes more pronounced as the number of qubits is increased.

\subsection{Generalized entanglement}

A different route for studying the multipartite entanglement is paved by 
the general observable based entanglement (see Section~\ref{measure-multipartite}).
It was first pursued by Somma {\em et al} for the LMG and the quantum $XY$-model. 
In the realm of solid state systems an experimental protocol 
to measure many-fermion entanglement based on this concept
has been proposed in~\cite{kindermann}.
A nice connection which emerges from the work of Somma {\em et al}  is the one 
between the  generalized entanglement and the quantum fluctuations of the 
magnetization which are responsible for the quantum phase transition~\cite{Somma04}.  
Later, Montangero and Viola considered the dynamical 
behavior of generalized entanglement in disordered systems~\cite{montangero06}.
As remarked in~\cite{Somma04}, it is important to choose the appropriate subset 
of observables in order to see
the critical behavior in the entanglement. 

\begin{figure}
\begin{center}
\includegraphics[width=7cm]{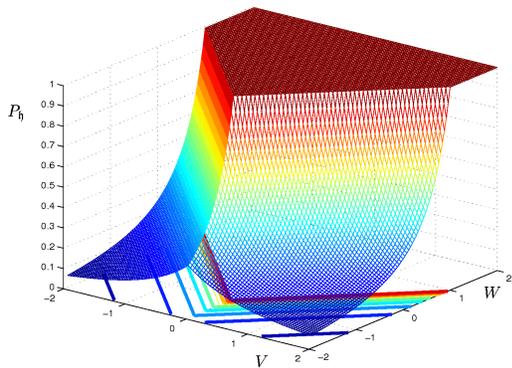}
\end{center}
\caption{Purity relative to the observable $S^z$ in the ground state 
of the LMG model. [From \protect\onlinecite{Somma04}]}
\label{puritylmg}
\end{figure}

In the case of the  LMG model a natural choice for the local observables was 
to consider the purity relative to the spin-$N/2$ representation of the angular momentum
$\displaystyle{
  P_{S} = \frac{4}{N^2}\left[ \langle S^{x}\rangle ^2 + \langle S^{y}\rangle ^2 +
          \langle S^{z}\rangle ^2 \right]}
$. 
Somma {\em et al.} also consider the purity relative to the single observable $S_z$: 
$\displaystyle{
P_{S^z} = \frac{4}{N^2}\left[\langle S^{z}\rangle ^2 \right] \; .
}$ 
With this last choice the relation between the multipartite entanglement and the 
order parameter becomes evident. The result is shown in Fig.\ref{puritylmg}.

\begin{figure}
\begin{center}
\includegraphics[width=7cm]{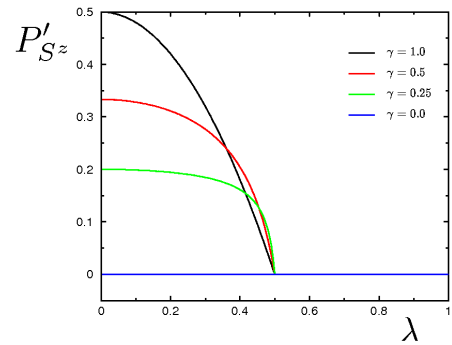}
\end{center}
\caption{The purity $P'_{u(N)}= P_{u(N)} - (1+\gamma)^{-1}$ of an $XY$ model 
in a transverse field as a
function of $\lambda$  for different anisotropies $\gamma$.The constant part has been 
subtracted in order to make the connection to the {\em disorder parameter}.
[From \protect\onlinecite{Somma04}]}
\label{purityxy}
\end{figure}

Further interesting results emerge from the $XY$ model. 
By choosing the following subset of operators expressed in terms of the 
spinless fermions of the
Jordan-Wigner transform 
$
u(N)={\rm span}\{c^{\dagger}_i c_{i'} + c^{\dagger}_{i'} c_{i},
i(c^{\dagger}_i c_{i'} - c^{\dagger}_{i'} c_{i}), \sqrt{2}(c^{\dagger}_i c_i -1/2)\}
$
it is possible to show that the associated purity may be considered as a 
{\em disorder parameter}, i.e. it is non-zero in the symmetric phase only. 
A transparent way 
to express the purity in this case is to relate it to the variance of the 
number fermion operator $N_f = \sum_i c^{\dagger}_i c_{i}$. The result is plotted in 
Fig.\ref{purityxy} for different values of the anisotropy parameter $\gamma$.
Two considerations are in order at this stage. First of all it is clear from the 
previous example that important properties of entanglement appear when one moves 
away from the conventional picture of partitioning in real space. Secondly the 
purity shows, differently from other measures as the concurrence, direct 
indications of the critical behavior (and not only in the derivatives).

\subsection{Renormalization group for quantum states}

We continue our discussion on multipartite entanglement with a
recent work ~\cite{verstraete06re} where the
method of renormalization group is applied to quantum states and not,
as traditionally done, to the Hamiltonian. The Renormalization Group is based 
on a recursive transformation which leads to an effective description of 
the low-energy (long distance) physics of a given system where all the 
effects of high energy modes are included in a renormalization of certain 
coupling constants of the model Hamiltonian. A study of the flow of these 
recursive equations, their fixed point(s) and the behavior of the flow close
to the fixed points allows to determine the critical behavior of the system 
under consideration.
Contrary to the conventional renormalization group approach, 
the idea of Verstraete {\it et al.} is to analyze how 
the quantum states change under this coarse graining and to classify their 
fixed points.
Given a system characterized by $N$ sites and a $D$-dimensional local Hilbert 
space. The steps of this {\em real space} renormalization procedure are the 
following.\\
\noindent 
(i) Coarse graining, in which clusters of neighboring sites are merged into
      one site of a new lattice.\\
\noindent
(ii) Rescaling of the local variables associated to the new sites.\\
\noindent
(iii) Identification of the states which are invariant under local unitary 
      transformation (long distance behavior should not depend on the choice 
      of the local basis).\\
\noindent 
(iv) Rescale the distances in order to have again a unit lattice constant.\\
In this way a representative of a given class of quantum states, invariant under 
local transformations, is transformed at each step. The irreversibility of the 
transformation is crucially related to the step (iii) of the transformation as one 
loses track of the unitary transformation performed (which need to be local over 
the cluster).

Verstraete {\em et al} carry out explicitly the renormalization group transformation for 
Matrix Product States (MPS), see Eq.(\ref{mps}). The decimation step in which
two neighboring sites, say $i$ and $i+1$, are merged together 
is performed here by means of merging the corresponding matrices
$A^{p_i}$ and $A^{p_{i+1}}$ into 
$\tilde{A}^{p_i, p_{i+1}}_{\alpha \beta}= \sum^{\mbox{min}(D_{MPS}^2,D^2)}_{\gamma=1}
A_{\alpha \gamma}^{p_i}A_{\gamma \beta}^{p_{i+1}}$. By means of a singular value 
decomposition of $\tilde{A}$ it is  possible to find the representative of the 
new state (see the step (iii)) and therefore to iterate the renormalization group map. 
In the case of $D_{MPS}=2$, Verstraete {\em et al.} provide a complete classification of 
the fixed points of the transformation which contains product, GHZ and domain wall 
states.  
A similar analysis in the case $D_{MPS} > 2$ and/or in higher dimensional systems 
may turn out to be very useful also for the classification of multipartite entanglement 
in many-body systems.

\subsection{Entanglement distribution for Gaussian states}

It has been observed in Section \ref{entharmon} that
the symmetry groups of admissible local transformations 
of Gaussian states and qubits are isomorphic.
This suggests to look for further analogies with the qubit case
or to search for deviations from it.
A striking feature of qubit entanglement is
the monogamy as far as entanglement distribution along
chains of qubits is concerned  (see \ref{measure-pairwise-mixed}).
For continuous variable entanglement such an inequality was originally
analyzed analytically for fully symmetric 
Gaussian states and numerically for randomly chosen
Gaussian states in~\cite{Adesso06}. 
The general proof of monogamy has been obtained very 
recently in~\cite{hiroshima07}. In this Section we use  
the particular case of a symmetric state as a guideline to 
discuss the monogamy for Gaussian states. A more detailed discussion 
and the general proof can be found in the review by~\onlinecite{illuminatirev}.  

A Gaussian state is called {\em fully symmetric} if and only if
its covariance matrix is invariant with respect to permutations
of the modes. Its covariance matrix can then be written as~\cite{Adesso04}
\beq\label{SymmGaussV}
V_{symm}=\Matrix{cccc}{
\hat{\beta}&\hat{\eps}&\dots&\hat{\eps}\\
\hat{\eps}&\hat{\beta}&\dots&\hat{\eps}\\
\vdots&\vdots&\ddots&\vdots\\
\hat{\eps}&\hat{\eps}&\dots&\hat{\beta}}
\eeq
where $\hat{\beta}$ and $\hat{\eps}$ are $2\times 2$ matrices.
Both can be diagonalized by means of local symplectic transformations 
in phase space, such that $\hat{\beta}=\diag\{b,b\}$ and
$\hat{\eps}=\diag\{\eps_1,\eps_2\}$.
This leads to a highly degenerate symplectic spectrum where
$n-1$ doubly degenerate symplectic eigenvalues are equal to
$\nu=(b-\eps_1)(b-\eps_2)$ and the remaining eigenvalue is 
$\nu_n=(b+(n-1)\eps_1)(b+(n-1)\eps_2)$.
After a partial transposition, $n-2$ symplectic eigenvalues $\nu$
remain unaffected. The negativity is then determined solely by $\tilde{\nu}_-$,
which is the smaller one of the remaining two affected eigenvalues 
$\nu_{\pm}$~\cite{Adesso04}.
This is due to the fact that the uncertainty leads to 
a lower bound $\hbar^2/4$ for the product $\nu_+ \nu_-$

For a single Gaussian mode with covariance matrix $V_1$, the tangle
is given by 
$
\tau_{1}=2(1-\det V_1)/\sqrt{\det V_1}
$
The quantity which is analog to the pairwise tangle has been identified
as the square of the logarithmic negativity
\beq\label{G2-tangle}
\tau^{2}(\tilde{\nu}_-)=\left[\ln\tilde{\nu}_-\right]^2
\eeq
and has been termed {\em contangle} in~\cite{Adesso06}. The identification 
of the square negativity as the continuous variable tangle is 
crucial for the demonstration of the monogamy inequality.
Extensions of these measures to mixed states are understood to be given
by the corresponding convex roofs~\cite{Uhlmann00}, where the average
of the pure state measure of entanglement has to be minimized over
all pure state decompositions of the density matrix in consideration.
A restriction to decompositions purely out of Gaussian states
gave rise to the notion of the 
{\em Gaussian entanglement measures}~\cite{Wolf04}
and the Gaussian entanglement of formation. It clearly establishes 
an upper bound for the entanglement of formation. 

Negativity and Gaussian entanglement measures 
have been analyzed with considerable detail in (\onlinecite{Adesso05})
for two-mode Gaussian states with particular focus onto symmetric 
Gaussian states.
One important result is that negativity and Gaussian measures
lead to different ordering of Gaussian states with respect to their
entanglement. For symmetric Gaussian states instead, both
give the same ordering.
This result must be handled with care, since it is not obvious
what precisely the restriction to Gaussian decompositions entails
for the ordering of states. Believing into the striking similarity 
to qubit systems, one might be tempted to conjecture that
for symmetric Gaussian states a purely Gaussian
optimal decomposition always existed.\\
There are two extremal classes of two-mode Gaussian states
considered: for fixed local and global purity,
those states that maximize the negativity are termed 
{\em Gaussian Maximally Entangled Mixed States} (GMEMS) introduced 
by~\onlinecite{AdessoPRA04},
whereas those states minimizing the negativity have been
termed GLEMS, which are states whose covariance matrix has one
symplectic eigenvalue equal to $1$ (mixed states with partial
minimal uncertainty).
The symmetric two-mode Gaussian states have been singled out as
those states containing minimal Gaussian entanglement at fixed
negativity~\cite{Adesso05}.

The entanglement sharing inequality for  Gaussian states
assumes the same form as for qubits
$\tau_{1,i}\ge\sum_{j\neq i} \tau_{2,ij}$
where the indices are the numbers of the various distinguishable modes
in a multi-mode Gaussian state.
This inequality has been proved for multi-mode Gaussian states
by~\onlinecite{hiroshima07}. \\
In the particular case of symmetric 
states all $\tau_{1,i}\equiv\tau_{1,1}$ and $\tau_{2,ij}=\tau_{2,12}$
for all indices $i,j$.

As for qubit systems one can define a residual contangle by the difference
$\tau_{n}:=\tau_{1,i}-\sum_{j\neq i}^n \tau_{2,ij}$
as a quantifier of the multipartite 
entanglement present in a Gaussian state. In particular is 
the residual contangle an entanglement monotone under Gaussian LOCC.
An important difference, however, arises as compared to the
residual tangle for qubits: not even for three modes is
the residual tangle invariant under permutations of the modes
and its minimum respect to the common mode $i$ has to be taken. 
Even for symmetric Gaussian states, where this anomaly is absent,
a {\em promiscuous} nature of continuous variable entanglement
is encountered, in sharp contrast to the {\em monogamy} inherent 
to qubit entanglement~\cite{Coffman00,chineseCKW,Osborne06}.
The term {\em promiscuous} is an interpretation of the fact that
the maximal residual contangle $\tau_{3}$ in a symmetric Gaussian 
three-mode state without pairwise contangle $\tau_{2}$ is smaller 
than the maximum possible
residual contangle without this restriction.
Having in mind the entanglement sharing inequality, 
this implies that the local contangle $\tau_{1}$ is larger, when
$\tau_{2}$ and $\tau_{3}$ coexist.
It is worth noticing at this point that the peculiarity of Gaussian
states is to be completely described by two-point correlation functions.
Consequently, all types of multipartite entanglement are inextricably related
in that the same type of correlations are responsible for either type
of quantum correlation.
This is not the case for non-Gaussian states and in particular not for
qubit systems or general higher-dimensional local Hilbert spaces.

\section{Dynamics of entanglement}
\label{dyn}
The interest in studying the properties of entanglement in many-body
system has been recently directed also to the understanding of its 
dynamical behaviour. Entanglement dynamics has been  studied from 
different perspectives. In a spirit similar to the study of propagation 
of excitations in condensed matter systems, several works analyzed the 
propagation of entanglement starting from a given initial  state where 
the entanglement has been  created in a given portion of  the many-body 
system. One can imagine for example to initialize a spin chain such that 
all the spins are pointing upwards except two neighboring spins which are 
entangled. Due to the exchange interaction the entanglement, initially 
localized on two neighboring sites of the chain will spread. This propagation 
is ballistic in clean systems. A ``sound velocity'' for the excitations 
results in a finite speed for the propagation of excitations.
If some weak disorder is present one might expect diffusing propagation. 
Entanglement localization and chaotic behaviour can eventually 
also be observed. An alternative approach is to start with the ground state 
of an Hamiltonian $H_0$ and the let the Hamiltonian change in time. 
Most of the attention up to now has been devoted to the case of sudden quench, 
i.e. after the preparation the Hamiltonian suddenly changes to $H_1$. 
Moreover since we are mostly dealing with interacting systems, entanglement can be 
also generated  or it can change its characteristics during the dynamical 
evolution. 
 
Another important aspect of entanglement dynamics 
is the possibility to generate entangled states with given 
properties by taking advantage of interaction present in a many-body 
systems. This is the natural generalization of the case where a 
Bell state can be  obtained by letting two qubit interact for fixed time by 
means of an exchange coupling of $XX$ symmetry. In the same spirit one can 
think to generate three-bit entangled GHZ or W states or multipartite 
entangled states by tailoring the appropriate exchange couplings 
in spin networks.

\subsection{Propagation of entanglement}		
\label{prop}

\subsubsection{Pairwise entanglement}

The simplest situation, which we consider first, is the propagation of 
entanglement in the one-dimensional $XX$-model, i.e. $\gamma=0$ and $\Delta=0$
in Eq.(\ref{general-spin})~\cite{amico04,subrahmayam04}. 
Suppose that the initial state of the chain is 
\begin{equation}
	| \Psi_{\pm} (t=0) \rangle \equiv
	\frac{1}{\sqrt{2}}(\sigma^{x}_i \pm \sigma^{x}_j)|0, \dots 0 \rangle \, ,
\label{wavefunction}
\end{equation}
namely all the spin are in a fully 
polarized state except two, at positions $i$ and $j$, which are in one of the 
two Bell states $|\psi_{\pm}\rangle = 2^{-1/2} (|01\rangle \pm |10 \rangle)$
In this case the problem  is amenable of a simple analytical solution. 
The reason is that, since the total magnetization is conserved, the evolution 
is confined to the sector where only one spin is up. In this sector the dynamics
is completely described by the states 
$| {\boldsymbol l}\rangle \equiv | 0, \dots 0, 1, \dots 0 \rangle$ ($l=1,\ldots,N$)
which represents a state of the chain where the $l$th spin is prepared 
in $|1\rangle$ while all the others $N-1$ ones are in $|0\rangle$.
At later times the state of the chain is
to be
\begin{equation}
	|\Psi_{\pm} (t) \rangle = \sum_l w^{(i,j)}_{\pm,l}(t) | {\boldsymbol l} \rangle \;.
\label{waveevolv}
\end{equation}
In the thermodynamic limit, $N\rightarrow \infty$, the coefficients can be 
expressed in terms of Bessel functions $J_n(x)$ as
$
	w^{(i,j)}_{\pm,l}(t) = \frac{1}{\sqrt{2}} \left \{
	J_{i-l}(4J t) \pm \,
 	(-i)^{j-i} J_{j-l}(4Jt) \right \} \; .
$
Eq.(\ref{waveevolv}) together with the coefficients defined previously 
allows to study various kinds of entanglement measures for this particular case.

As a first indication of the entanglement propagation we analise the dynamical 
evolution of  the concurrence between two sites, located at positions $n$ and $m$,  
(initially the entangled state is between the sites $i$ and $j$). The concurrence
reads  
\begin{equation}
	C^{i,j}_{n,m}(\pm,t) = 
	2 \Bigl | w^{(i,j)}_{\pm,n}(t) w^{(i,j)\star}_{\pm,m}(t) \Bigr | \; .
\label{concnm}
\end{equation}
\begin{figure}
	\includegraphics[width=.75\linewidth]{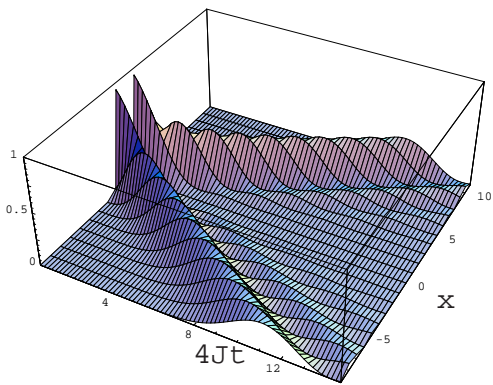}
	\caption{Concurrence between sites $n=-x, m=x$, symmetrically
	placed with respect of the state where the singlet was initially created. 
        from their initial position $i=-1$ and $j=1$.
	[From \protect\cite{amico04}].} 
\label{treconc}
\end{figure}
The function $C^{i,j}_{n,m}(t)$, plotted in Fig.\ref{treconc}, shows for sites 
which are symmetrical respect to the initial position of the Bell state 
$|\psi_{\pm}\rangle$. The time evolution dictated by the $XX$ Hamiltonian, amounts to a 
propagation of the single flipped spin through the chain. The speed of the 
propagation is set up by the interaction strengths. The information exchange or 
entanglement propagation over a distance of $d$ lattice spacings, approximately 
takes a time $t \sim \hbar d/J$. The external field $h$ does not enter in 
Eq.(\ref{concnm}), since all of the components of the state are in the same sector 
where one spin has been flipped and therefore it contributes only to 
an overall dynamical phase to the state evolution.

Having in mind the use of spin chains as quantum channels, the preparation scheme 
described above does not lead to faithful entanglement transfer. The most natural 
way to perform this task is to prepare the entangled state and then let only one 
of the qubit to propagate along the spin chain (thereby following the protocol 
originally proposed by Bose~\cite{bose03} or its modified version~\cite{christandl04}  
to achieve perfect transfer with modulated chain). A detailed analysis in this 
direction was recently performed by Hartmann {\em et al}~\cite{hartmann06}. They 
considered initial maximally entangled states and used the chain to transfer the 
state of one of the two qubits, found a relation between a measurement of the 
entanglement fidelity at the fidelity of state transfer and concluded that there 
are possibility to have perfect entanglement transfer. 

If the chain is initially prepared in $| \Psi_{\pm} (t=0) \rangle$, given in 
Eq.(\ref{wavefunction}), the entanglement propagates maintaining its original 
characteristics. This is not the case if, for example, the initial states of the 
two entangled sites is of the type $|\Phi_{\pm}\rangle = (1/\sqrt{2})(00 
\rangle \pm |11\rangle )$. These states are superpositions of components 
belonging to different magnetization sectors. The entanglement propagates 
with the same velocity as before, however, under certain conditions, it turns 
out that, the propagating quantum correlations change their character. After some 
time a singlet-like entangled state propagates even if the initial state was 
not a singlet. 
A different set of initial states has also been considered. In~\cite{hamieh05}
the chain was initialized in a separable state by means of a set of projective 
measurement (in particular they considered all the spins aligned in the z-direction 
except one prepared in an eigenstate of $S_x$). The evolution can be described 
using the same approach as outlined above. The new ingredient now is the creation of 
entanglement during the dynamics. Hamieh and Katsnelson showed that the oscillations
of the entanglement wave has the same periodicity, but out of phase, with the 
oscillation of the (non-equilibrium) average magnetization. The distribution of 
entanglement in the chain has been analyzed after launching a single excitation 
from the central site of a $XX$ chain in~\cite{fitzsimons05}.
It was found that the second-order moment of the spatial extent of the concurrence 
grows much faster (with a rate increasing as $t^{5/2}$) if some disorder is present 
in the central portion of the chain (in the ordered case the increase goes as $t^2$).

\begin{figure}
	\includegraphics[width=0.95\linewidth]{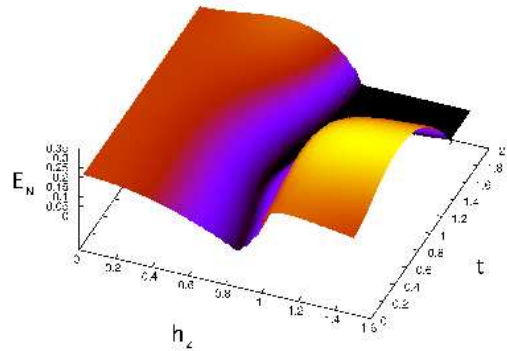}
	\caption{The nearest neighbor logarithmic negativity \(E_N\) 
	as function of  the initial transverse field \(h_z\) and time \(t\) for the anisotropy 
	\(\gamma = 0.5\).
	At short times $t \le 1.8$ the dynamical phase transition is a point of 
	reentrance for the entanglement. At later times the state remains 
	separable for values of the field above the critical value.	
	For the \(t=0\), entanglement  vanishes, as it 
	should, as \(h_z\) grows. 
	The transverse magnetization of the evolved state does not 
	show a similar critical behavior as a function of \(h_z\).
	[From \protect\cite{sende05}].} 
\label{dpt_sen}
\end{figure}
Additional features emerge in the quantum $XY$ with $\gamma \ne 0$. In this case 
the magnetization is not any longer a constant of motion (two spins can be flipped 
simultaneously). The calculations were done analytically~\cite{amico04} resorting to 
the exact calculation of the necessary set of out of equilibrium correlation 
functions~\cite{Amico-Osterloh}. 
The most notable difference in the two-site entanglement, studied 
is an entanglement production from the vacuum. 
This occurs uniformly along the 
chain and it superimposes to the entanglement wave associated discussed before. The 
velocity of propagation of the entanglement is almost independent on the anisotropy 
parameter $\gamma$. What is strongly dependent on $\gamma$ is the damping coefficient 
of the entanglement wave. As the anisotropy approaches $\gamma=1$, the Ising point, the 
wave is strongly damped and vanishes approximately after few $\sim J^{-1}$. 
Furthermore in the anisotropic case it is possible to observe a non-trivial dynamics 
when the external magnetic field is time-dependent~\cite{huang05,sende05}. 
In~\cite{sende05} the system is initially prepared in the ground state;
the evolution is analyzed after the magnetic field is (suddenly) switched off. 
Sen(de) {\em et al.} analyzed the two-site entanglement by means of the logarithmic 
negativity as shown in Fig.\ref{dpt_sen} and demonstrated the existence of a dynamical 
phase transition, not observable in the magnetization. As it can be seen in 
Fig.\ref{dpt_sen} 
at a fixed time $t$ the entanglement vanishes (and then grows again) for certain 
values of the initially applied magnetic field. For  $t \le 1.8$ the critical 
field is almost independent on the time $t$ and is $h \sim 0.8$. 
A remarkable different situation occurs for $t \le 1.8$, here a dynamical phase 
transition occurs where the entanglement vanishes for any value of $h \ge h_c$.
For values of the initial  field in the region near the phase transitions, entanglement 
behaves non monotonically with respect to temperature of the initial  equilibrium state. 

The two-site entanglement is non ergodic~\cite{sendeer}. The evolution of
two-site entanglement of a $XY$ chain has studied after a sudden 
change of the external magnetic field. The evolved state does not approach
 its equilibrium value. That is, entanglement by itself does not relax 
to its equilibrium value in the absence of further contact with reservoirs.
Therefore entanglement in such systems cannot be described
by equilibrium statistical mechanics. 

The entanglement dynamics has been studied, to a large 
extent analytically, also in the  LMG model~\cite{vidal04,latorre05}. 
They considered the 
dynamical evolution of entanglement by analyzing both the one-tangle $\tau_1(t)$ and 
the concurrence $C(t)$ for the cases in which the initial state is fully polarized 
either in the z- or in the x-direction. Because of the symmetries of the model 
and since the initial state belongs to the sector with maximal spin, $S=N/2$, both 
these quantities do not depend on the spins which are selected. This means that they 
can be expressed solely in terms of the average value of the total spin 
$\langle S^{\alpha}\rangle$ and its correlation functions $\langle S^{\alpha} 
S^{\beta}\rangle $ ($\alpha = x,y,z$). An interesting feature of this model is that its 
level spacing is finite even in the thermodynamic limit. Contrary to the expectations,
however, the Poincar\'e  time is {\em dependent} on the number of spins $N$ and this  
has important consequences on the entanglement dynamics.

In the case in which the spins are prepared in the state
$|0 \dots 0 \rangle$ analytical results can be obtained in the limit of zero 
and very large magnetic field.
In the limit of zero external field the tangle and the concurrence can be expressed as 
and 
$
	\tau_1(t) = 1 - c(t)^{2(N-1)}
$
and
$
	C(t) =(1/4)(c(t)^{(N-2)} -1 + {[c(t)^{(N-2)}- 1]^2+ [4 c(t)^{(N-2)} s(t)^{(N-2)}]^2}^{1/2})
$,
with $c(t)=\cos(4Jt/N)$, $s(t)=\sin(4Jt/N)$, and show perfect anti-correlation in time.
In the opposite limit of a strong applied field it is possible to resort to a semiclassical 
approximation of the Heisenberg equation of motion~\cite{law01}, $\dot{S}_x=2hS_y$, 
$\dot{S}_y=-2hS_x +(2J/N) [S_z,S_x]_+$, $\dot{S}_y=-(2J/N) [S_z,S_y]_+$ by noting that, 
for $N \gg 1$, the z-component of the magnetization has negligible fluctuations 
($S_z(t) \sim S=N/2$). Therefore the set of 
equations governing the dynamics of the system becomes linear and can be easily solved. 
The concurrence as a function of time for positive values of the external field evolves
as 
$
	C(t) = 1 - c_h(t)^2 - (J^2/4h_z^2) s_h(t)^2
$
with $c_h(t)^2=\cos^2(\sqrt{h(h-J)} t)$ and $s_h(t)^2=\sin^2(\sqrt{h(h-J)} t)$.
Dynamics of two-site entanglement was discussed also in the context of spin-boson like models.
In~\cite{ciancio05}  the negativity for a two-modes Jaynes-Cummings was analyzed with particular 
emphasis on the entanglement between the two bosonic modes mediated by the qubit. The relaxation 
dynamics of the entanglement, quantified trough the concurrence, in the spin-boson model was 
discussed in~\cite{lim06}

We finally mention a study where it was noticed  that the 
entanglement encoded in the states caused by the splitting 
of the degeneracy  determined by the transverse field in the quantum $XY$ model
is not preserved by an adiabatic perturbation. Separable states can become 
entangled after the field is varied adiabatically, and viceversa~\cite{Cao06}.

\subsubsection{Dynamics of the block entropy}

The dynamical behaviour of the block entropy was first considered in~\cite{calabrese05}  
in the general framework of conformal field theory and via an exact solution of the quantum Ising
model. Later D\"ur {\em et al} considered Ising models with long range interaction and 
De Chiara {\em et al}~\cite{dechiara06a} performed numerical simulations of the $XXZ$ chain. 

\begin{figure}[ht]
\begin{center}
\begin{tabular}{cc}
\includegraphics[width=0.50\linewidth]{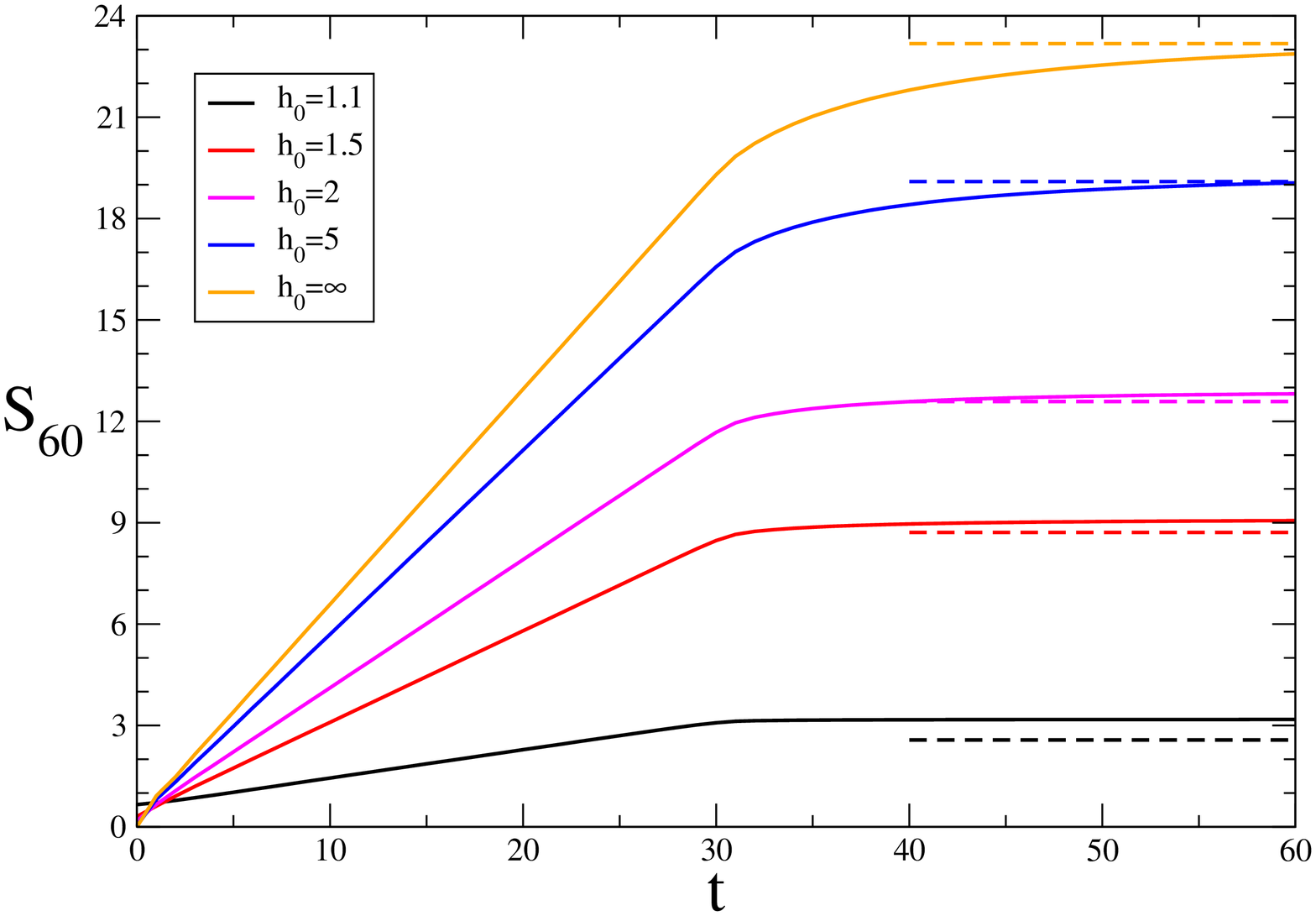}&
\includegraphics[width=0.50\linewidth]{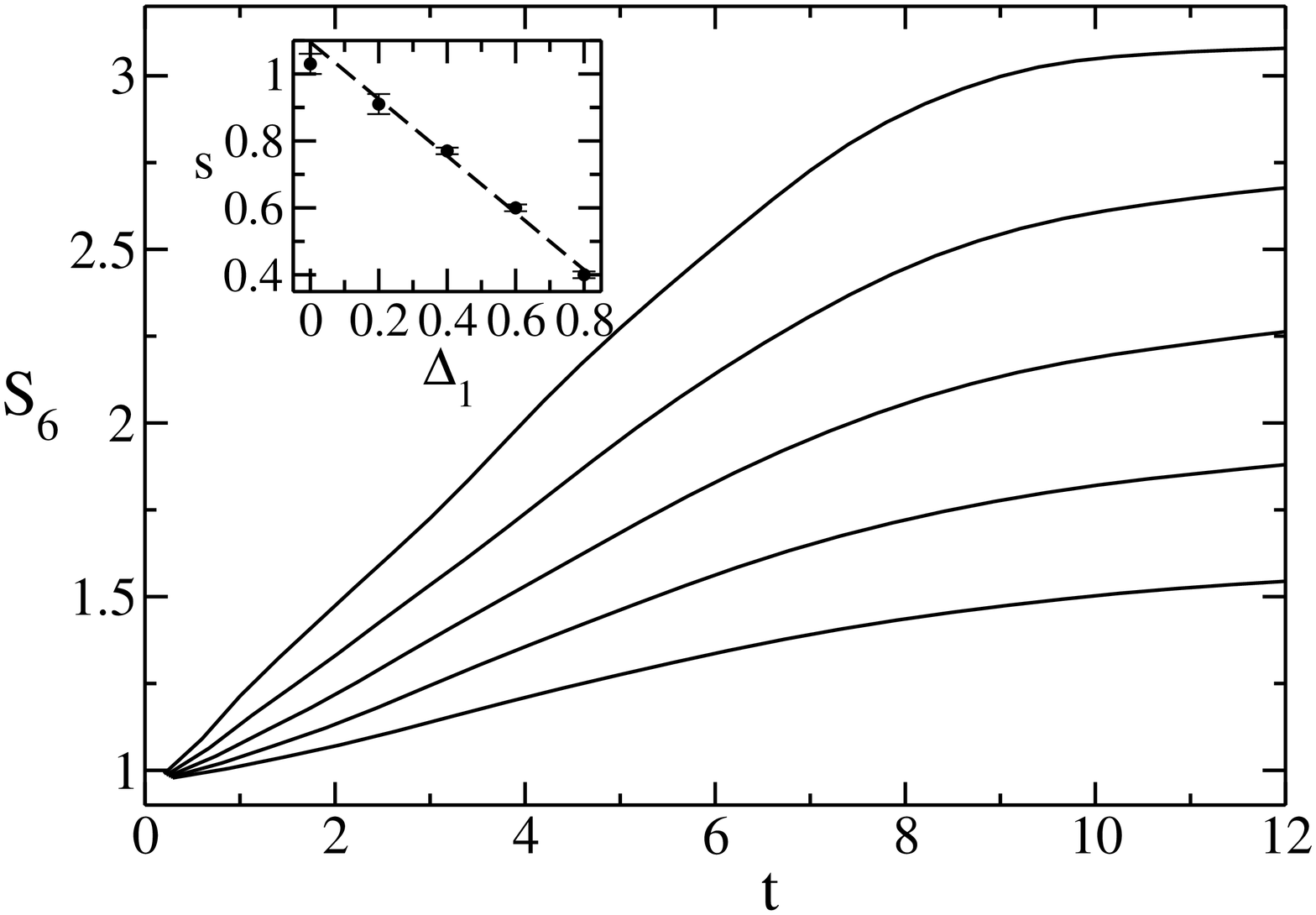}
\end{tabular}
\end{center}
\caption{Evolution of the entropy for different types of quenches in the case of 
	the Ising (left panel) [from \protect\cite{calabrese05}] and Heisenberg (right 
        panel) [from \protect\cite{dechiara06a}]models.  The block are of 60 and 6 sites in 
        the Ising and the Heisenberg cases respectively. In the Ising case the quench is 
	obtained by changing the external magnetic field from $h_0$ to $h_1=1$.
	In the Heisenberg model the anisotropy parameters is used instead, 
	$\Delta_0=1.5$ while $\Delta_1=0.0, 0.2, 0.4, 0.6, 0.8$. The time is expressed in 
        units of the spin wave velocity. The inset of the left panel shows the behaviour 
        of the initial slope as a function of the final value of the anisotropy}
\label{entprod1}
\end{figure}

\begin{figure}[ht]
\begin{center}
\includegraphics[width=0.50\linewidth]{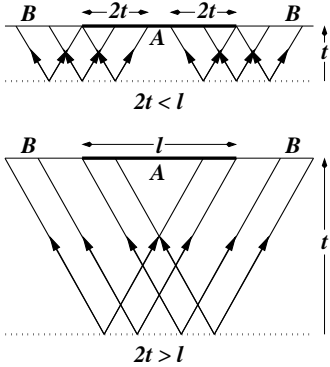}
\end{center}
\caption{Schematic representation of the dynamics of block entropy. Entangled 
particles are emitted from the region $A$, they will contribute to the block 
entropy as long as one of the two particles ends in the region $B$. 
[from \protect\cite{calabrese05}]}
\label{entprod2}
\end{figure}

Calabrese and Cardy showed that a quench of the system from a non critical to a critical 
point leads the block  entropy first to increase linearly in time and then to saturate.
For periodic boundary conditions and given a block of dimensions $\ell$, the time at 
which the entropy saturates is given by $t^*=\ell/(2v)$ 
where $v$ is the spin wave velocity.
\begin{equation}
 	S_{\ell}(t) \sim  
        \left\{ 
          \begin{array}{cc}
            t     &   \;\;\;\;\;  t \le t^* \\
            \ell  &   \;\;\;\;\;  t \ge  t^*
          \end{array}
          \right.
\end{equation}
Thus there is an arbitrary large entanglement entropy in the asymptotic state, contrarily 
to the ground-state case, where the entropy diverges only at critical point. An example 
of the time dependence of the block entropy for the Ising and $XXZ$ models is shown in 
Fig.\ref{entprod1}.
%
Calabrese and Cardy proposed a simple interpretation of this behaviour in terms of 
quasiparticles excitations emitted from the initial state at $t=0$ and freely 
propagating with velocity $v$. 
The argument goes as follows.  The initial state has a very high energy relative to 
the ground state of the Hamiltonian which governs the subsequent time evolution, 
and therefore acts as a source of quasiparticle excitations. Particles emitted from 
different points are incoherent, but pairs of particles moving to the left or right from
a given point are highly entangled (see Fig.\ref{entprod2}).   
%
%
The idea is that a point $x_A$ in the region $A$ will be entangled with a point $x_B$ in 
the region $B$ if a pair of entangled particles emitted at an earlier time arrive 
simultaneously at $x_A$ and $x_B$. In this picture the block entropy is simply 
proportional to the length of the interval where this can happen. The saturation is reached  
when the most distant quasiparticles (which started in the middle of the block) have 
already got entangled. In the presence of disorder the dynamical behaviour is strikingly
different. This picture applies in the more general context of the study of dynamical correlation 
functions after a quench as recently discussed in~\cite{calabrese06}.

Possible evidence of localization in the block entropy has been discussed 
in~\cite{dechiara06a}. 
The case of Ising models in zero magnetic field with long range interaction the dynamics 
was analyzed in the case of factorized initial states~\cite{dur05}.
The general picture is confirmed also in this case although in the short time limit 
additional oscillations appear probably due to the various different phases related to 
the different couplings.

We finally mention that the entropy in the case of a finite time quench has been considered 
in \onlinecite{cherng06} and by \onlinecite{cincio07}. Very recently rigorous bounds for the time evolution of the block 
entropy were obtained by Osborne and Eister~\cite{eisert06} and by Verstraete 
{\em et al}~\cite{verstraete06b}.

\subsubsection{Chaos \& dynamics of entanglement}

The evolution of entanglement is very sensitive to the different properties of the 
spectrum in the transition to chaos as in the case of a quantum computer with 
imperfections.  A small inaccuracy in the coupling constants induces errors in 
gating or a unwanted time evolution if the Hamiltonian cannot be 
switched exactly to zero.  If the  imperfection strength increases, new phenomena 
occur and above a certain threshold the core of the computer can even ``melt'' 
due to the setting in of chaotic behavior~\cite{georgeot00,benenti01}. 
In addition to the understanding of the behaviour of the fidelity as an indicator 
to measure the stability of the  quantum memory (see \onlinecite{Gorin.review06} for a review), a  more complete characterization 
has recently included the behavior of entanglement on approaching the transition to 
the chaos either by considering a dynamics of a 
disordered~\cite{montangero03,montangero06,sendedis} or time-dependent Ising 
model~(\onlinecite{lakshminarayan05}; see  \onlinecite{prosen07} for a recent 
review on the dynamical complexity analysed on the kicked Ising model) 
and by studying 
the dynamics of a quantum map~\cite{rossini04,bettelli03,miller99,bandyopadhyay02,ghose04}.    
In particular the disordered Ising model has been proposed~\cite{georgeot00} to describe the 
hardware of a quantum computer, in which system imperfections generate 
unwanted inter-qubit couplings and energy fluctuations. Three  different regimes 
appear depending on the variance of the fluctuations of the exchange couplings. 
At a first critical value $ \delta J_c$ the systems crosses from the perturbative to the 
chaotic regime, while at the second point$ \delta J_E$ the system goes into an ergodic 
state characterized by a Gaussian local density of states. All these dynamical 
regimes have been detected in the dynamics of entanglement 
in~\cite{montangero03,montangero06}. 

We finally mention the interesting connections~\cite{boness06} found for the properties 
of the entanglement in a Heisenberg chain with pulsed magnetic field with the localization 
behaviour of the quantum kicked rotator.

\subsection{Generation of entanglement}
\label{gen}

The Hamiltonians discussed so far can be also used to generate particular entangled
states. 

The simplest case is to consider the $XX$-model in the sector in which 
only one spin is flipped. The fact that the Hilbert space is spanned by the basis set 
$| {\boldsymbol l}\rangle \rangle$ hints to the fact that W-states can be generated. 
For short chains, $N=3,4$, generalized W-states of the type
$
| W \rangle = \frac{1}{\sqrt{4}}
[| 1000 \rangle -i  | 0100 \rangle -| 0010 \rangle - i| 0001 \rangle ]
$
appear at discrete times is the initial state of the chain is fully 
polarized~\cite{Wang01}. 
This simple scheme cannot be extended easily to an arbitrary number of qubits. The reason 
is related to the fact that for generic terms the various frequencies appearing in the 
dynamical evolution of the state are not commensurate.

An interesting example of generation of entangled states is that of 
cluster states~\cite{briegel01} 
which has been shown to be essential for the one-way 
quantum computation~\cite{raussendorf01b,raussendorf03} 
(see ~\onlinecite{proceedhein06} for a review).
Remarkably, they can be generated by certain spin Hamiltonians. 
In fact, the dynamical evolution of an Ising model in zero magnetic field, $H = \sum_{i,j} J_{ij} 
\sigma_i^{z}\sigma_j^{z}$, is equivalent to a series of conditional phase shifts. In the 
case of a regular lattice this Hamiltonian generates  cluster states which are readily 
generalized to graph states for an Ising model defined on a simple graph.
It is easy to realize that the evolution operator at time $Jt = \pi/4$ 
can be written as 
$
	U(\pi/4) = \prod_{i,j} \frac{1}{2}(1+ \sigma_i^{z}\sigma_j^{z})
$  
If the initial state is a product state with all the spin pointing along the x-direction, 
than at time $Jt=\pi/4$ the state is a graph state characterized by a maximal bipartite 
entanglement and by maximal entanglement persistence. 
An example of cluster state generate by an Ising chain with nearest-neighbour couplings,
for $N=4$, is given by
$
| \Psi \rangle_{cl} = | 0000 \rangle + | 1100 \rangle +| 0011 \rangle -| 1111 \rangle 
$
(for $N=2,3$ cluster states coincide with Bell and GHZ states respectively).
The Ising model is not the only case when graph states can be created. Borhani 
and Loss~\cite{borhani05} showed how to generate them using Heisenberg interaction 
while Clark {\em et al}~\cite{clark05} 
considered the $XX$-model Hamiltonian.

An appropriate tailoring of the initial state or the spin graph can lead to the
production of properly tailored entangled states~\cite{koniorczyk05}. 
In a $XX$-model in a star network it is possible to control  the concurrence 
between two spins by varying the initial state of the central spin only. Such a spin mediates the interaction between the outer ones as also discussed in~\cite{hutton04}. 
The pairwise entanglement can be  maximized by choosing all the outer spin down and 
the central spin up. These dynamically generated states saturate the CKW 
inequality Eq.(\ref{ckw})~\cite{Coffman00} and hence have the maximal possible two-site 
entanglement.
Koniorczyk {\em et al} analyzed also the concurrence of assistance, i.e. the maximum 
amount of entanglement which could be concentrated on two qubits by means of 
assisted measurements on the rest of the system. Depending on the initial system, 
the behaviour of the concurrence and the concurrence of assistance can be quite different. 

\subsection{Extraction of entanglement}

The entanglement naturally contained in a many-body state can in principle be extracted 
and therefore used for quantum information processing. This means that such entanglement can be
transferred to a pair of particles and subsequently used, in principle, for quantum 
computation or to test the violation of the Bell's inequalities. De Chiara {\em  et al} proposed 
a scheme of entanglement swapping by means of scattering  of a pair of particles with a
spin chain or an optical lattice~\cite{dechiara06b}. To this end one sends simultaneously 
a pair of probe particles toward the entangled spin chain in such a way that each probe 
interacts with a different spin. The entanglement between the probes has been extracted 
from the spin chain and cannot exist without entanglement in the chain. This is a genuine 
non local process between the two probes like in the case of entanglement swapping.
In practice the scattering interaction between probes and spins in the chain must be 
capable of (at least
partially) swap their state. This is the case of very common interactions like the Heisenberg or 
the $XY$ ones. 
The most natural way to extract entanglement from entangled electron spins in solids would be to 
scatter pairs of neutrons off the solid. Another possibility would be to realize
Hamiltonians of entangled spin chains or ladders that 
can be realized using trapped cold 
atoms~\cite{garcia-ripoll03,garcia-ripoll04,duan03} 
and, as probe particles one can use marker qubits~\cite{calarco04}.

\subsection{Time evolution of the entanglement in Gaussian states}
\label{harmdyn}
The dynamics of Gaussian states has first been discussed in~\cite{Eisert04}, where 
essentially two aspects have been highlighted: the creation of entanglement
from an initially disentangled state
and the propagation of an entangled state
on top of a disentangled background, 
both induced by Hamiltonian dynamics. 

The initially entangled two-mode state is characterized by the 
{\em squeezing parameter} $r$, which appears in the co-variance matrix
as $V_{\xi_\alpha,\xi_\alpha}=\cosh r$ for all phase space variables
of the zeroth and the first oscillator mode, whereas
$V_{q_0,q_1}=V_{p_0,p_1}=\sinh r$; in absence of further off-diagonal elements,
all other diagonal elements are equal to $1$.
Two different types
of nearest neighbor couplings of the oscillator have been considered: 
ideal springs obeying Hooke's law, and its Rotating Wave Approximation (RWA). 
In this approximation both 
the kinetic and potential energy terms assume the same form~\cite{Eisert04}.
The RWA conserves the number of bosonic excitations in the system, 
whereas the ideal spring coupling does not.
The initial entanglement is encoded in a zeroth oscillator,
itself decoupled from the chain of oscillators, and one oscillator
within the chain as described above. In this case, the Hamiltonian for the 
harmonic lattice (with an appropriate choice of the matrix $\mathbb{U}$, see 
Eq.(\ref{HamEisert})) reads
\beq\label{Hspring}
\hspace*{-0.2cm}H=\frac{1}{2}\left(q_0^2+p_0^2+\sum_{k=1}^N (q_k^2+p_k^2+ K_1(q_{k+1}-q_k)^2)
\right)
\eeq
After discarding the terms
$\hat{a}\hat{a}$ and $\hat{a}^\dagger\hat{a}^\dagger$, 
in RWA, the~\eqref{Hspring} can be written (up to a constant) as~\cite{Eisert04}
\beq\label{Hrwa}
H_{\rm RWA}=\hat{a}^\dagger_0\hat{a}^{}_0
+(1+K_1)\sum_{k=1}^N \hat{a}^\dagger_k\hat{a}^{}_k 
- K_1(\hat{a}^\dagger_{k+1}\hat{a}^{}_k+\hat{a}^\dagger_k\hat{a}^{}_{k+1}) 
\eeq
For both  
(\ref{Hspring}) 
and (\ref{Hrwa}) the time evolution of 
the position and momentum operator can be evaluated analytically.

The initially entangled state of two oscillators --
a decoupled and one harmonic oscillator within a periodic chain -- 
is released at time $t=0$ 
into the background of initially disentangled oscillators, 
all being prepared in their ground state. 
In regard to the pairwise entanglement we can observe that 
the n-th and the zeroth oscillators will become entangled 
after a finite time $t_c(n)$, which is essentially given by the velocity of
sound of the underlying model Hamiltonian. After this 
entanglement wave arrives, the $n$-th oscillators' entanglement exhibits
damped oscillations with the characteristic time scale of the model.
For the model (\ref{Hspring}), the velocity of sound has been 
determined as $v=K_1/\sqrt{1+2K_1}$.
Within the RWA this velocity appears enhanced: $v_{\rm RWA}=K_1$. 
The height ${E_N}_f$ of the first local maximum in the 
logarithmic negativity has been used to define the transmission 
efficiency $T_{\rm eff}$ for entanglement in the chain as 
$T_{\rm eff}={E_N}_f/{E_N}_i$,
where ${E_N}_i$ is the logarithmic negativity initially
prepared between the oscillators zero and one.

In both models ${E_N}_f$ is observed to saturate when cranking up $r$ and hence the initial entanglement.
As expected, the saturation value is a decreasing function of the 
distance $n$. The main difference is that for the generic coupling
the saturation value also depends on the coupling strength $K_1$
and is the smaller, the stronger is the coupling. Within the
RWA it is independent of the coupling strength.
Given the initial logarithmic negativity
${E_N}_i=-2 r/\ln 2$, the transmission efficiency
behaves differently. For the generic model it is zero for
zero squeezing, has a maximum at a finite squeezing $r_{max}$ and anishes for $r\to\infty$.
In sharp contrast, in RWA $T_{\rm eff}$ is a monotonically decreasing function
of $r$.
Interestingly enough, for the model in RWA, the oscillator frequencies and
coupling strengths can be optimized as to have perfect
transmission from one end of an open chain of $M$ sites
to the other end. This is achieved by virtue of
$
\mathbb{U}_{j,j+1}=\mathbb{U}_{j+1,j}=K_1\sqrt{n(M-n)}$, $\mathbb{U}_{j,j}=1$,
$\omega_n=1-K_1\sqrt{n\bar{n}}-K_1\sqrt{(n-1)(\bar{n}+1)}$
where we defined $\bar{n}=M-n$.
The same turns out to be impossible for the generic model
and $M>2$. The only possibility is to choose the couplings and
frequencies as in the RWA case in the limit of $K_1\to 0$ (where 
the RWA is exact).
In this way, the transmission efficiency can be pushed 
arbitrarily close to perfect transmission but with a transmission speed
tending to zero.  
The transmission of quantum information has been found to 
be robust against noise in $K_1$ 
for both models~\cite{Eisert04}. 

Another effect occurring in the generic model is
entanglement creation from a disentangled state, which is not an eigenstate.
This can be realized by an abrupt change of the coupling strength. 
As in the spin case (see Section \ref{prop}) no entanglement 
creation can be generated within the RWA. 
In an open chain the oscillators at the end points become entangled
after half the time a signal needs to travel through all the chain.
This indicates that this initial pairwise entanglement is mediated from
the center of the chain. Actually this is the first signal that possibly
can arrive when essentially pairwise entanglement is created or
contributes to the eventual pairwise entanglement of the boundary oscillators.
Raising a finite temperature for a thermal state, the arrival time
is slightly enhanced and the signal is subject to damping.
However, this effect turned out to be much more sensitive to
noise in the coupling strength than the propagation of 
quantum information~\cite{Eisert04}.


\section{Conclusions and outlook}
\label{concl}
During the last years it became evident that quantum information offers 
powerful instruments to grasp the properties of complex many-body systems. 
This is the reason why this area of research is undergoing an impressive 
expansion. In this review we touched only one particular aspect of this 
activity, namely the properties of entanglement in many-body systems. 

As mentioned in the introduction, there are several important aspects which,
for space limitations, were ignored. In this respect we should certainly 
mention the increasing interest towards the optimization of numerical 
simulations of quantum systems. There were already a number of breakthroughs
(see also the introduction) that, for example, lead to the development of the 
time-dependent DMRG. The design of 
variational methods to study the ground state and finite temperature properties 
of many-body Hamiltonians has been exploited in numerous interesting works.  
We already mentioned in the introduction the large body of activity on quantum 
state transfer in spin chains. Here the knowledge of the low-lying excitations 
of the chain (spin-waves) has helped in finding new quantum protocols. 
More is expected to come in the next future.

It is very tempting, although very difficult, to try to envisage the 
possible evolution of this line of research. The study of the topological 
entanglement entropy is a very important aspect that will be investigated 
in the coming future. Adiabatic quantum computation and the one-way quantum 
computation will benefit from the study of entanglement in complex systems.  
The study of the topological entanglement may also have remarkable spin off to 
understand many puzzling phenomena in condensed matter physics amoung which   
high $T_c$ superconductors  and heavy fermions are of paramount 
importance~\cite{coleman}. Furthermore will the analysis of new measures for
multipartite entanglement provide with additional insight necessary for
understanding the role of entanglement in such complex phenomena, which might
also reveal valuable information, e.g. on how to simulate these systems 
efficiently on a computer.

Many interesting results have been already obtained, but the overlap between 
quantum information and  many-body physics is not yet fully unveilled. 
There is a number of open questions which provide a fertile ground for a 
field of lively exciting research.

\acknowledgments

We acknowledge very fruitful discussions with J. Anders, G. Benenti, C. Brukner,
P. Calabrese, A. Cappelli, G. De Chiara, P. Facchi, G. Falci, A. Fubini, 
V. Giovannetti, H. Frahm, L. Heaney, J. Hide, F. Illuminati, D. Kaszlikowski, 
A. Maugeri, V. Korepin, A. Montorsi, S. Montangero,      
G. Palacios, G.M. Palma, E. Paladino, D. Patan\`e, F. Plastina, S. Pascazio, 
M. Rizzi, D. Rossini, G. Santoro, J. Siewert, W. Son, V. Tognetti, G. T\'oth, P. Verrucchi, 
L. Viola, P. Zanardi, and A. Zeilinger. 
We are very grateful to A. Anfossi, P. Calabrese, A. Fubini, V. Giovannetti, 
F. Illuminati, E. Jeckelmann, V. Korepin,  A. Montorsi, I. Peschel, 
A. Silva, P. Verrucchi, and L. Viola 
for all their comments and suggestions to improve the manuscript.  
This work has been supported by European Community through grants 
EUROSQIP (R.F. and L.A.), 
SPINTRONICS (R.F.),  by Ministero dell'istruzione, Universita' e Ricerca (MIUR) through PRIN (R.F. and 
L.A.), by the Royal Society, the Wolfson Foundation and by  Engineering and Physical 
Sciences Research Council, as well as the National Research Foundation 
                     and Ministry of Education, in Singapore (V.V.). 
The present work has been performed within the "Quantum Information" research 
program of Centro di Ricerca Matematica ``Ennio De Giorgi'' of Scuola Normale 
Superiore (R.F.).

\bibliographystyle{apsrmp}

\end{document}